# Optical trapping with structured light: A review

**Yuanjie Yang [a,*], Yu-Xuan Ren [b,*], Mingzhou Chen,[c,*] Yoshihiko Arita[c,d] , Carmelo Rosales-Guzmán [e,f,*]**

[a]University of Electronic Science and Technology of China, School of Physics, Chengdu, China
[b]Department of Electrical and Electronic Engineering, The University of Hong Kong, Pokfulam Road, Hong Kong SAR 999077, China
[c]SUPA, School of Physics and Astronomy, University of St Andrews, St Andrews, KY16 9SS, UK
[d] Molecular Chirality Research Center, Chiba University, Chiba, Japan
[e]Centro de Investigaciones en Óptica, A.C., Loma del Bosque 115, Colonia Lomas del campestre, C.P. 37150 León, Guanajuato, Mexico
[f]Wang Da-Heng Collaborative Innovation Center for Quantum manipulation & Control, Harbin University of Science and Technology, Harbin, China.

E-mail: dr.yang2003@uestc.edu.cn; yxren@ustc.edu.cn; mingzhou.chen@st-andrews.ac.uk; carmelorosalesg@hrbust.edu.cn

**Abstract**. Optical trapping describes the interaction between light and matter to manipulate micro-objects through momentum transfer. In the case of 3D trapping with a single beam, this is termed optical tweezers. Optical tweezers are a powerful and non-invasive tool for manipulating small objects, which have become indispensable in many fields, including physics, biology, soft condensed matter, amongst others. In the early days, optical trapping were typically used with a single Gaussian beam. In recent years, we have witnessed the rapid progress in the use of structured light beams with customized phase, amplitude and polarization in optical trapping. Unusual beam properties, such as phase singularities on-axis, propagation invariant nature, have opened up novel capabilities to the study of micromanipulation in liquid, air and vacuum. In this review, we summarize the recent advances in the field of optical trapping using structured light beams.

## 1 Introduction

The light-matter interaction has a long history both in physics and astronomy. About 400 years ago, Kepler observed the deflection of a comet's tail away from the sun, which may constitute the first reported conjectures of the radiation force[1-3]. In the 1700s, John Michell attempted to measure radiation pressure[4], while Euler hypothesized that light beams induce pressure on illuminated bodies[5]. In the 1800s, Maxwell predicted that light was an electromagnetic wave[6], which was confirmed by the first demonstration of a radiation force originating from thermal light sources by Lebedev[7,8] and Nichols *et al*[9] in 1901. We now know that light beams can be considered as a large collection of photons, each carrying a quantized amount of momentum, which can be transferred to matter. However, as derived by Poynting in 1906, the radiation pressure is so minute that it only affects for small bodies[10]. Shortly afterwards, Mie[11] and Debye[12] proposed exact physical models to calculate scattering force and radiation pressure of light in 1908 and 1909, respectively. At that stage, no one could imagine any practical value of radiation pressure with it being too weak to overcome frictional forces in most circumstances. For a considerable time, researchers focused their attention on the use of radiation pressure in space, e.g., solar sail propulsion systems[13] due to the absence of appreciable friction in space.

This situation did not change significantly until the invention of the laser in 1960. In the years following this discovery, Ashkin demonstrated, for the first time, laser trapping of micrometer-sized dielectric particles with two counter-propagating beams[14]. To get stable three-dimensional (3D) confinement of particles, the scheme of two weakly focused beams with opposing radiation pressure was adopted. Optical trapping underwent a revolution after Ashkin et al found that even a single, tightly focused laser beam can form a 3D stable optical trap – optical tweezers[15]. Since then, the broad field of optical manipulation or the so-called "micromanipulation" has found a tremendous range of applications in many fields, such as biomedicine[16-18], physics[19-21] and chemistry[22], and has been noted in three Nobel Prizes in Physics: firstly in 1997 for the development of methods to laser cooling and trapping atoms, for the achievement of Bose-Einstein condensation in 2001 and for optical tweezers and their application to biological systems in 2018.

It is noted that although the annular (high-order LG) beam had ever been used to study the self-focusing and self-trapping of a laser beam in artificial Kerr media[23,24] in 1981, the fundamental (or $TEM_{00}$) transverse Gaussian mode of a laser beam was exclusively used for optical trapping experiments in the early decades of laser development. It was only in the last two decades that structured light beams have been adopted widely in optical tweezers. Structured light beams have added new dimensions and functionalities to optical manipulation as well as provided new insights into light-matter interactions where trapped particles can act as probes of structured light fields. Optical trapping and structured beams therefore, help each other to understand and improve these two important areas. Structured light beams can be of two distinct forms, either scalar or vector. In the scalar form, a structured light beam can be tailored by its amplitude and phase while polarization is only modified in the vectorial form of the beam. Structured light beams have been widely used in optical manipulation, due to their unique properties, such as optical vortices carrying orbital angular momentum[25] (OAM) and propagation invariant beams [26].

Undoubtedly, the ability to tailor the optical properties of a trapping beam is crucial in the development of novel optical trapping techniques. Thus far, structured light beams with customized phase and amplitude have been successfully applied to drive the optical transport of particles in 3D trajectories by exerting optical forces arising from high intensity and phase gradients. Recently, there are several excellent review articles on this topic[27-33], including the optical pulling force[28], optical transport of small particles[29], optomechanics with levitated particles[30], acoustic and optical trapping for biomedical research[31], and so on[32]. More recently, the advanced optical manipulation using structured light was reviewed[33], which focused on the manipulation of transparent dielectric particles. In this manuscript, we review the breadth of structured beams and discuss the recent advances in optical manipulation employing structured beams, notably for both scalar and vectorial forms. We provide an overview of seminal contributions that have changed the landscape of optical tweezers, with an extensive reference list. This emerging field is being continuously and rapidly reshaped by new approaches, and we hope this paper will appeal to a broad audience with an interest in optical manipulation techniques. Our aim is to offer an up-to-date status of the field of optical manipulation with structured light.

## 2 Principle of optical tweezers

Optical tweezers are a powerful technique to hold and move microscopic particles or biological specimens with a single tightly focused laser beam, akin to normal tweezers[15]. Here, we briefly describe the principle of optical tweezers. A more detailed discussion can be found in Refs [2, 21, 34-39] and references therein.

### *2.1 Optical gradient and scattering forces*

Depending on the relative size of spherical particles to the laser wavelength, optical forces can be described in three regimes: the Rayleigh regime[40], the intermediate regime[41] and the Ray optics regime[42]. The size parameter of the particle $\xi$ is defined as $\xi = k_m a$, where $k_m = 2\pi n_m/\lambda_0$, $\lambda_0$ is the wavelength of the trapping beam in vacuum, $a$ is the radius of the spherical particle. The refractive index of the particle and the surrounding medium are $n_p$ and $n_m$ respectively. When $\xi \gg 1$, it is in the ray optics regime where the force can be described by a ray optics model. When $\xi \ll 1$ and $\xi * n_p/n_m \ll 1$, it corresponds to the Rayleigh regime where the particle can be approximated as a dipole. For particles of size between the above two, this is the intermediate regime where the Lorenz-Mie theory can be used to investigate the optical force.

First, we use the simple ray optics model to explain how optical tweezers work. Figure 1(a) shows a laser beam with two light rays (white arroes a and b) passing through a dielectric spherical particle located off-axis of the laser beam. The light rays will change their propagation directions due to the refraction, resulting in a change in their momentum. As shown in Fig.1 (a), the central portion of the beam with higher intensity (indicated by a thick arrow a, is refracted to the left, which means a change in the laser's momentum to the left. Based on the conservation of momentum at the particle boundary, the particle will feel a momentum kick to the right and therefore has a force $F_a$ towards the center of the beam (dotted line). Analogously, the light ray b with lower intensity will change its momentum to the right and exert an optical force $F_b$ on the particle away from the beam centre. Since the light ray represented by **a** is much stronger than the ray represented by **b**, the net force will push the particle to the right. Conversely, if the particle is located on the right side of the beam axis, the optical force will push it to the left. As such, the light field intensity gradient always causes a gradient force ($F_{grad}$) on the particle towards the maximum intensity of the beam. Besides, the rays reflected from the particle surface can produce forward scattering forces ($F_{scat}$) along the beam propagation direction. Figure 1(b) shows the longitudinal gradient force pushing the particle down towards the focal plane. and *vice versa,* in a highly focused laser beam. Therefore, the net force pushes the particle to the focus of the beam, as shown in Fig.1(c). In the case of $F_{grad} >> F_{scat}$, a stable three-dimensional trap can be formed in a tightly focused laser beam spot.

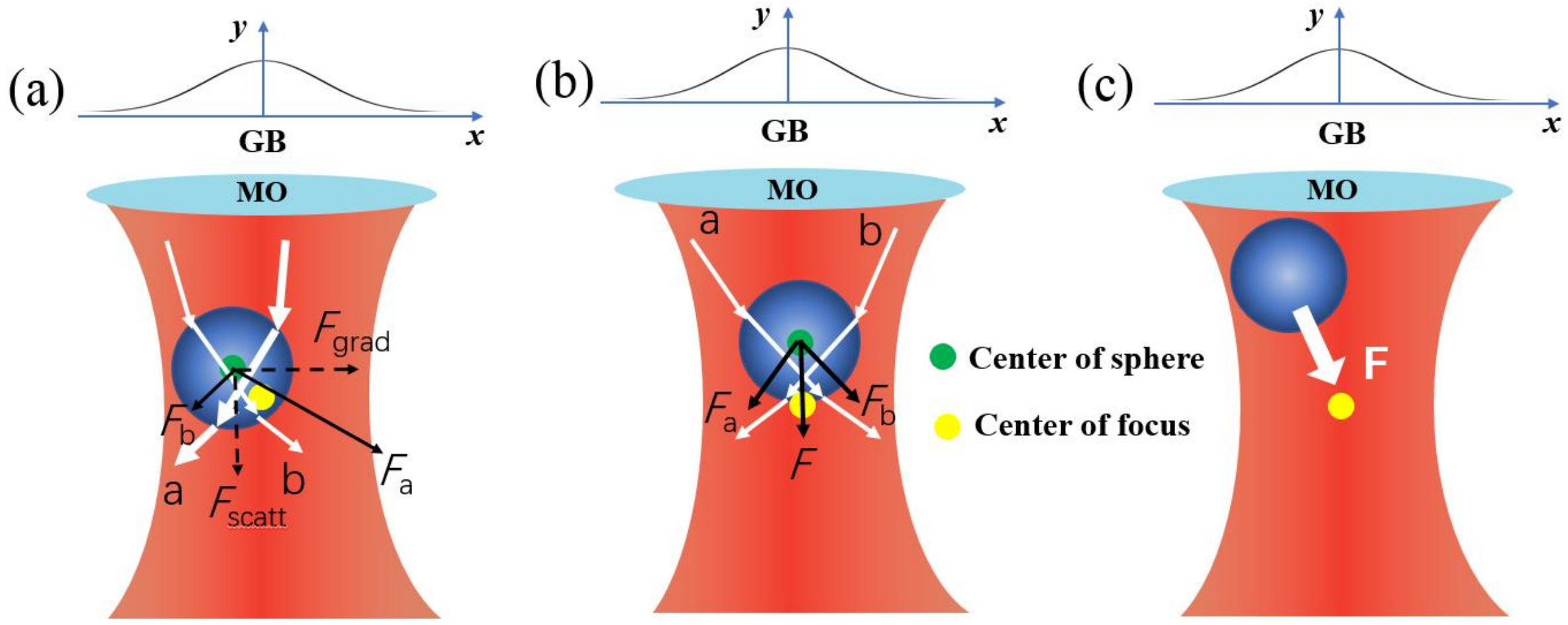

**Fig. 1** Schematic diagram of optical tweezers. (a) when the particle is away from the center of the beam focus, (b) when the particle is slightly above the center of the beam focus, and (c) net force acting on the dielectric sphere. $F_a$ and $F_b$ are the forces produced by the refracted rays a and b, respectively. $F_{grad}$ and $F_{scat}$ denote the gradient force and scattering force, respectively. GB: Gaussian beam; MO: Microscope Objective.

In the Rayleigh regime, the ray optics is not applicable to explain the optical forces as the particle size is smaller than the wavelength of light. Here, the particle needs to be considered as a point electric dipole and the optical forces can be written as[42]

$$\langle F\rangle = \frac{1}{4}\mathrm{Re}(\alpha_p)\nabla|E|^2 + \frac{\sigma(\alpha_p)}{2c}\mathrm{Re}(E\times H^*) + \sigma(\alpha_p)c\nabla\times\left(\frac{\epsilon_0}{4\omega i}E\times E^*\right), \qquad (1)$$

where, $\alpha_p$ is the polarizability of the particle and *E*, *H* are the electric field and magnetic field, respectively. $\sigma(\alpha_p)$ denotes the total particle cross-section. The first term in Eq. (1) indicates the gradient force ($F_{\mathrm{grad}}$), and the last two terms represent the scattering force ($F_{\mathrm{scat}}$). The second term in Eq. (1) is normally called the scattering force, and the third term, so-called spin curl force [42].

Equation (1) indicates that each of these three optical forces on a particle depends on the polarizability, which can be expressed as [42]

$$\alpha_p = \frac{\alpha_0}{1 - i\alpha_0 {k_0}^3/6\pi\epsilon_0}, \tag{2}$$

where $\alpha_0 = 4\pi {n_m}^2 r^3 \epsilon_0 \left(\frac{\eta^2-1}{\eta^2+2}\right)$ is the Clausius-Mossotti relation, *r* is the particle radius, $\eta = n_p/n_m$ is the relative refractive index of the particle to the surrounding medium. Therefore, we can play with the polarizability to optimize the optical traps. Then, the optical gradient force and optical scattering force can be rewritten as[2,43]

$$\boldsymbol{F}_{grad} = -\frac{2\pi n_m r^3}{c}\left(\frac{\eta^2-1}{\eta^2+2}\right)\nabla I(\boldsymbol{r}), \tag{3}$$

and

$$\boldsymbol{F}_{scat} = \frac{8\pi n_m k^4 r^6}{3c}\left(\frac{\eta^2-1}{\eta^2+2}\right)^2 I(\boldsymbol{r})\hat{z}, \tag{4}$$

where *c* is the speed of light, $\hat{\boldsymbol{z}}$ is the unit vector along the *z* direction, $I$ is the intensity of light and *k* is the wavenumber.

Both the ray optics and the dipole theory are powerful tools to study the optical forces in their respective size regimes, which have all given good physical pictures of the trapping mechanism. However, particles with their size lying between these two regimes make these approaches no longer valid. Instead, the Lorenz-Mie theory can be used to calculate the optical forces for such particle sizes, which are the exact solutions of the Helmholtz equations[41,44]. If the particles are not spherical, or the incident beam is not a plane wave, the generalized Lorenz-Mie theory can be used [45]. There are also many other full numerical methods to calculate the optical force, such as Finite Difference Time Domain (FDTD) and Finite Element Method (FEM). For instance, the FDTD analysis allows the numerical simulation of the scattered light field over an arbitrary particle. The optical force would be the surface integral of the Maxwell stress tensor. It should be noted that non-dielectric materials, e.g., metals and semiconductors and non-spherical particles, require different approaches. Proper choice of the force evaluation method relies on particle size, geometry and the structure of the light field[45,46].

A conventional optical tweezers system is schematically shown in Fig.2. First, a collimated laser beam is expanded to overfill the back aperture of a microscope objective. A dichroic mirror ($DM_1$) reflects the laser beam to the high numerical aperture (NA) objective lens, which focuses the beam to a tightly focused diffraction limited spot inside a sample chamber for trapping. To image trapped particles on a CCD camera through $DM_1$, an LED white light source is typically used to illuminate the sample. A condenser lens collects forward scattered light from the trapped particles and projects an image onto a quadrant photodetector (QPD) using back focal plane interferometry [34]. The balanced photodetection provided by the QPD allows the precise measurement of the motion of the trapped particles. The axial position z can be measured by the sum of the intensity of the four quadrants [34].

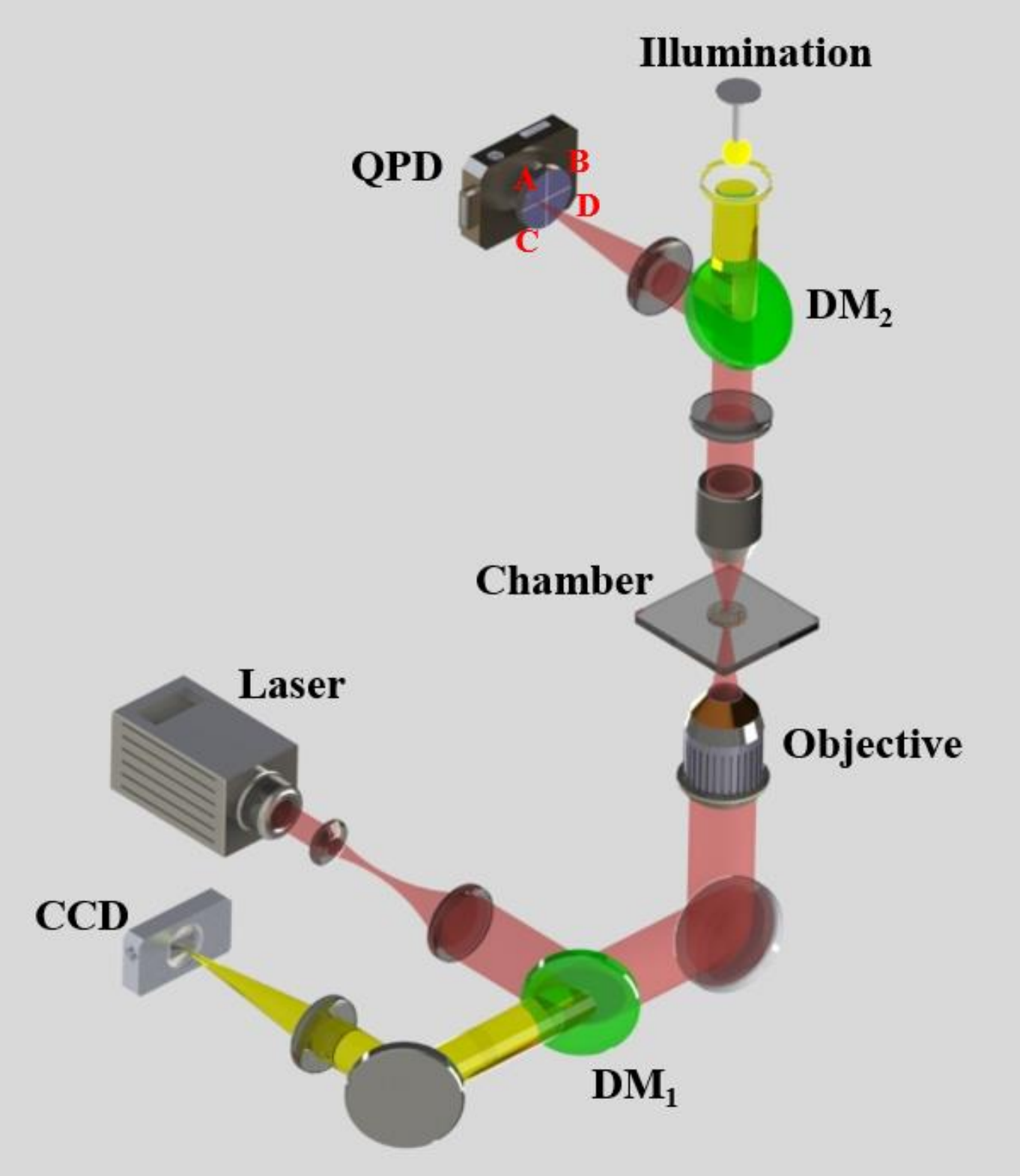

Fig.2 Experimental configuration of a conventional optical tweezers. A simple telescope is used to expand the laser beam to overfill the back aperture of the objective. The expanded laser beam, reflected by a dichroic mirror ($DM_1$), is coupled into the objective. The laser beam is focused by the objective and form an optical trap. The quadrant photodiode (QPD) is placed in a conjugate plane of the condenser, to collect the forward scattered light that reflected by the dichroic mirror ($DM_2$). The trapped particles are imaged with a CCD camera. The lateral (x, y) position of the particle can be measured by the normalized output voltage signals from the four quadrants, namely, $x = \frac{(A+D)-(B+C)}{A+B+C+D}$ and $y = \frac{(A+B)-(C+D)}{A+B+C+D}$.

### *2.2 Langevin equation*

The motion of a Brownian particle in a fluid can be described by the Langevin equation[47]. In the overdamped case, where particles are immersed in a viscous medium e.g. water, the one-dimensional motion (x-direction) of an optically trapped Brownian particle will be described by the equation:

$$\Gamma_0\dot{x} = -\kappa_0 x + F_{th}, \tag{5}$$

where the thermal random force $F_{th}$ drives the Brownian motion through collisions with surrounding molecules of the fluid, and $\Gamma_0 = 6\pi\eta a$ is the Stokes drag coefficient ($\eta$ is the viscosity of the fluid and $a$ is the radius of the particle). The trap stiffness $\kappa_0 = m\Omega_0^2$, where $m$ is the mass of the particle and $\omega_0$ is the trap frequency, determines the magnitude of the optical restoring force depending on its position relative to the trap center. For a silica particle of radius $a = 1\mu m$ ($m \approx 2 \times 10^{-8}$ kg) in water at room temperature, the resonant frequency of the harmonic oscillator is $10^3 \leq (1/2\pi)\sqrt{\kappa_0/m} \leq 10^4$ Hz for typical stiffnesses in the range of $0.5 \leq \kappa_0 \leq 50$ pN/$\mu$m.

The power spectral density (PSD) of the particle's motion can be used to characterize the trap. At equilibrium, the PSD is Lorentzian [48]. For the particle's x-displacement,

$$S_x(f) = \frac{4\Gamma_0 k_B T/\kappa_0^2}{1+f^2/f_c^2}, \tag{6}$$

where $f_c = \kappa_0/(2\pi\Gamma_0)$ is the corner or roll-off frequency that verifies the trap stiffness. The integral of the PSD in units of $m/\sqrt{Hz}$ yields the position variance or the mean square displacement of the particle, which verifies the equilibrium temperature of the particle:

$$\int_0^\infty S_x(f)\,df = \langle x^2 \rangle = \frac{k_B T}{\kappa_0}. \tag{7}$$

The position variance $\langle x^2 \rangle$ directly measured by a CCD or a QPD (see Section 2.1) can also verify the trap stiffness, which benchmarks the PSD method.

### *2.3 Optical torques for rotation*

Rotation of micro and nano-objects, caused by the transfer of angular momentum from light beams, is of great interest due to its potential applications in optically driven micromachines, motors, actuators or biological specimen. A beam of circularly polarized light carries spin angular momentum (SAM), which was derived by Poynting[49] in the early 1900s. The first experimental observation of SAM was performed by Beth[50] in 1936. In this experiment, he observed a mechanical torque on a double refracting slab due to a change in the circular polarization based on the conservation of angular momentum. It is now well understood that SAM of $\pm\hbar$ per photon is associated with the circular polarization of light, where the sign depends on its handedness. SAM of light has been used for rotation of both elongated, birefringent and absorbing particles as well as particle cluster in optical tweezers [34]. Notably, with birefringent particles[51], a maximal torque efficiency of $2\hbar$ per photon can be achieved, contrary to the $0.05\hbar$ per photon achieved with elongated particles. In addition, optically trapped micron-sized birefringent particles were rotated by a circularly polarized beam at rotating rates as high as 350 Hz in water[39] and 10MHz in vacuum[40]. Crucially, in recent time it was demonstrated that SAM can also be used for the selective 3D trapping of chiral micro and nanoparticles[52, 53]. It was shown that under appropriate conditions, a light beam with SAM can induce non-restoring or restoring forces on chiral microparticles[52]. Similar results have been also reported on the interaction of chiral nanoparticles with chiral optical fields [53].

In 1992, Allen *et al.* introduced the concept of a light beam possessing OAM[25], which is in addition to any SAM, and characterized by an azimuthally varying phase of $\exp(i\ell\varphi)$, where $\ell$ is an integer value, termed the topological charge, and $\varphi$ is the azimuthal angle. The first experimental demonstration of OAM transfer from light to matter, in the context of optical tweezers, was performed by the group of Rubinsztein-Dunlop[54,55], who demonstrated the optically induced rotation of absorptive particles. This was followed by the demonstration of the simultaneous transfer of both SAM and OAM to the same absorptive particles with circularly polarized $LG_0^\ell$ modes[56]. In these experiments, the rotation rate was shown to be proportional to the sum of SAM and OAM, i.e. the total angular momentum[57]. Other pioneering experiments include the transfer of SAM and OAM to a birefringent particle, which causes the particle to spin about its own axis as well as to rotate about the beam axis[58,59], as shown in **Fig. 3**. While these previous demonstrations were performed with particles confined in 2-Dimensions, with the particle being pushed against a microscope slide, a key experiment that demonstrated the transfer of SAM and OAM to particles in 3-Dimensions was performed by Simpson et al[60]. Besides the high-index particles, the 3D optical trapping of low-index particles has been studied as well[61,62], it was shown that the low-index particles can be trapped stably on the axis and slightly above the focal plane of a strongly focused optical vortex beam.

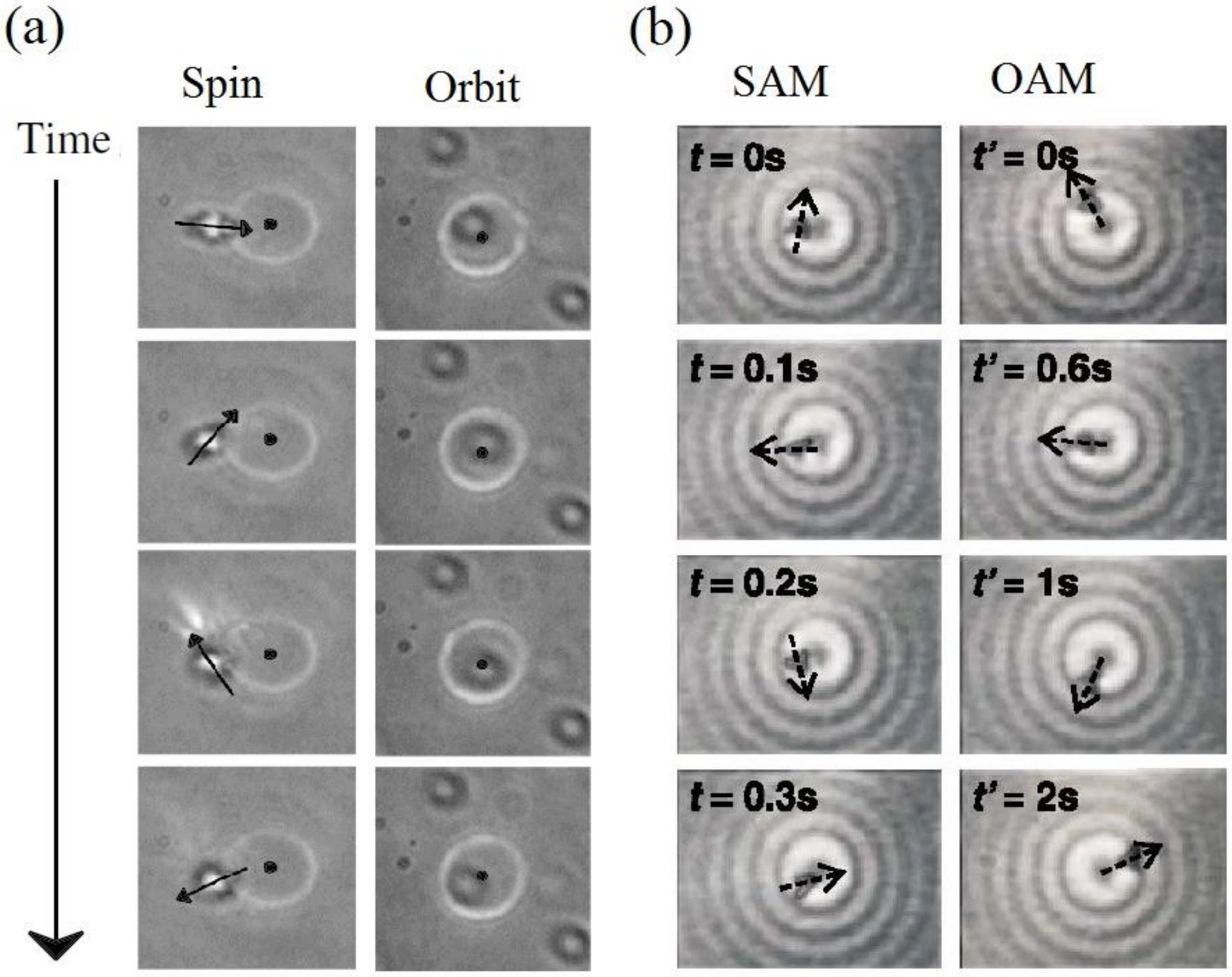

**Fig. 3** Optical trapping of birefringent microparticles that showing the transfer of OAM and SAM. (a)The trapped particle is spinning counterclockwise about its own axis (left column) and is orbiting clockwise about the beam's axis (right column) separately. Adapted from [58]. (b) The particle rotates around its own axis (left column) and the beam's axis (right column) simultaneously. Adapted from [59].

## 3 Overview of structured light beams

Light is an electromagnetic wave, as such, it can be characterized by its wavelength (color), amplitude and phase or polarization, the later associated with the direction of oscillation of the electromagnetic field in space, transverse to the direction of propagation. For unpolarized light, the direction of oscillation is random. On the contrary, for polarized light this can take distinctive forms including linear, circular or elliptical, which are the preferred configurations for beam shaping. In what follows, we will restrict to the case of polarized light, and first discuss the case of homogeneous polarization (scalar fields), followed by the non-homogeneous case (vector beams). Inevitably, any discussion about structured light fields, involves the Helmholtz equation, either in its exact or paraxial forms. As such, we will start our discussion by reviewing some of the most relevant solutions to the Helmholtz equation that have played a crucial role in the development of optical tweezers with structured light as we know them today.

### *3.1 Laguerre-Gaussian beams*

Laguerre-Gaussian (LG) modes are a set of solutions to the paraxial wave equation in cylindrical coordinates. Their normalized transverse profile can be defined as[25],

$$LG_p^\ell(\rho,\varphi,z) = \frac{\omega_0}{\omega(z)}\sqrt{\frac{2p!}{\pi(|\ell|+p)!}}\left(\frac{\sqrt{2}\rho}{\omega(z)}\right)^{|\ell|} L_p^{|\ell|}\left[2\left(\frac{\rho}{\omega(z)}\right)^2\right]\exp[i(2p+|\ell|+1)\xi(z)]$$

$$\times\exp\left[-\left(\frac{\rho}{\omega(z)}\right)^2\right]\exp\left[-\frac{ik\rho^2}{2R(z)}\right]\exp(i\ell\varphi), \qquad (8)$$

where, $L_p^{|\ell|}(x)$ is the associated Laguerre polynomials with $\ell\in\mathbb{Z}$ and $p\in\mathbb{N}$ are the azimuthal and radial indices, respectively. A set of the parameters $\xi(z)$, $\omega(z)$, $R(z)$ is defined as $\omega(z) = \omega_0\sqrt{1+(z/z_R)^2}$ , $R(z) = z\sqrt{1+(z/z_R)^2}$ and $\xi(z) = \arctan(z/z_R)$, respectively. Here, $z_R$ is the Rayleigh range and $\omega_0$ is the beam waist. Importantly, LG beams carry OAM $\ell\hbar$ per photon, associated with the phase term $\exp(i\ell\phi)$. This phase term results in a spiral azimuthal phase that creates an inclined wavefront. The Poynting vector has therefore a non-zero azimuthal component that is at the origin of the angular momentum. **Figure** 4 shows four examples of the transverse intensity profiles of LG modes.

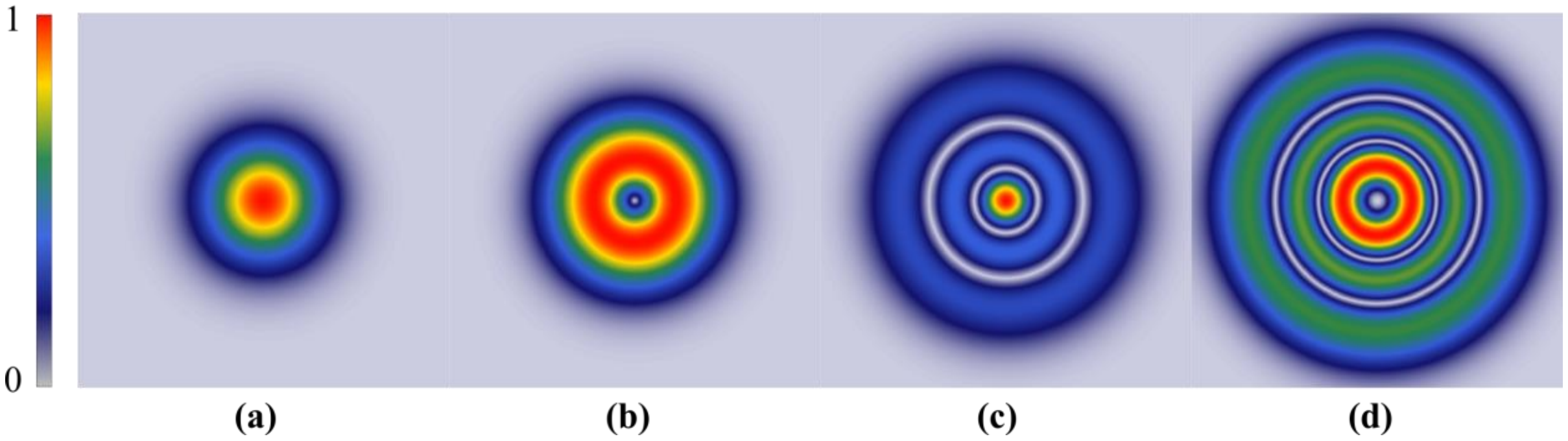

**Fig. 4** Transverse intensity profiles of Laguerre-Gaussian modes with (a) $LG_0^0$ , (b) $LG_0^1$, (c) $LG_2^0$ , (d) $LG_2^2$. The color denotes the normalized intensity distribution.

### *3.2 Bessel beams*

Another set of solutions to the Helmholtz equation in free space, when solved in cylindrical coordinates $(\rho,\varphi,z)$, is the Bessel modes[63-65],

$$B(\rho,\varphi,z) = E_0 J_\ell(k_t\rho)\exp[ik_z z]\exp[i\ell\varphi], \qquad (9)$$

where $J_\ell(x)$ represents the $\ell$-th order Bessel polynomial, and $\ell\in\mathbb{Z}$ is associated with orbital angular momentum $\ell\hbar$ per photon carried by the beam. Moreover, $k_t$, $k_z$ are the transverse and longitudinal components of the wave vector **k**, respectively, obeying the relations $|\boldsymbol{k}| = k = 2\pi/\lambda$ and $|\mathbf{k}|^2 = k_t^2 + k_z^2$.

A more intuitive way of describing a Bessel beam is by considering these as the result of a superposition of plane waves propagating on a cone, where each of them undergoes identical phase shifts, $k_z \Delta z$ over a distance $\Delta z$. This interpretation can be observed in the angular spectrum of the Bessel beam, which takes the form of a ring or radius $k_r$ in the k-space. Therefore, the optical Fourier transform of the Bessel beams is a ring and vice versa, the optical Fourier transform of a ring will result in a Bessel beam. Therefore, Bessel beams can be generated using a ring-slit aperture[64]. Using this method, Durnin *et al* produced a Bessel beam experimentally for the first time[63]. **Figure 5** illustrates this concept schematically, in **5(a)** we show the transverse intensity profiles of a Bessel beam with topological charge $\ell = 0$ and its Fourier transform, while in **5(b)** we show that of a Bessel beam of topological charge $\ell = 1$ along with its Fourier transform.

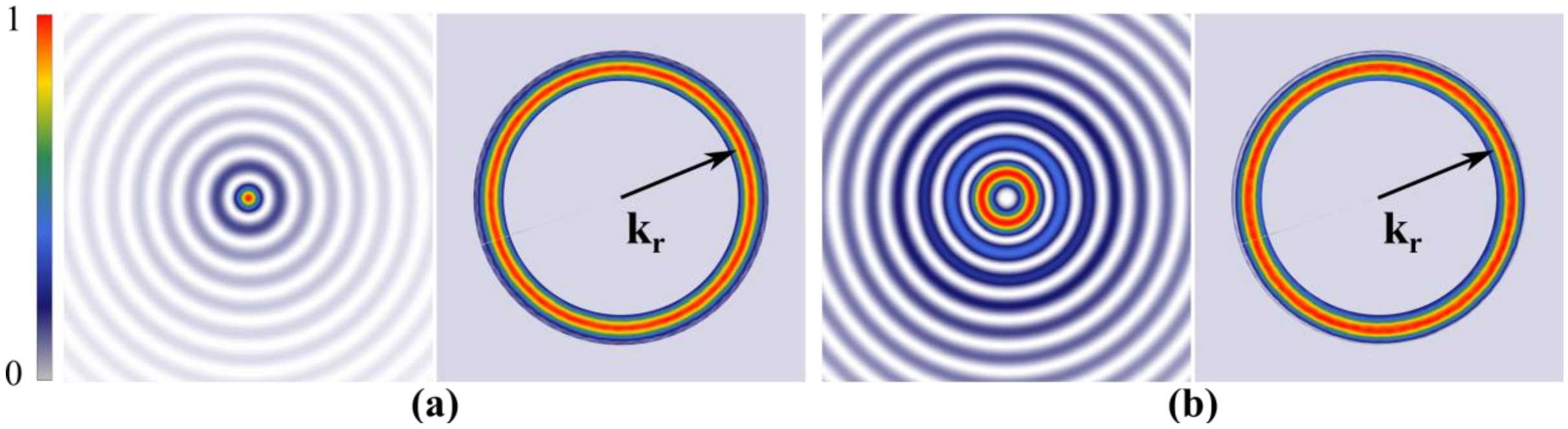

**Fig. 5** Transverse intensity profiles of Bessel beams and its Fourier transform with (a) $\ell = 0$ and (b) $\ell = 1$. The color denotes the normalized intensity distribution

Two of the most prominent properties of Bessel modes are, on one hand, their tendency to maintain an invariant intensity profile, namely, $I_B(\rho, \varphi, z \geq 0) = I_B(\rho, \varphi, 0)$ and on the other, their tendency to recovery its original form when an opaque obstruction is placed in its path[64]. Such property can be explained by invoking again the plane waves propagating on a cone approach as detailed next. When the opaque object or radius $a$ is placed in the center of the Bessel beam, some of the waves that create the beam are blocked by this object, creating a shadow region. Nonetheless, some other plane waves can pass the object unaffected, which ultimately are the ones that reconstruct the intensity profile of the beam after a certain propagation distance[65,66]. As mentioned earlier, Bessel beams can be generated by placing a slit aperture in front of a ring-slit aperture, none the less, this is a very inefficient way since most of incident beam's intensity is blocked. A far more efficient way to produce a Bessel beam is using an axicon[64]. The on-axis intensity is formed by conical wavevectors that propagate on the surface of a cone.

As we will discuss later, the zeroth-order Bessel beam has demonstrated its great relevance in optical trapping for the study of multi-particle arrangements along the beam axis[67,68]. Since Eq. (9) shows that these modes have infinitely extending sidelobes and carry an infinite amount of energy, their experimental realization is approximated by adding a Gaussian envelope. These modes are known as Bessel-Gauss modes, which at $z = 0$ take the form [69,70],

$$B(\rho, \varphi, 0) = J_\ell(k_t \rho) \exp\left[-\left(\frac{\rho}{\omega_0}\right)^2\right] \exp[-i\ell\varphi]. \tag{10}$$

### *3.3 Perfect vortex beams*

The concept of "perfect vortex beam" was proposed by Ostrovsky *et al.*[71] in 2013, whose intensity profile is independent of its topological charge. Its complex amplitude at a given propagation distance can be expressed as[71],

$$V_\ell(\rho, \theta) = \exp\left(-\frac{(\rho - \rho_0)}{\Delta\rho^2}\right) e^{i\ell\theta}, \tag{11}$$

where, $\rho_0$ is the radius of the bright ring, $\Delta\rho$ is the width of the ring, and in general, $\Delta\rho \ll \rho_0$. Contrary to LG modes, in which the radius of the ring-like transverse intensity profile scales up with the topological charge, in perfect vortex modes it remains constant. Even though, recent reports suggest that the width of such modes experience a small scaling with the topological charge [72]. This is illustrated in **Fig.** 6, where the transverse intensity profiles of a set of perfect vortex beams with topological charges, $\ell = 1$, 4, 10 and 15 are shown. Vaity *et al*. pointed out that a perfect vortex beam is actually a Fourier transform of a Bessel beam[73], and it can be generated by employing a phase hologram whose transmittance is the phase of a Bessel beam[71-74]. It is worth noting that although the concept of "perfect vortex beam" was proposed in 2013, such beam had been introduced and used for 3D optical trapping and transport of particles by Y. Roichman and D. Grier [75] in 2007. Roichman *et al* had pointed out that the radius of this 3D ring trap is independent of the topological charge, and the first experimental demonstration of optical trapping using the "perfect vortex beam" has been reported under the name of holographic ring trap[76]. The perfect vortex has been used by the optical trapping community as a particular case of the structured light beam to study the dynamics of driven particles in the form of optical matter[77-80].

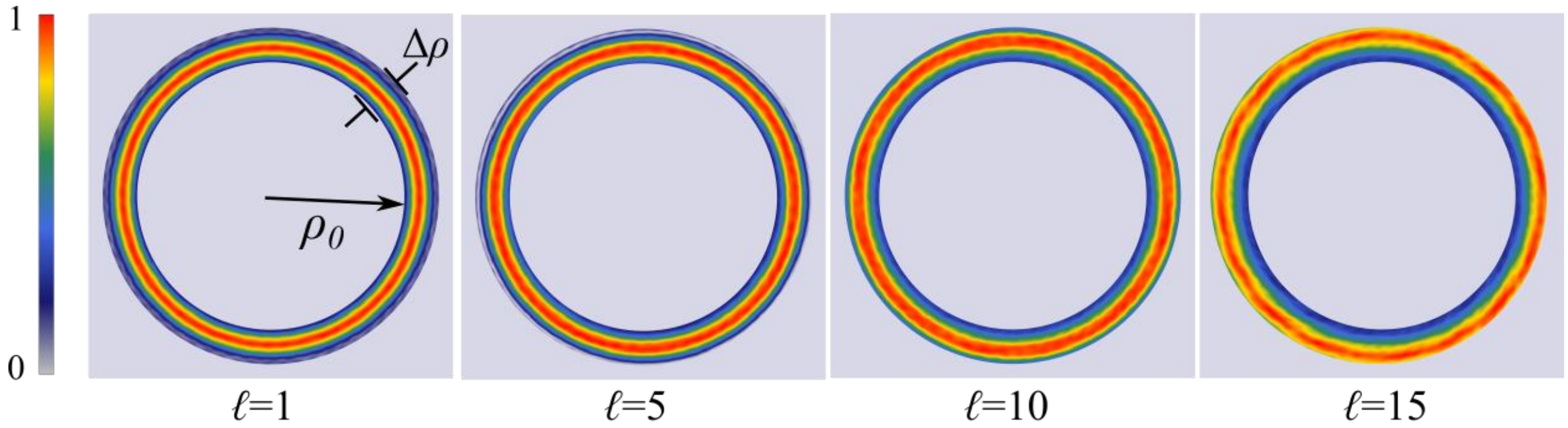

**Fig.** 6. Transverse intensity distribution of perfect vortex beams with topological charge $\ell$ = 1, 4, 10 and 15. Here, $\rho_0$ is the radius of the ring-like intensity profile and Δρ of its width. Notice that the intensity profile remains constant as $\ell$ increases. The color represents the normalized intensity distribution

### *3.4 Mathieu-Gauss beams*

Another interesting set of vector modes are the Mathieu Gauss beams, which are obtained when the Helmholtz equation is solved in elliptical cylindrical coordinates [59]. In such coordinates, defined by $x = f\cosh\xi\cos\eta$, $y = f\sinh\xi\sin\eta$ and $z = z$, where $\xi\epsilon[0,\infty)$ is a radial coordinate and $\eta\epsilon[0,2\pi)]$ is the angular coordinate, the Helmholtz equation can be separated into a longitudinal and a transverse part. The former is easily solved by having a solution of the form $\exp(-i\,k_z z)$ and the latter is found as a solution of the equation [81, 82],

$$\left[\frac{\partial^2}{\partial\xi^2}+\frac{\partial^2}{\partial\eta^2}+\frac{f^2k_t^2}{2}(\cosh 2\xi-\cos 2\eta\,)\right]u_T(\xi,\eta)=0 \qquad (12)$$

Here, the semi-focal distance is represented by $f$ and is given in terms of the major and minor as $f^2 = a^2 - b^2$ $a$ rand $b$, represented, and is related to the eccentricity $e$ as $e = f/a$. Further, the parameter $k_t$ is the transverse component of the wave vector **k**. Equation 12 can be split into the radial and the angular Mathieu equations, using the well-known separation of variables technique, to obtain the non-diffracting Mathieu beams [83],

$$M_m^e(\xi,\eta,z;q) = C_m Je_m(\xi;q)ce_m(\eta;q)\exp(ik_z z) \qquad (13)$$

$$M_m^o(\xi,\eta,z;q) = S_m Jo_m(\xi;q)se_m(\eta;q)\exp(ik_z z). \qquad (14)$$

In the above $C_m$ and $S_m$ are normalization constants, whereas $Je_m, Jo_m$ are the even and odd radial Mathieu functions, respectively, and $ce_m$, $se_m$ are the even and odd angular Mathieu functions. For even modes the sub-index $m = 0,2,3,\ldots$ while for odd modes it takes the values $m = 1,2,3,4,\ldots$ Again, the non-diffracting Mathieu beams carry an infinite amount of energy and cannot be realized experimentally. Nonetheless, a finite-energy version can be realized using a Gaussian envelope, which is known as the Mathieu-Gauss beam. Such modes retain the non-diffracting properties of the ideal Mathieu beams over a finite propagation distance. The Mathieu-Gauss modes are described mathematically as [83],

$$MG_m^e(\tilde{\xi},\tilde{\eta},z;q) = \exp\left(-\frac{ik_t^2}{2k}\frac{z}{\mu}\right) M_m^e\left(\tilde{\xi},\tilde{\eta},z;q\right) \exp\left(-\frac{r^2}{\mu\omega_0^2}\right)\frac{\exp(ikz)}{\mu} \quad (15)$$

$$MG_m^o(\tilde{\xi},\tilde{\eta},z;q) = \exp\left(-\frac{ik_t^2}{2k}\frac{z}{\mu}\right) M_m^o\left(\tilde{\xi},\tilde{\eta},z;q\right) \exp\left(-\frac{r^2}{\mu\omega_0^2}\right)\frac{\exp(ikz)}{\mu} \quad (16)$$

Using the new definitions, $x = f_o(1 + iz/z_R)\cosh\tilde{\xi}\cos\tilde{\eta}$ and $y = f_o(1 + iz/z_R)\sinh\tilde{\xi}\sin\tilde{\eta}$, where now $f_0$ is the semifocal separation at $z = 0$. The parameter $\mu$ is defined in terms of the Rayleigh range $z_R = k\omega_0^2/2$ as, $\mu = 1 + iz/z_R$, where $\omega_0$ is the waist radius of a Gaussian beam. **Figure 7(a)** shows the transverse intensity distribution of a set of even, while **Fig. 7(b)** shows those of odd Mathieu-Gauss beams, given by the parameters, $z = 0$, $k_t = 6$, $e = 0.9$ and $f_0 = 0.9$. Importantly, Mathieu-Gauss beams are another class of "non-diffracting" optical fields, which are a variant of superposition of uniform conical waves, i.e. Bessel beams. Therefore, they have a similar capability of self-reconstruction after an opaque finite obstruction.

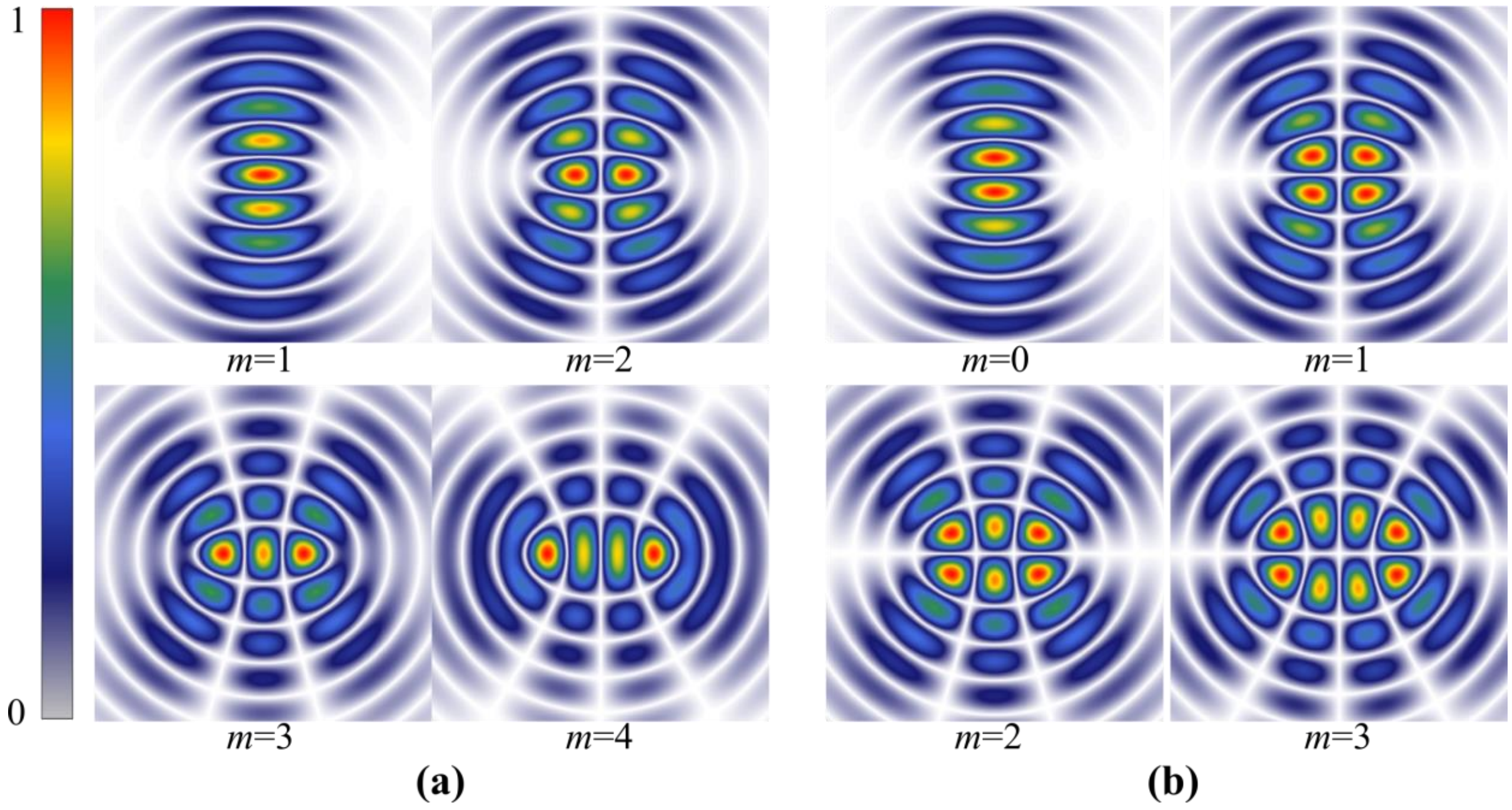

**Fig. 7**. Transverse intensity pattern of a truncated zero-order Mathieu beam with even (a) and odd (b) modes. The color represents the normalized intensity distribution.

### *3.5 Airy-Gaussian beams*

Another important solution to the paraxial wave equation is the Airy beam, which is given in terms of Airy functions Ai(x)[84,85],

$$A(s_x, s_y, \xi) = \mathrm{Ai}\left[s_x - \left(\frac{\xi}{2}\right)^2\right]\mathrm{Ai}\left[s_y - \left(\frac{\xi}{2}\right)^2\right]\exp\left[\frac{i\xi}{2}\left(s_x + s_y - \frac{\xi^3}{3}\right)\right]. \quad (17)$$

where $s_x = x/x_0$ and $s_y = y/y_0$ are dimensionless coordinates in the transverse plane and set by the scale parameters $x_0$ and $y_0$. Moreover, $\xi = z/(kx_0^2)$ represents a normalized propagation distance. Similar to the Bessel modes, Airy beams exhibit unique properties of self-acceleration, "non-diffraction" and self-reconstruction. Among these, its tendency to accelerate in the transverse plane following a parabolic trajectory has attracted considerable interest [See **Fig. 8 (a)** and **(b)**]. Since Airy beams also carry an infinite amount of energy, its experimental realization with finite-energy can be approximated by Airy-Gaussian modes [See **Fig. 8 (c)** and **(d)**], and given by,

$$A_0(s_x, s_y, \xi) = \mathrm{Ai}\left[s_x - \left(\frac{\xi}{2}\right)^2 + ib\xi\right] \mathrm{Ai}\left[s_y - \left(\frac{\xi}{2}\right)^2 + ib\xi\right]$$
$$\times \exp\left[b(s_x + s_y) - b\xi^2 + ib\xi^2 - \frac{\xi^3}{6} + \frac{i\xi(s_x+s_y)}{2}\right], \quad (18)$$

where $b < 1$ is a negative parameter that limits the energy of the Airy beam. As a result, the non-diffracting property can only persist for a finite distance. The inverse Fourier transform of $A_0$ at $\xi = 0$ yields a product of a Gaussian beam and a cubic phase [84],

$$\mathcal{F}^{-1}\{A_0(s_x, s_y)\} \propto \exp[-b(k_x^2 + k_y^2)]\exp\left[\frac{i(k_x^3+k_y^3)}{3}\right]. \quad (19)$$

where, $k_x$ and $k_y$ are the transverse components of the inverse Fourier transform. Therefore, we can experimentally generate an Airy beam by modulating a Gaussian beam with a cubic phase in the Fourier domain.

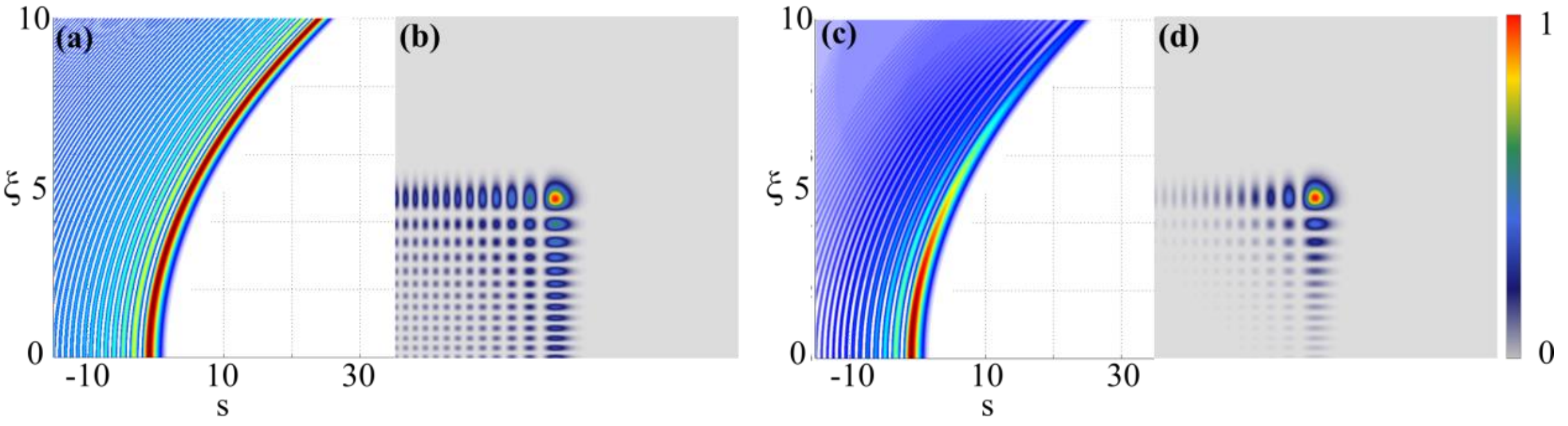

**Fig. 8** Airy beam profiles. (a) Parabolic trajectory and (b) transverse intensity profile of an Airy beam with infinite energy compared with those of a finite energy Airy beam in (c) and (d). The color represents the normalized intensity distribution

### *3.6 Ince-Gaussian beams*

The Ince-Gaussian (IG) modes are another important family of orthogonal solutions to the paraxial wave equation, which can be described as[86, 87],

$$\mathrm{IG}^e_{p,m}(\boldsymbol{r}, \epsilon) = \frac{C\omega_0}{\omega(z)} \mathrm{C}^m_p(\mathrm{i}\xi, \epsilon)\mathrm{C}^m_p(\eta, \epsilon)\exp\left(\frac{-r^2}{\omega^2(z)}\right)\exp i\left[kz + \frac{k^2}{2R(z)} - (p+1)\xi(z)\right], \quad (20)$$

$$\mathrm{IG}^o_{p,m}(\boldsymbol{r}, \epsilon) = \frac{S\omega_0}{\omega(z)} \mathrm{S}^m_p(\mathrm{i}\xi, \epsilon)\mathrm{S}^m_p(\eta, \epsilon)\exp\left(\frac{-r^2}{\omega^2(z)}\right)\exp i\left[kz + \frac{k^2}{2R(z)} - (p+1)\xi(z)\right], \quad (21)$$

where $\mathrm{IG}^{e}_{p,m}$ and $\mathrm{IG}^{o}_{p,m}$ represent the even and odd solutions of order $p$ and degree $m$, with $c$ and $s$ being normalization constants, $C_p^m$ and $S_p^m$ being even and odd Ince polynomials, respectively. $\epsilon = 2f_0^2/\omega_0^2$ together with $f_0$, $\omega_0$ are the scale parameters related to the geometric size of the mode. We note that IG modes are a continuous transition from LG to Hermite–Gaussian modes. **Figure** 9 shows the intensity profiles of even (a) and odd (b) IG modes for $\epsilon = 2$.

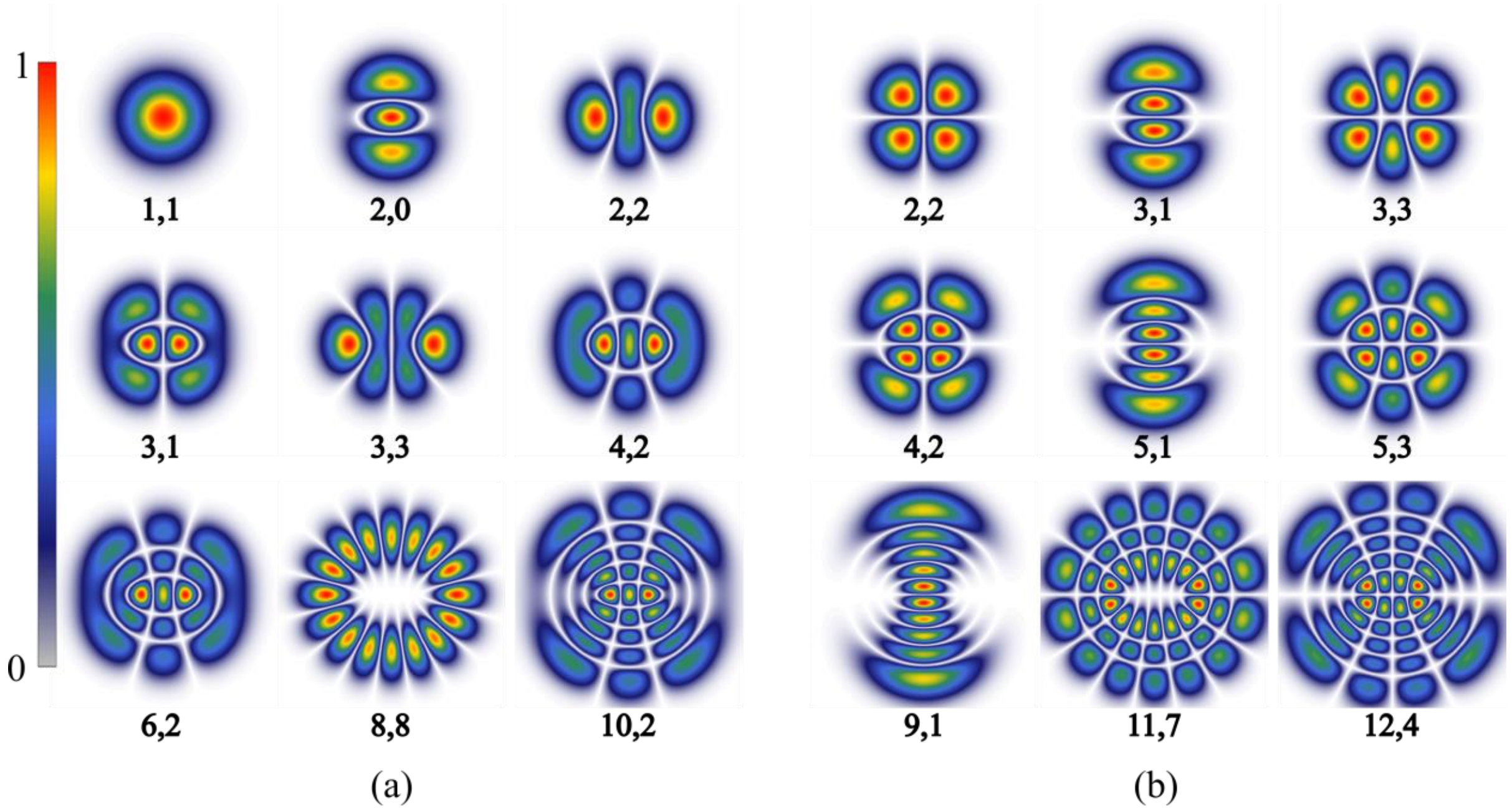

**Fig. 9** Transverse intensity distribution of low-order (a) even and (b) odd Ince-Gaussian modes with $\epsilon = 2$, z=0 and $\omega_0 = 1\ mm$. The color represents the normalized intensity distribution.

### *3.7 Helico-conical beams*

Helico-conical (HC) beams, contrary to the above optical modes, have radial phase dependence[88,89]. The phase in HC beams is unique and it is given by a product of both the radial and the azimuthal coordinates[66],

$$\psi(r,\varphi) = \ell\varphi(K - r/r_o), \qquad (22)$$

where $\ell$ is the topological charge, $K$ is either 0 or 1 and $r_o$ is a normalization constant in the radial coordinates. As a result of this phase dependence, upon propagation the optical field exhibits a helical geometry with anomalies in both phase and amplitude. **Figure** 10 shows the transverse intensity profiles of the HC beam comparing experimental and simulation results. Interestingly, these beams also exhibit a self-healing property, as demonstrated recently[67].

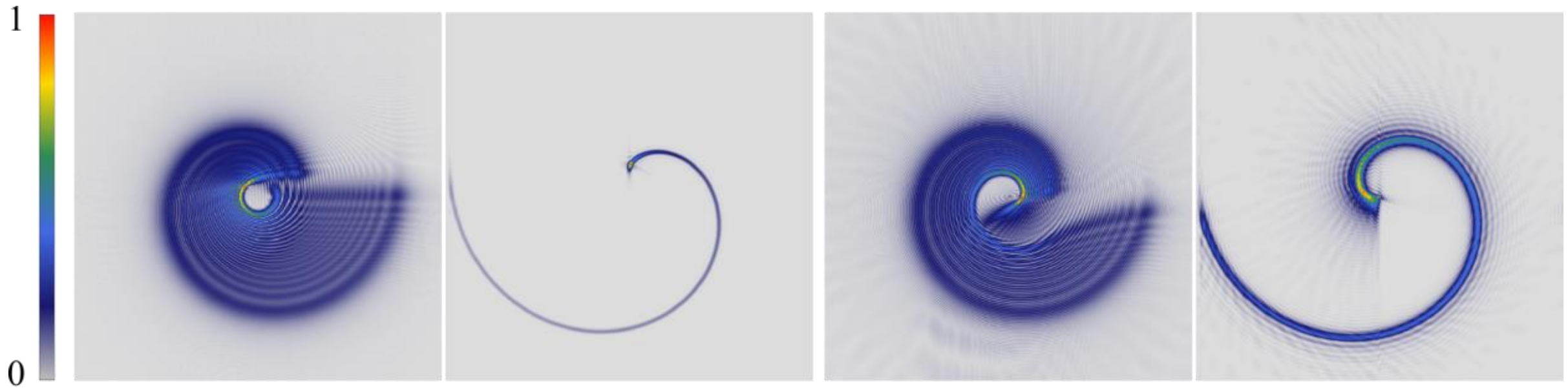

(a) (b) (c) (d)

**Fig. 10** Intensity distribution of a Helico-conical beam from numerical simulations. (a) and (c) show the near-field intensity distribution while (b) and (d) the far-field. In (a) and (b) K=0, while in (c), (d) K=1. In all cases $\ell = 50$. The color represents the normalized intensity distribution.

### *3.8 Vector light fields*

In the previous sections, we considered a set of structured light fields, in which the polarization distribution in the transverse plane was homogeneous. Here we will now look into a more general class of light fields with spatially inhomogeneous polarization, commonly known as vector beams, which arise naturally as solutions to the vectorial Helmholtz equation[90, 91]. These modes are commonly regarded as non-separable superpositions of spatial modes and polarization[92],

$$U(\boldsymbol{r}) = u_1(\boldsymbol{r})e^{i\delta_1}\hat{e}_R + u_2(\boldsymbol{r})e^{i\delta_2}\hat{e}_L, \tag{23}$$

where the spatial degree of freedom is represented by the orthogonal functions $u_1(\boldsymbol{r})$ and $u_2(\boldsymbol{r})$ and the polarization degree of freedom is represented by the orthogonal unitary vectors $\hat{e}_R$ and $\hat{e}_L$ in the circular polarization basis. Moreover, the parameters $\delta_1$ and $\delta_2$ are intermodal phases that introduce a phase delay between both polarization components. Notice that the spatial degree of freedom spans an infinite space, and the number of combinations between the spatial and polarization degrees of freedom is also infinite, giving rise to an infinite set of vector modes. In principle, the spatial degree of freedom can be any of the scalar modes described above. In the $LG_p^\ell$ basis, equation (23) can be rewritten as[91, 92],

$$U(\boldsymbol{r}) = \frac{1}{\sqrt{2}}\left( LG_{p_1}^{\ell_1}e^{i\delta_1}\hat{e}_R + LG_{p_2}^{\ell_2}e^{i\delta_2}\hat{e}_L\right). \tag{24}$$

In principle, $\ell$ and $p$ can take any integer value. Here we will consider the simplified case of $\ell_1 = -\ell_2 = \ell$, $p_1 = p_2 = 0$ and $\delta_1 = 0, \delta_2 = \delta$, in which case, Eq. (24) can be expressed as,

$$U(\boldsymbol{r}) = \frac{1}{\sqrt{2}}\left( LG_0^\ell\hat{e}_R + LG_0^{-\ell}e^{i\delta}\hat{e}_L\right). \tag{25}$$

By substituting $\ell = 1$, the following set of orthogonal vector modes can be obtained,

$$U_{TE}(r) = \frac{1}{\sqrt{2}}(\, LG_0^1 \hat{e}_R + LG_0^{-1} \hat{e}_L), \tag{26}$$

$$U_{TM}(r) = \frac{1}{\sqrt{2}}(\, LG_0^1 \hat{e}_R - LG_0^{-1} \hat{e}_L), \tag{27}$$

$$U_{HE}^{o}(r) = \frac{1}{\sqrt{2}}(\, LG_0^1 \hat{e}_L + LG_0^{-1} \hat{e}_R), \tag{28}$$

$$U_{HE}^{e}(r) = \frac{1}{\sqrt{2}}(\, LG_0^1 \hat{e}_L - LG_0^{-1} \hat{e}_R). \tag{29}$$

These cylindrical vector modes are commonly known as Bell states, which are eigenmodes of both free-space and optical fibers[93]. Both the $U_{TE}(r)$ and $U_{TM}(r)$ modes with radial [Fig. 11 (a)] and azimuthal [Fig. 11 (c)] polarization have been used in optical tweezers due to their unique transverse and longitudinal force components in the trapping plane. This will be discussed in more detail in a later section.

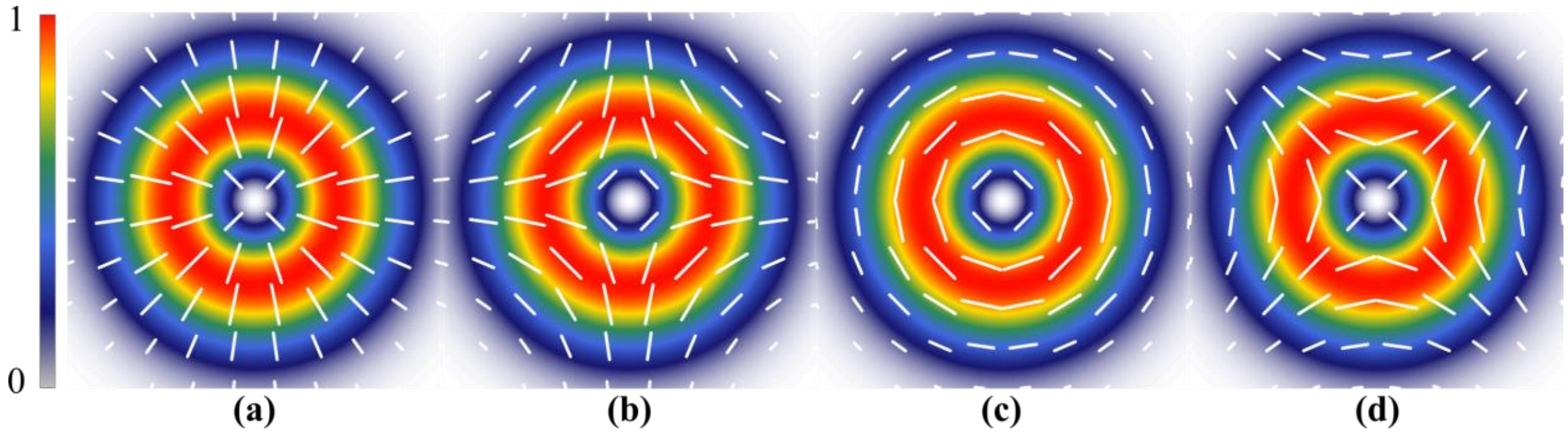

**Fig. 11** Intensity and polarization distribution of the fundamental cylindrical vector beams. (a) Radial, (b) hybrid odd, (c) azimuthal and (d) hybrid even modes. The color represents the normalized intensity distribution, while the lines are associated with the polarization orientation of the electric field.

## 4 Optical trapping with structured light beams

Experimentally, these complex beams described above can be generated in a wide variety of ways, which include interferometric arrays[94-96], glass cones[97,98], liquid crystal wave plates[99,100], metamaterials[101], Spatial Light Modulators (SLMs) [102-105] and in recent time also Digital Micromirror Devices (DMDs) [106-108]. Noteworthy, SLMs enable the generation of complex 2D and 3D shaped light beam patterns, which have significantly advanced the configurable optical trapping of particles. Since these devices are typically utilized in the Fourier plane of an optical system, optical trapping systems using holographically-generated light beams are known as holographic optical tweezers (HOT). In this section, we overview the use of complex light fields in optical tweezers.

### *4.1 Optical trapping with propagation invariant beams*

In this section, we will discuss some of the key demonstrations of HOT within the domain of single optical traps.

*4.1.1 Optical tweezers with Laguerre-Gaussian beams*

As have discussed in Section 2.2, the LG beam is one of the most common beams used for optical trapping. Since the high-order LG beams can carry OAM, which can be used to achieve the orbital rotation of particles. Pioneering experiments demonstrated a full 3D rotational control of microspheres and biological specimens using $LG_0^\ell$ modes [113,114]. The $LG_0^\ell$ modes were interferometrically combined with a reference $LG_0^0$ mode (Gaussian mode) to produce a spiral structure. While the phase profile of the $LG_0^0$ mode can be considered to be flat (spherical more precisely), the $LG_0^\ell$ modes are featured with a more intricate structure, resembling a helical staircase. Hence, the interference of an $LG_0^\ell$ mode with a Gaussian mode produces a new structured azimuthal phase with an intensity pattern containing $\ell$ spiral arms. Importantly, the spiral structure can be rotated in any direction and at arbitrary rates by introducing a phase delays in one of the arms of the interferometer. In this way, particles trapped in the bright lobes of the interference pattern can be rotated in any direction around the optical axis of the beam. By simply changing the azimuthal index $\ell$, it is possible to manipulate objects with different shapes or to trap many objects simultaneously[109]. Importantly, this pioneering experiment showed for the first time the potential application for the control of biological organisms, in this case, the authors trapped and rotated a Chinese hamster chromosome.

Another experiment[110] was performed using a collinear interferometric superposition of $LG_0^\ell$ modes, with opposite topological charges $\ell$ and $-\ell$, to produce the intensity patterns with petal-like structures of $2\ell$ bright lobes. Similarly, the structure could be rotated by introducing phase delays in one of the arms. **Figure** 12 (a) and (b) shows 3D manipulation and rotation of particles trapped in intensity patterns produced by the superposition of $LG_0^\ell$ and $LG_0^{-\ell}$. Figure 12 (c) shows the rotation of the 3D microstructure of particles. Finally, Figure 12 (d) shows a schematic representation of the 3D structures that were created in this experiment by stacking more particles on each bright spot.

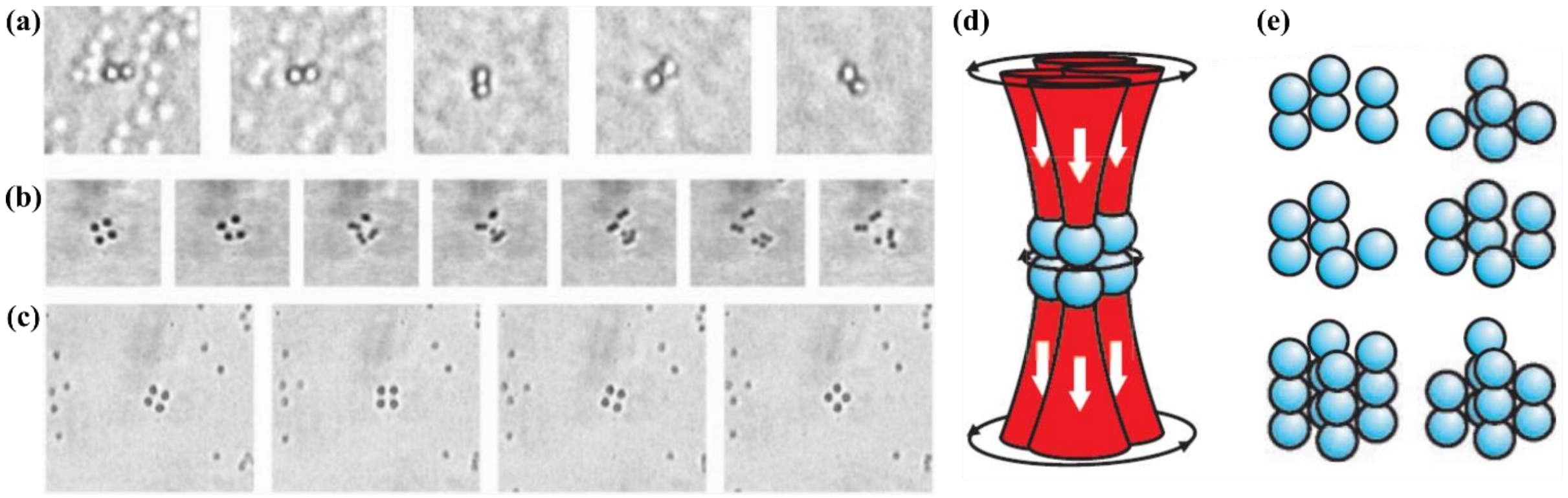

**Fig. 12** Optical trapping and control of 3D structures using superpositions of $LG_0^{\ell}$ and $LG_0^{-\ell}$ as the trapping beam. (a) Two microspheres trapped at the two bright spots created by the superposition of the modes $LG_0^1$ and $LG_0^{-1}$. (b) Trapping and release of eight microspheres trapped along the beam's propagation axis by the intensity pattern generated by superposed modes of $LG_0^2$ and $LG_0^{-2}$, as schematically shown in (d). (c) Rotation of the eight-microsphere cubic structure. (e) Schematic representation of the generation of 3D structures containing a larger number of microspheres. Adapted from Ref. [110].

More recently, Laguerre-Gaussian beams were used to explore the manipulation of single or multiple silicon nanowires [111]. It was demonstrated the orbiting of silicon nanowires around the optical vortex aligned parallel to the propagation axis of the beams. Figure 13 a) shows the position of the particles as a function of time, evincing a clear orbital motion, the right inset shows a schematic representation of the trapped nanowire overlapped with the intensity pattern of the trapping beam. Additionally, the author demonstrated silicon nanorods oriented perpendicular to the propagation axis of the beam can be used as light-driven nanorotors, resulting from the transfer of SAM.

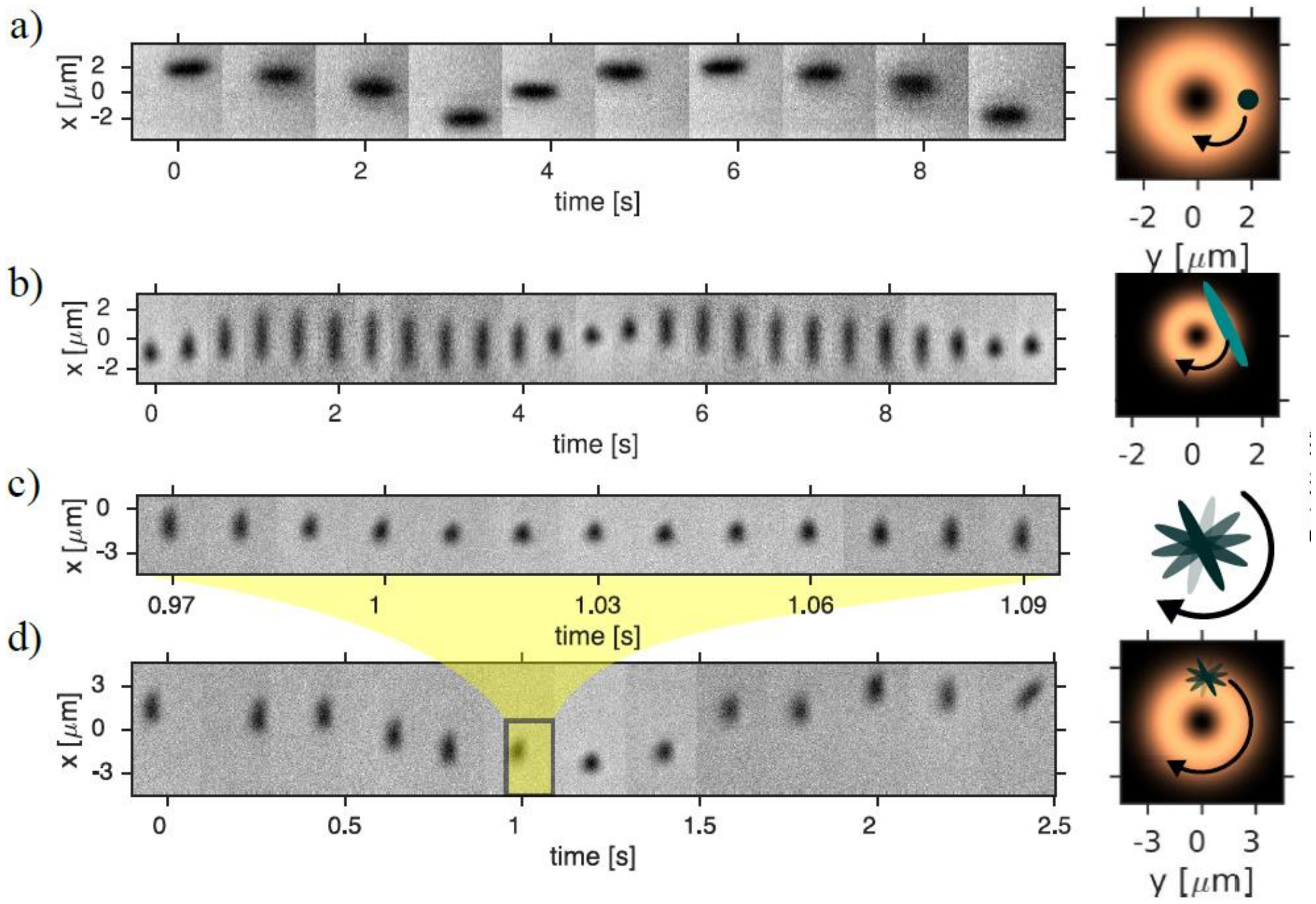

**Fig. 13** Optical trapping and rotations in counter-propagating circularly polarized Laguerre-Gaussian beams of silicon nanowires aligned (a) parallel and (b) perpendicular to the beam propagation axis, where orbiting and orbiting-reorientation are shown, respectively. The simultaneous spinning and orbiting of a shorter nanowire is shown in (c) and (d). Adapted from Ref. [111].

It is noted that the small particles can be trapped in the bright ring of LG vortex beams, while if the size of the particle is comparable to the waist size of the incident structured beam, the particle is predicted to be trapped at the center of the beam with the proper choice of the beam parameters such as vortex charge and polarization[112]. Moreover, a very recent study showed that when the LG beam is strongly focused, the rotation direction of the trapped particles might be in the opposite direction[113], which is much different for the case of paraxial LG beams. Figure 14 shows the anomalous motion of a particle trapped in a strongly focused LG beam. From Fig.14 (c) we can see that in the region β, the direction of the optical torque is opposite to the that of OAM of the LG beam.

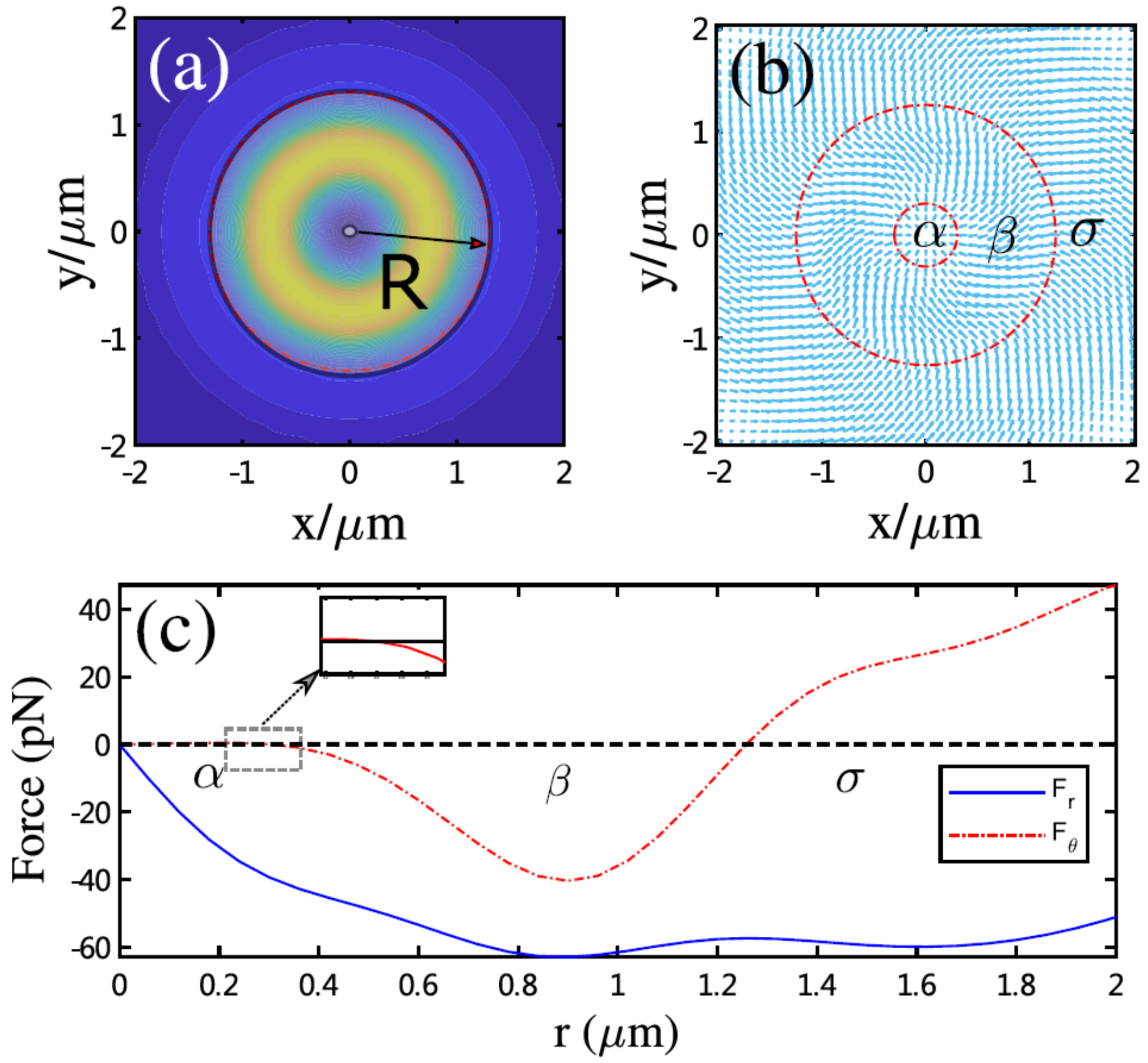

**Fig. 14** Anomalous motion of a particle trapped in a strongly focused high-order LG beams. (a) The intensity profile of the LG beam with topological charge 3. (b) The plot of the distribution of the radiation force exerted on the trapped particle for different center-of-mass radii. (c) The radial (blue line) and azimuthal (red broken line) components of the radiation force for different radii of the trapped particles. Adapted from Ref. [113].

The trapping geometry and the particle shape can have a great influence on the angular momentum transfer. Jesacher *et al.* reported an observation of particles orbiting in a reverse direction with respect to the OAM of the incident light field[114]. Irregular-shaped glass microparticles were trapped at an air-water interface with an LG mode of topological charge $\ell = \pm 30$ based on the holographic tweezers setup. Figure 15 shows the rotation of a polystyrene bead and a glass sliver depending on the sign of the topological charge. Intriguingly, only the polystyrene bead changes its direction of rotation by inverting the topological charge.

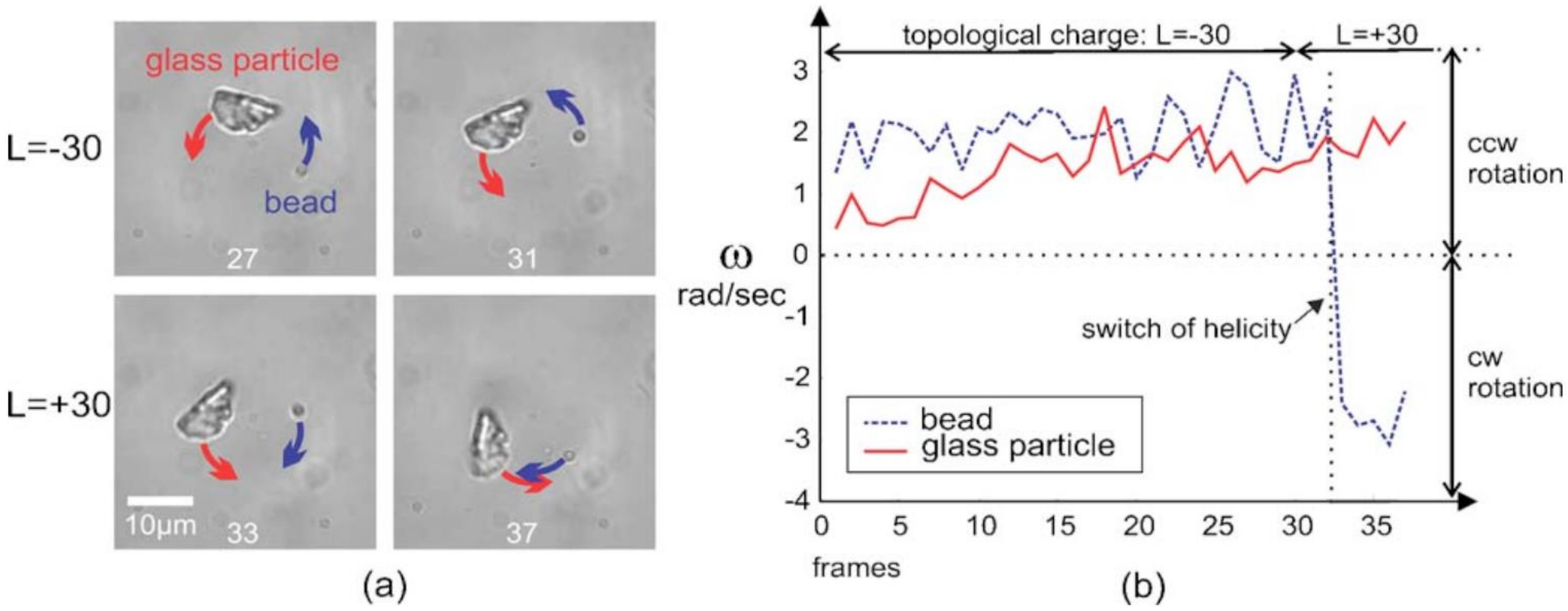

**Fig. 15** Rotation of a polystyrene bead and a glass sliver trapped with an LG mode of $\ell = \pm 30$. (a) Sequential images of these particles showing their direction of rotation. (b) Angular velocity of each particle with respect to the topological charge. Adapted from Ref. [114].

A simple ray optics model for a wedge-shaped glass particle placed at the air-water interface indicates that reflection and refraction of light rays on the particle surfaces can cause the asymmetric object to move against the direction of the OAM of the incident light field. Importantly, this demonstration suggests that the shape-anisotropy of the particles could be optimized for optical momentum transfer and thus for efficient optical nanotransport techniques. Other demonstrations of negative optical torques include reversed orbiting and spinning with a Bessel light beam[115], elliptically polarized beams[116], and circularly polarized Gaussian light beams[117].

*4.1.2 Optical trapping with Bessel-Gaussian beams*

Since the non-diffracting property, the Bessel beams have great advantages for optical manipulation. The first experiment using zeroth-order Bessel beams demonstrated the trapping of multiple particles along the beam axis, as well as the transport of these particles over long distances (~5 mm)[26]. Subsequent experiments[68] showed the manipulation of microparticles in multiple optical planes along the beam axis, using two sample cells (100 $\mu m$ in thickness) separated by 3 mm, as shown in Fig. 16(a). Crucially, even though the Bessel-Gauss beam is distorted by the particles in the first cell, its self-healing properties allow recovering its original intensity profile after a certain propagation distance, which coincides with the separation distance of the cells. Hence, the beam is capable to trap again microparticles in its multiple rings inside the second cell. We note that in this experiment, the radius and non-diffracting propagation distance of the Bessel-Gaussian beam are $5\mu m$ and ~4 $mm$, respectively. It is worth mentioning that such a beam can also align multiple rods along the beam propagation direction[68]. Along the same line, counterpropagating Bessel beams have been also used to manipulate submicron particles in 3 dimensions. This optical conveyor belt is based on the generation of a standing wave, created from the on-axis superposition of two counterpropagating Bessel beams, which allows to confine and deliver over long distances (hundreds of micrometers) and with high precision multiple submicron particles (Fig. 16 (b)). The delivering of particles was achieved by dynamically changing the phase of one of the beams, which in turn causes the whole structure of nodes and antinodes to shift along the optical axis of the standing wave[118]. In 2008, it was demonstrated a practical application where microspheres trapped with Bessel beams served as objective lenses to focus a laser onto a surface to enable near-field direct writing with nanometer resolution[119]. This technique was used to demonstrated nanopatterning of arbitrary patterns with dimensions in the order of 100 nm and positioning accuracy in the order of 40 nm.

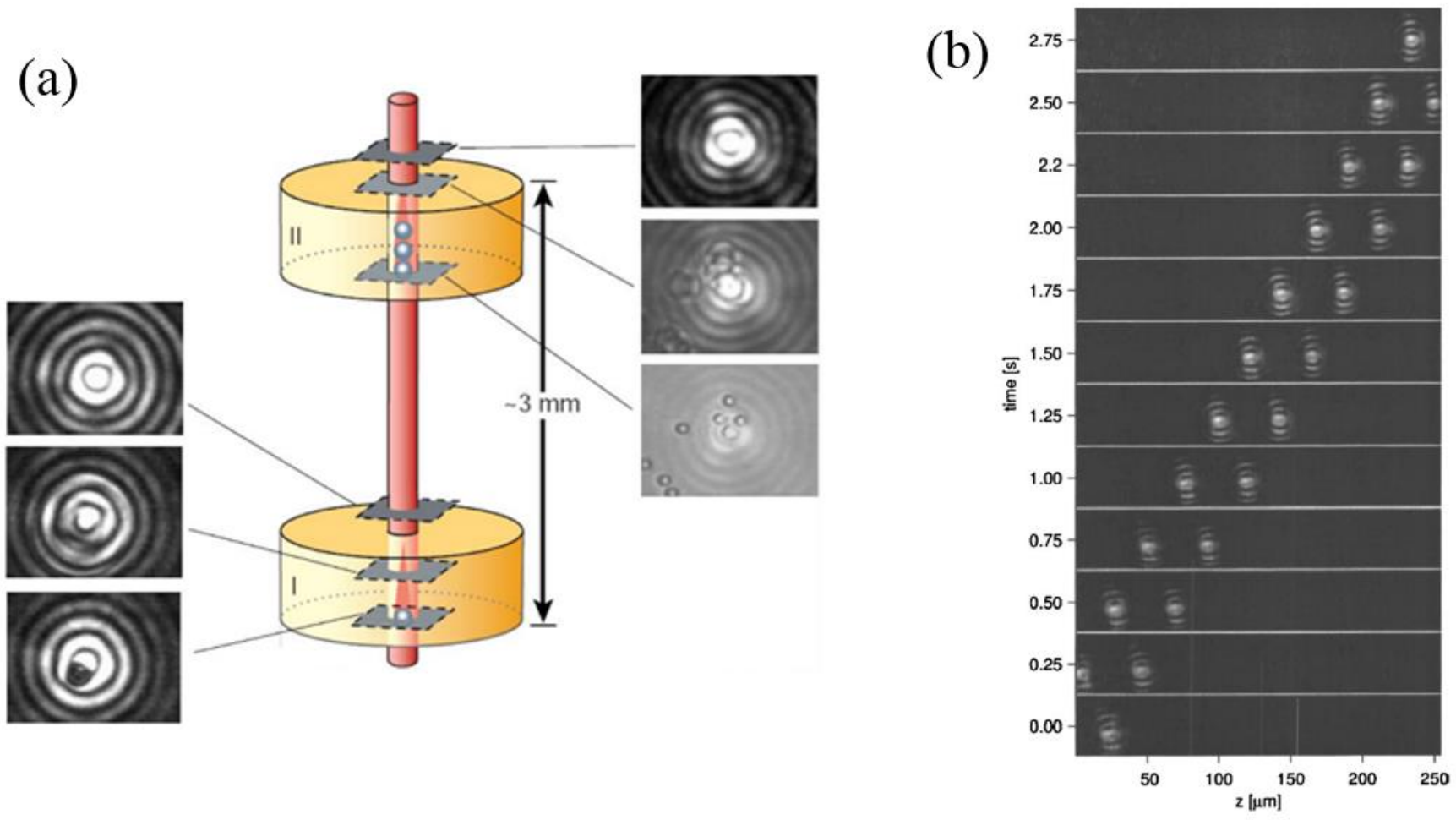

**Fig. 16** Optical trapping with Bessel beams. (a) Trapping of multiple particles at different optical planes. Adapted from Ref. [68]. (b) Trapping and delivering of two particles using a sliding Bessel standing beam. Adapted from Ref. [118].

Very recently, a superposition of multiple co-propagating Bessel beams, so-called frozen waves, was used to build stable optical trapping as well[120]. Besides demonstrating that such beams possess greater optical trapping stabilities, the authors also demonstrated their capabilities to trap in multiple parallel planes along the propagation direction, as schematically shown in Fig. 17 (a). Here, in order to observe the motion along the propagation direction, the authors implemented a digital propagation, in which, the observation plane is fixed and the different propagation planes are simulated holographically by changing the hologram that generates the frozen waves. An experimental sequence of frames of the beam at different propagation planes is shown in Fig. 17 (a1)-(d1) and (a2)-(d2) for the case on one and two trapped particles, respectively.

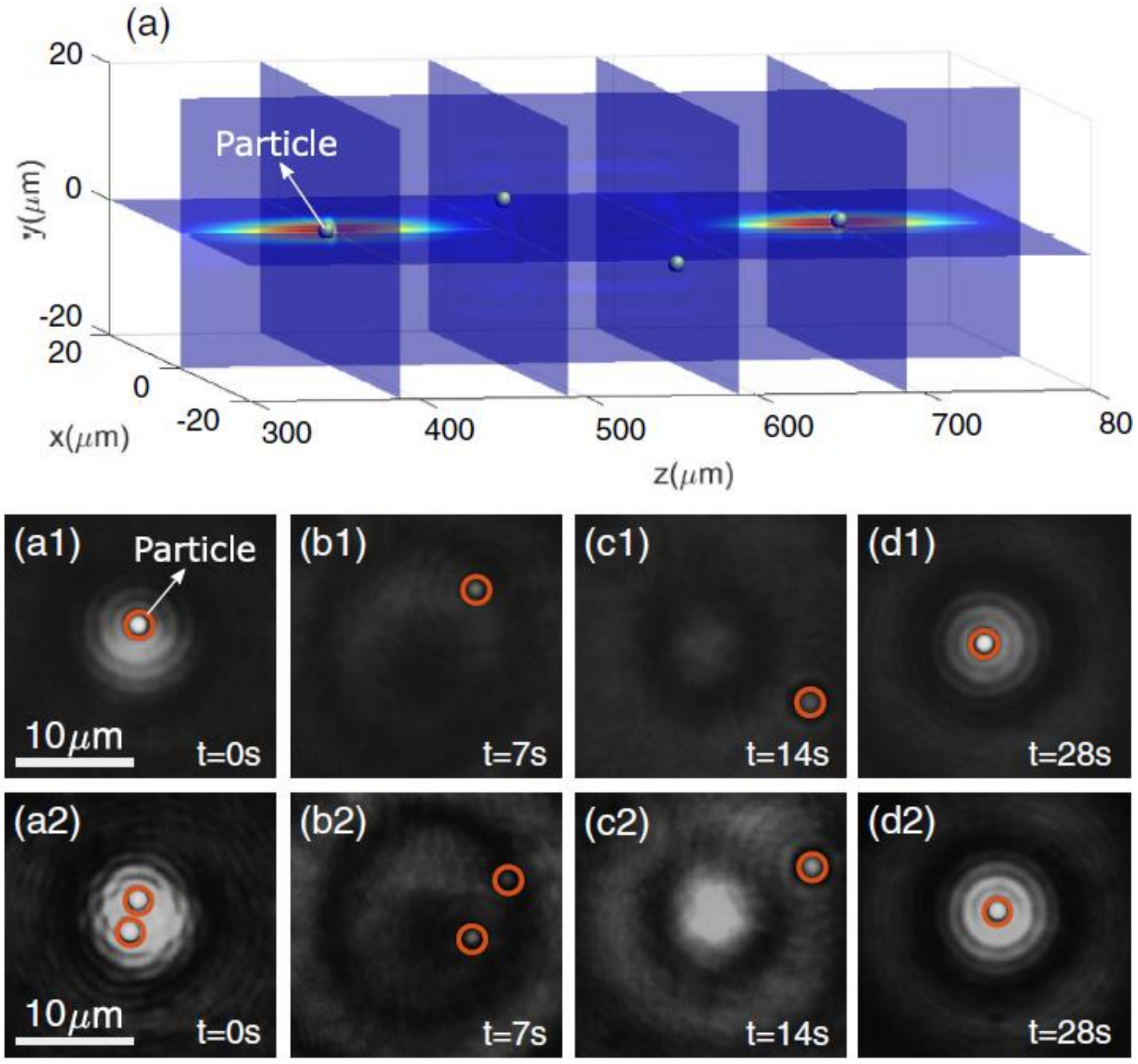

Fig. 17(a) Schematic representation of optical trapping with frozen waves in multiple parallel planes. Sequential of one (a1)–(d1) and two (a2)–(d2) microparticle (orange circle) trapped at different transverse planes along the propagation direction Adapted from Ref. [120].

If we replace the zeroth-order Bessel beams with higher-order Bessel beams, namely, Bessel vortex beams, the situation will change significantly. That is because the centre of the Bessel vortex beam has zero intensity, due to the phase singularity on the axis. Therefore, it was experimentally demonstrated the trapping of multiple transparent particles in the multiple rings of intensity maxima of higher-order Bessel beams[70], and the rotation of the trapped particles is observed as well. It is worth noting that the optical trapping behavior is dependent on the size of the particles[121]. If the size of the particle is much smaller than the radius of the vortex bright ring, the particle will be trapped in the bright ring, meanwhile, the trapped particle will orbit along the bright ring, as shown in Fig. 18(a). If the size of the particle increases and close to the radius of the vortex ring, then the particle will be trapped stably, without rotation, at certain off-axis position inner the bright ring, as shown in Fig. 18 (b). If the particle is bigger enough to cover the whole bright ring of the vortex beam, then the particle will be trapped in the centre of the vortex beam, and the particle does not rotate either Fig.18(c). As a comparison, Fig.18 (d) show the optical trapping of a silver nanowire with LG vortex beam, and it was shown that the nanowire rotates only if its length is longer than the size of the bright ring of vortex beam[122].

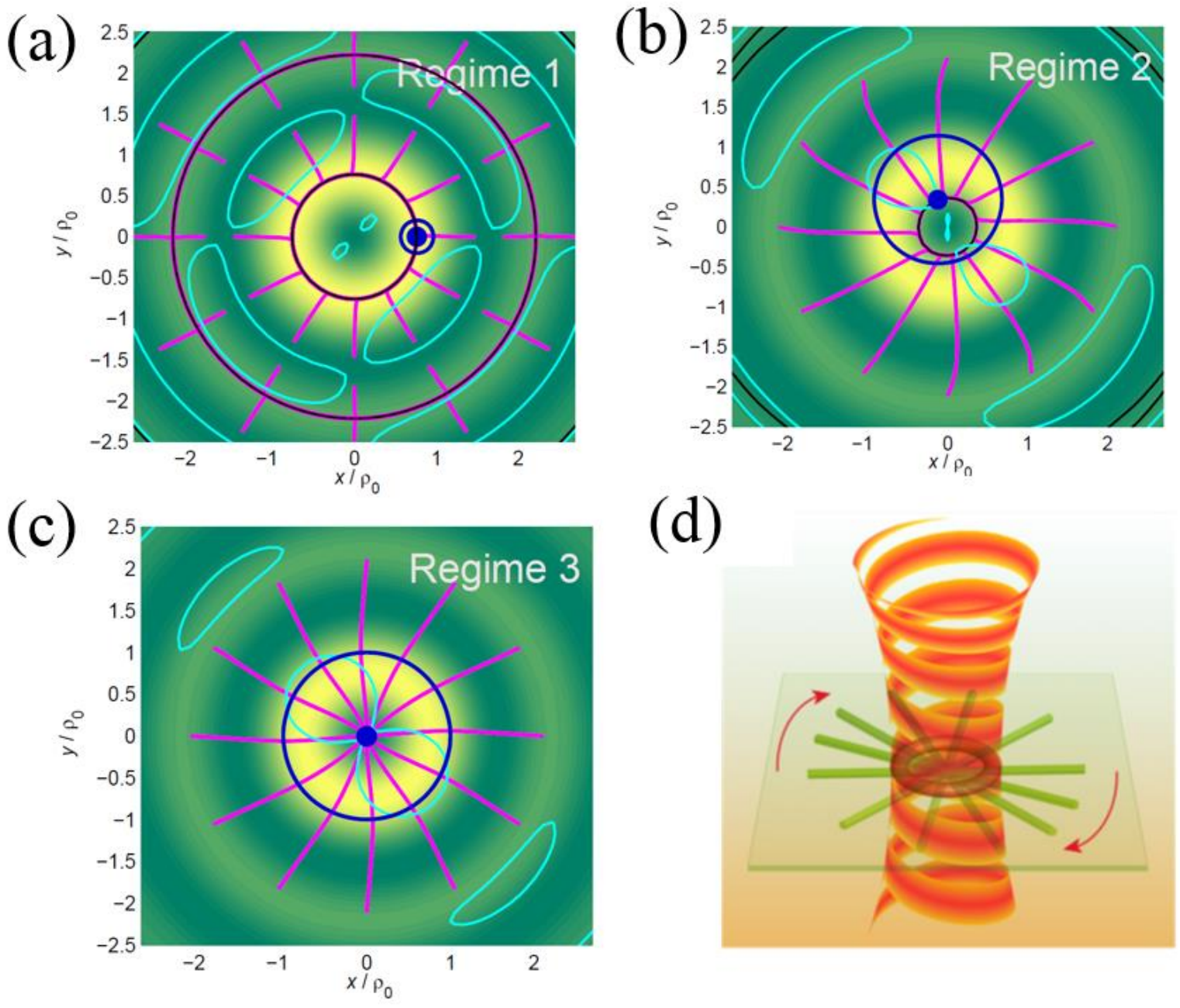

Fig. 18 Effect of the size of the trapped particle on optical trapping with vortex beams. (a)-(c) Optical forces of a particle with different sizes comparing to the radius of the bright rings of Bessel vortex beams. The blue circle and blue dot denote the edge and the centre of the trapped particle. The cyan contour denotes the zero forces azimuthal directions. The magenta lines represent the deterministic trajectory of a particle. (d) Illustration of a rotation of a single silver nanowire trapped by an LG vortex beam. (a-c) Adapted from Ref. [121]; (d) Adapted from Ref. [122].

### *4.1.3 Optical trapping with perfect vortex beams*

Optical manipulation using perfect vortex modes possessing annular intensity profiles independent of topological charges has been of great interest in recent years[123]. The motion of trapped particles is determined by the gradient and scattering forces resulting from the perfect vortex beam. The particles continuously move along the annulus due to the scattering force from the inclined wavefront, as shown in Figs. 19 (a) and (b). As the OAM density is well-defined in the perfect vortex beam, a linear relationship between the rotation rate and the OAM is guaranteed for both negative and positive topological charges, as shown in Fig. 19(c). The experimental realization of perfect vortex beams is rather challenging as they tend to exhibit azimuthal intensity variations due to optical aberrations. As a result, the rotation of a single particle is difficult to maintain at a constant velocity around the trap due to local intensity hot spots. Nonetheless, with intensity profile correction techniques, it is possible to eliminate these hot spots[123], in which a single trapped particle can continuously move along the vortex ring, as schematically illustrated in Fig. 19 (d). However, the trapped particle still exhibits a non-uniform angular velocity due to local variations of the OAM density[124], as shown in Fig. 19 (e). Based on the linear dependence of the particle rotation rate upon the OAM density, the angular velocity can be adjusted *in situ* by correcting the local OAM density with a phase correction mask. It is worth noting that experimentally, the perfect vortex beam can induce rotation of the trapped particle at a constant velocity with integer topological charges, while fundamentally this cannot be achieved with fractional topological charges[73].

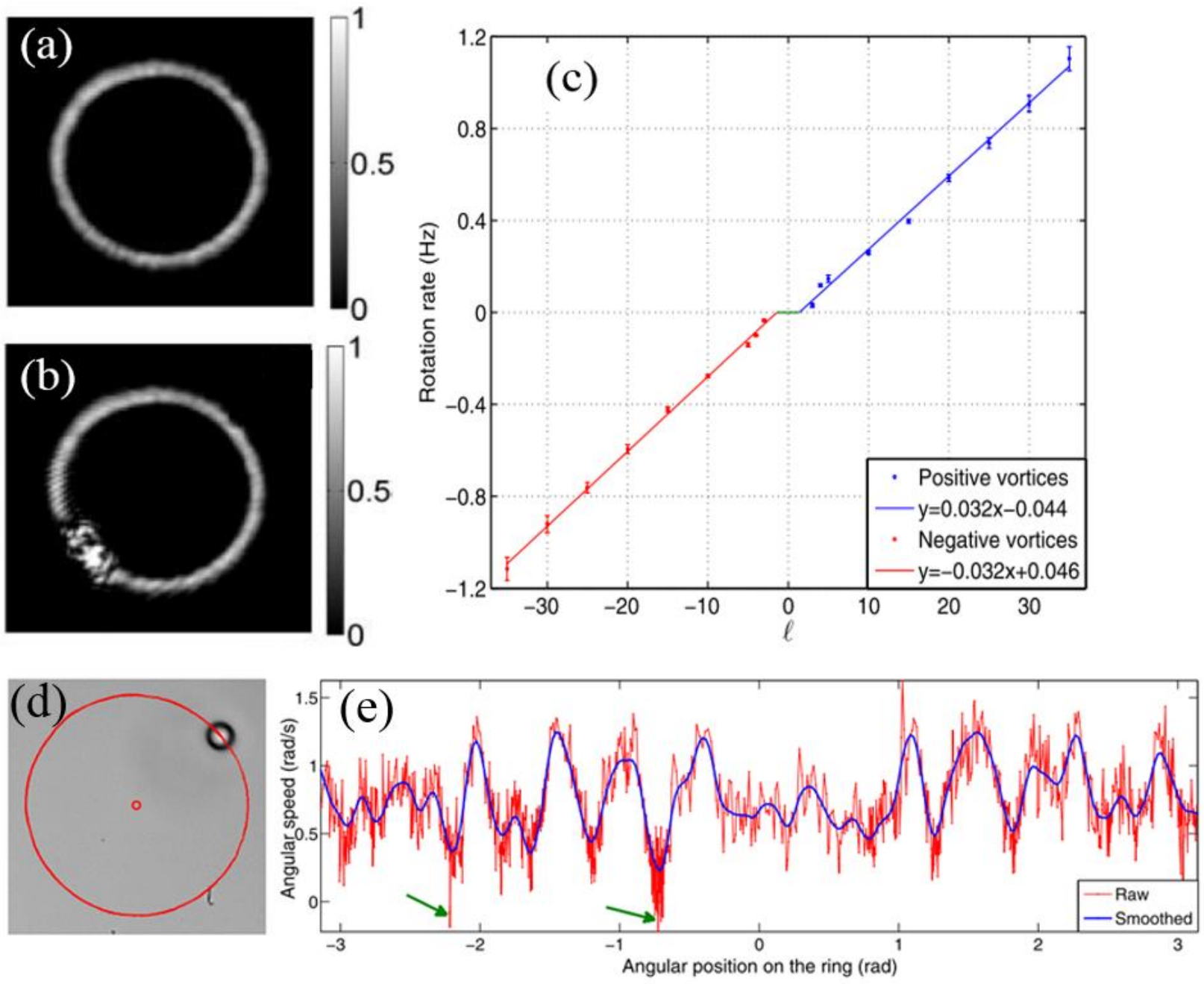

**Fig. 19** A (a) perfect vortex beam with $\ell$ =25 and (b )the beam with the scattered light from a single trapped particle. (c) Linear relationship between the particle rotation rate and the integer topological charge. (d) Snapshot of a trapped particle rotating around the circumference of a perfect vortex beam indicated by the red circle. (e) Angular velocity of the particle as a function of its angular position. Adapted from Ref. [123 and 124].

### *4.1.4 Optical trapping with Mathieu beams*

Mathieu beams are a family of propagation invariant beams with self-healing properties. Compared to Bessel-Gaussian beams, they are especially well-suited for the 3D arrangement of particles due to their rich variety of complex transverse mode distributions[125]. Notably, Mathieu beams allow to trap and orient elongated objects in the transverse *x*-*y* plane to the beam (*z*) axis, which is not normally possible with single optical tweezers. Figure 20 shows optical trapping of non-spherical particles with a fourth order ($m$ = 4) even Mathieu beam (a-c), which was rotated within the $x$-$y$ plane (d,e). Elongated particles (approx. 3×5 μm) were arranged with their long axis in the transversal plane (Fig. 20 (f,g)) and their orientation depends on the beam orientation.

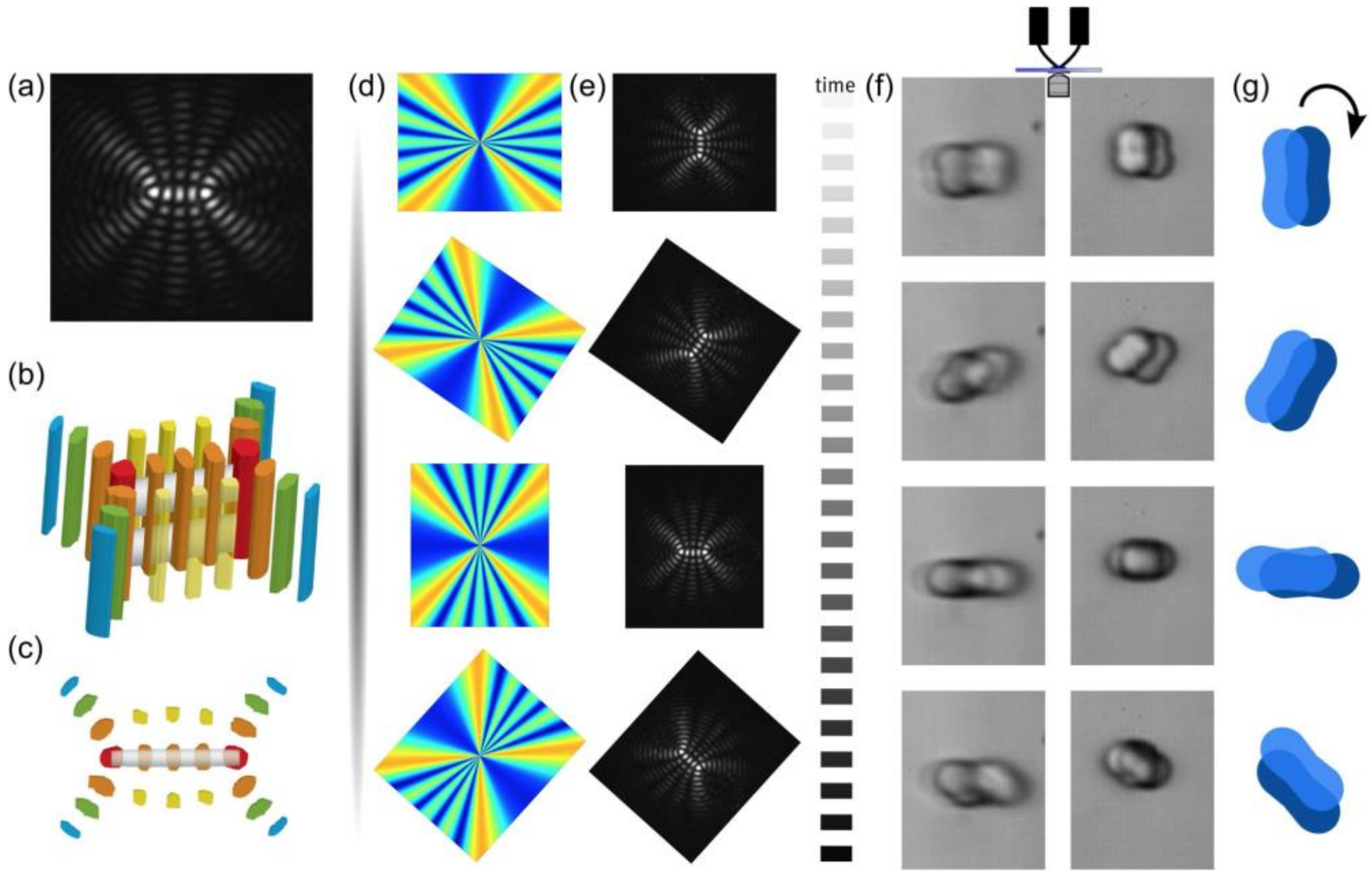

**Fig. 20** Optical trapping of non-spherical particles. (a) Mathieu beam with $m$ = 4. (b) 3D intensity distribution. (c) Particles orientation within transversal intensity distribution. (d) Rotating hologram and (e) corresponding Mathieu beam. (f) Time-lapse images of trapped particles depend on the orientation of the Mathieu beam and (g) their corresponding schematics. Adapted from [125].

We note that it is possible to assemble multiple particles in $z$-direction because of the self-healing property of Mathieu beams. Furthermore, Mathieu beams also offer superpositions of even and odd modes, resulting in helical Mathieu beams which provide continuous phase variations. Thus, these beams were also used to investigate the transfer of OAM[126] to particles in optical micromanipulation[127].

### *4.1.5 Optical trapping with Airy-Gaussian beams*

One of the most prominent properties of Airy beams is their ability to freely accelerate in the transverse plane resulting in a parabolic trajectory upon propagation. Exploiting this property, Baumgartl *et al.* transported particles along parabolic trajectories, as schematically represented in Fig. 21 (a)[128,129]. An Airy beam was shaped from a 25 mW Argon-ion laser using an SLM and focused down to a size of 10 μm into a chamber, containing an aqueous suspension of colloidal glass microspheres. The optical gradient forces attracted particles towards the main lobe of the Airy beam, where they were transported along the parabolic trajectory due to radiation pressure. The Airy beam shifted laterally over a distance of approximately 10 μm after only a propagation distance of 75 μm. Figure 21 (b) and (c) shows experimental results of the so-called “snowblowing” effect. The field of view was divided in four quadrants marked with different colors in order to see this effect clearly. Fig. 21 (b) shows that an Airy beam launched into the second (green) quadrant cleared particles in this section by carrying them into the third (purple) quadrant along the 3D parabolic trajectory. This process is reversible if the orientation of the Airy beam is inverted, as shown in Fig. 21 (c). Recently, a circular Airy vortex beam has attracted interesting due to its autofocusing property[130, 131], which is a hollow beam carrying OAM, as shown in Fig.21 (d). Since such a beam carrying OAM, which can be used to trap and rotate particles as well, as shown in Fig.21 (e). Moreover, it is shown that the rotational velocity of the trapped micro-particles first increases with the increasing topological charge and then decreases after reaches a maximum[131].

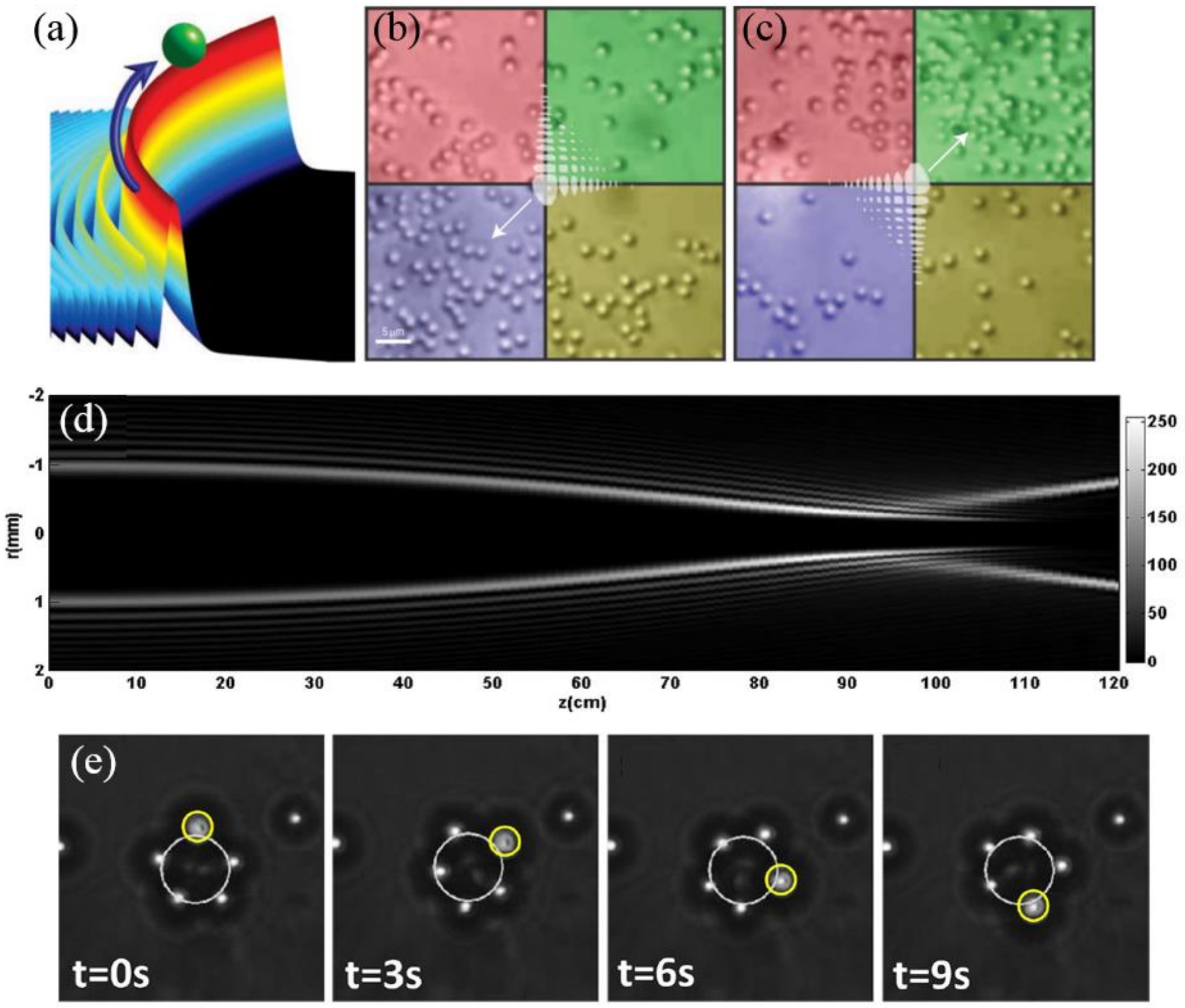

**Fig. 21** Optical manipulation with Airy beams. (a) Schematic representation of a microparticle being transported along a parabolic trajectory, adapted from Ref. [128]. Transporting particles (b) from quadrant two (green) to quadrant three (purple) and (c) from quadrant three to quadrant two, adapted from Ref. [129]. (d) The y-z plane intensity profile of a circular Airy vortex beam. (e) Rotation of the trapped silica particles on the primary ring of circular Airy vortex beam for topological charge 12. The white and yellow circles denote the vortex ring position and the position of a selected trapped particle and a different time. Adapted from Ref. [131].

### *4.1.6 Optical trapping with Ince-Gaussian beams*

As previously mentioned, IG modes appear as natural solutions to the paraxial wave equation in elliptical coordinates (see the previous section on Ince-Gaussian modes). A rich variety of transverse intensity profiles given by high-order assemblies of this family of modes allows micromanipulation of particles at each bright spot. Woerdemann *et al.* utilized these IG modes to create 3D structures of trapped microparticles[132]. Figure 22 shows the IG modes generated by an SLM (top row) and particles trapped in the bright parts of these IG modes (bottom row). Notably, microbeads can be trapped in 3D according to the 3D structures of these beam profiles.

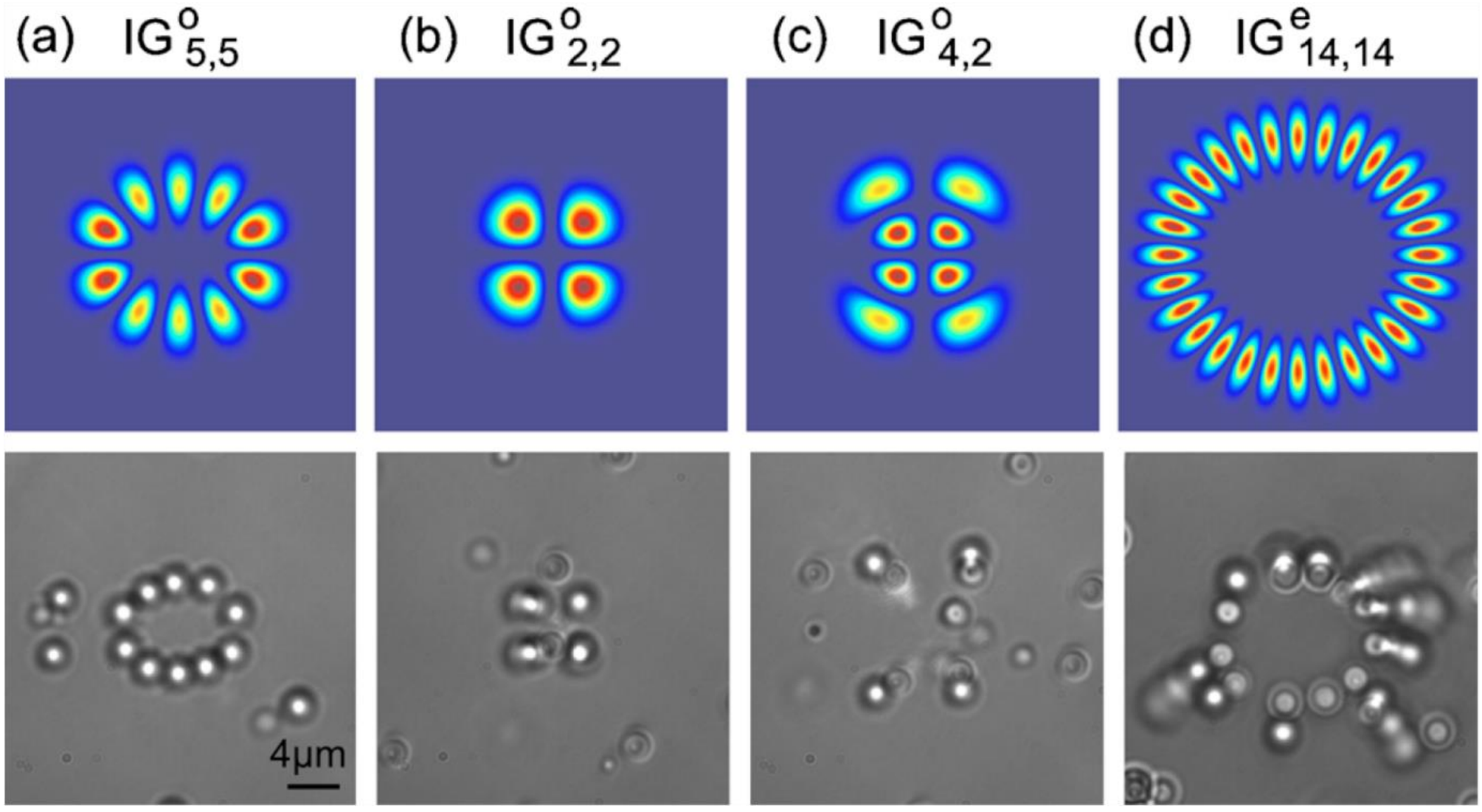

**Fig. 22** Micromanipulation with Ince-Gaussian beams. The top row shows the transverse intensity patterns of the beams, while the bottom row shows trapped microparticles with the corresponding beams. (a) $\mathrm{IG}^{o}_{5,5}$ mode. (b) $\mathrm{IG}^{o}_{2,2}$ mode with four columns of particles stacked along its beam axis. (c) $\mathrm{IG}^{o}_{4,2}$ mode where the four central petals close to each other shows no stable traps. (d) $\mathrm{IG}^{e}_{14,14}$ mode with particles trapped at certain locations. Adapted from Ref. [132].

### *4.1.7 Optical trapping with Helico-conical beams*

Another important class of OAM beams is the Helico-conical (HC) beams with spiral phase and intensity profiles[88,89]. A schematic representation of beam generation and its 3D beam profile upon propagation are shown in Fig. 23 (a). With this unique beam property, both linear momentum and OAM are to be transferred to microparticles[133]. In the experiment, 2 $\mu$m-diameter silica beads are dispersed in water in a chamber with a thickness of 100 $\mu$m. Fig. 23 (b) shows time-lapse images of a trapped bead moving upwards from the bottom along a spiral trajectory around the axis of the HC beam ($\ell = 20$).

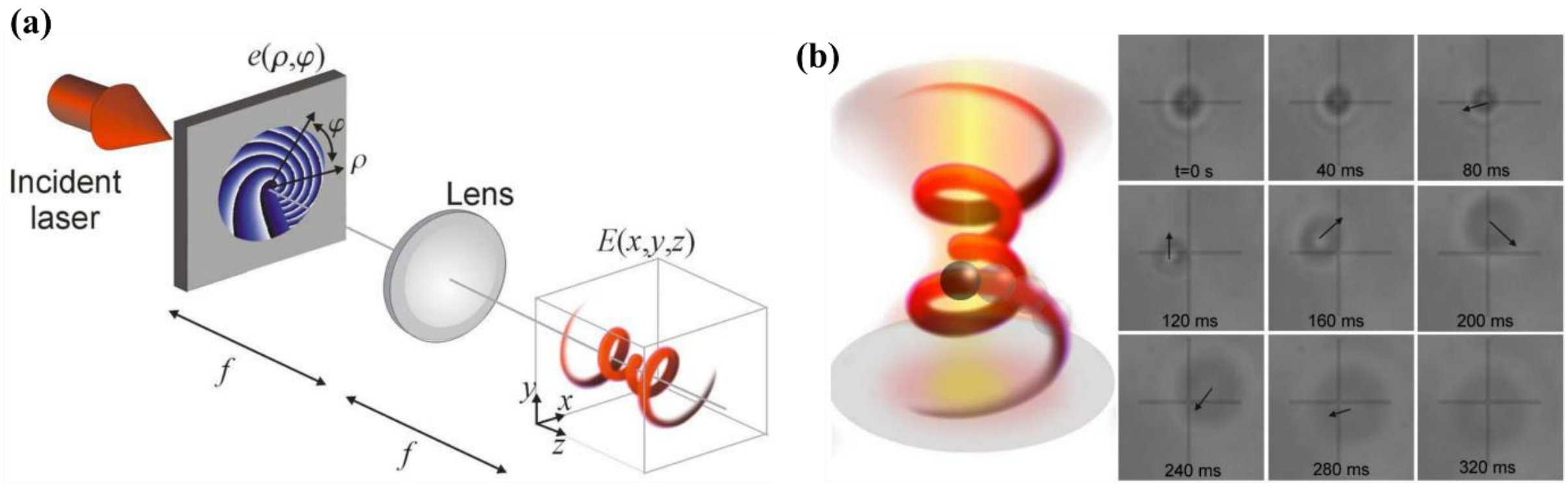

**Fig. 23** Optical manipulation with Helico-conical beams. (a) Schematic representation of the setup required for beam generation. (b) Time-lapse images of a microbead trapped and guided along with the maximum intensity of the beam, as illustrated on the left. Adapted from Ref. [133].

## *4.2 Optical trapping with holographic optical traps*

### *4.2.1 Holographic arrays of multiple optical traps in two and three dimensions.*

As we have shown in previous sections, SLMs have provided the ability to create 2D and 3D optical traps for guiding or transporting microparticles. Nonetheless, most of the previously discussed techniques rely on the use of common light beams that are solutions to the wave equation, except for Helico-conical light beams. In what follows, we will discuss a more general class of optical traps, which rely on arbitrary customized 2d and 3D light beams.

An immediate and almost obvious step forward to achieve the simultaneous trapping of many particles is through the use of galvo-scanning mirrors or acousto-optic beam deflectors (AOD)[134], in which a single beam can be time-shared at multiple locations in order to simultaneously trap micron-sized polymer spheres. This approach can be used in the study of the single molecule H-NS protein, for example[134]. An alternative approach to generate arrays of optical traps relies on the use of diffractive optical elements (DOE)[135], as shown in Fig 24. The key idea in this approach is based on the deflection of an input beam to a different angle defined by a grating period of a diffraction grating. Therefore, the generation of multiple trapping focal spots can be achieved by superimposing multiple diffraction gratings. A typical binary version of the DOE pattern for a hexagonal tweezers array is shown in Fig. 24 (b). The trapped microspheres in the array are shown in Fig. 24 (c).

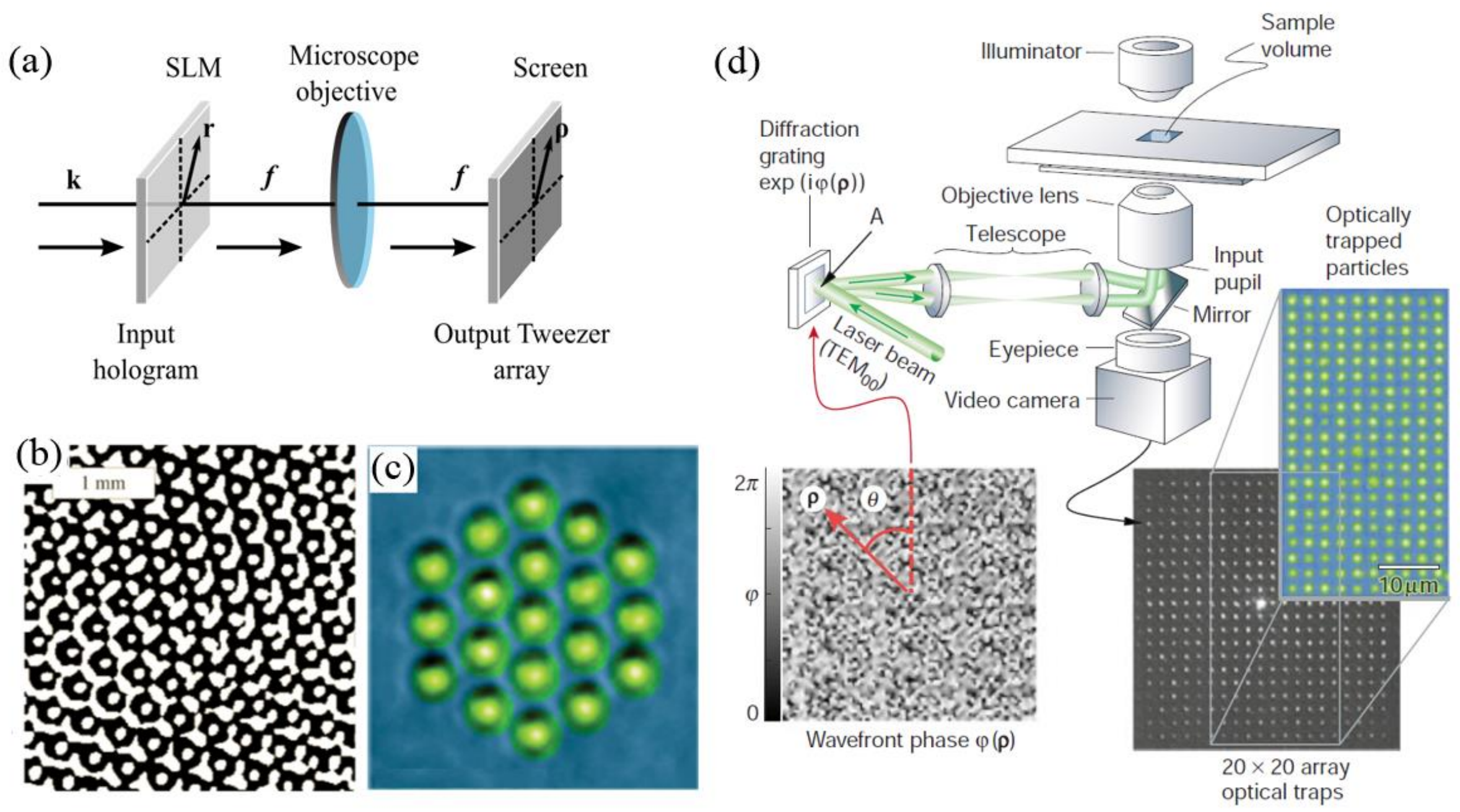

**Fig. 24** Optical tweezer arrays using computer generated holograms. (a) Schematic representation of the fields at the input hologram and output Fourier planes, where $\vec{k}$ is the wave vector. (b) Diffractive optical element (the black color represents a phase shift of $\pi$- radians) etched on a fused silica substrate for a hexagonal array of traps. (c) 19 Silica spheres (1-$\mu m$-diameter) trapped in the hexagonal array. Adapted from Ref. [135]. (d)Typical experimental setup for optical tweezers using computer generated holograms. A telescope relays the plane of the diffraction grating to the input pupil of the microscope objective. In this way, multiple beams generated by the diffraction grating can create multiple optical traps. The bottom left inset shows an example of a phase grating capable of generating an array of 20 by 20 optical traps. The bottom right inset shows the optical trapping of multiple polystyrene spheres (800 nm in diameter) in water. Adapted from Ref. [138].

SLMs can facilitate this task by adding more flexibility and new features to this field. Not only the wavefront of the incident beam can be modulated to desired optical traps but also the wavefront distortion (aberration) and attenuation can be corrected or compensated by SLMs[136-138]. In addition, SLMs enables dynamic and interactive control of holographic optical tweezers. Figure 24(d) schematically illustrates a typical holographic optical tweezers setup that incorporates the use of an SLM[138,139]. The phase pattern for generating an array of 20 by 20 traps (bottom-right inset with a zoom-in image for the trapped microspheres) is shown on the bottom left inset. Importantly, such holographic optical tweezers can be brought to a third dimension for the simultaneous manipulation of multiple particles in three dimensions, which are useful to create, for example, micro-crystal structures. Proposed techniques can create structures of several tens of microns capable to rotate dynamically about arbitrary axis[136,140,141].

A step forward in the generalization of customized light shapes to manipulate microparticles consist in the computation of the phase, in a reverse-engineered process, of the desired light beam, which we briefly explain next. In essence, a 2D phase distribution $\Phi^{\mathrm{in}}(\boldsymbol{r})$ of the beam wavefront is calculated from the desired intensity pattern at the trapping plane. Figure 24 (a) depicts the relationship of the electric field between the input and the focal planes[135]. The monochromatic plane wave is modulated by a phase $\Phi^{in}(\boldsymbol{r})$ at the input leading to $E^{in}(\boldsymbol{r}) = A^{in}(\boldsymbol{r})\exp[\Phi^{in}(\boldsymbol{r})]$, where the amplitude and imposed phase are real-valued functions. The electric field at the focal plane yields $E^{f}(\boldsymbol{\rho}) = A^{f}(\boldsymbol{\rho})\exp[\Phi^{f}(\boldsymbol{\rho})]$, which constitutes a Fourier transform pair with $E^{in}(\boldsymbol{r})$ and can be written as[135]:

$$E^{f}(\boldsymbol{\rho}) = \frac{k}{2\pi f} e^{i\theta(\boldsymbol{\rho})} \int d^2\boldsymbol{r} E^{in}(\boldsymbol{r}) e^{-ik\boldsymbol{r}\cdot\boldsymbol{\rho}/f} \tag{30}$$

where $f$ represents the focal length of the lens and $k = 2\pi/\lambda$ the wavenumber. Since there is no analytical solution for the phase distribution, usually an iterative algorithm, e.g., an adaptive-additive algorithm, is applied to search for an optimized phase[135].

So far, we have discussed the creation of static arrays of optical traps using an SLM. In the following sections, we will discuss novel approaches for the creation of complex light fields for micromanipulation.

*4.2.2 Optical trapping with modulated optical beams*

Optical trapping along arbitrary trajectories was first demonstrated in 2003[142]. The idea behind this technique is the fact that, for the family of LG transverse modes, the mode profile and radius of peak intensity vary with the topological charge $\ell$ of the beam[142],

$$R(\theta) = a\frac{\lambda}{NA}\left[1 + \frac{1}{\ell_0}\frac{d\varphi(\theta)}{d\theta}\right], \tag{31}$$

$$R_\ell \approx a\lambda/NA(1 + \ell/\ell_0), \tag{32}$$

where *NA* is the numerical aperture of the focusing lens and $a$ and $\ell_0$ are constants related to the radial amplitude profile of the beam. It provides a way to tailor the radius of the vortex beam via the function of $\theta$. For example, Lissajous patterns can be directly obtained with $\varphi(\theta) = \ell[\theta + \alpha\sin(m\theta + \beta)]$, where the constants $\alpha$ and $\beta$ are used to control the depth of modulation. **Figure 25 (a)** shows the intensity patterns for a constant $\alpha$ and a varying *m* (top row) and a constant *m* and a varying $\alpha$ (bottom row). Optical tweezers with beams generated in this way can drive nanoparticles along with closed intricate circuits. **Figure 25 (b)** shows two polystyrene nanospheres (400 nm in radius) suspended in water are transported along with a modulated optical vortex with $l$=60, $m$ = 3, and $\alpha = 0.1$, in which each particle completes a cycle within two seconds. It is worth noting that the optical forces driving the particles are larger when $R(\theta)$ is smaller as the light is most intense.

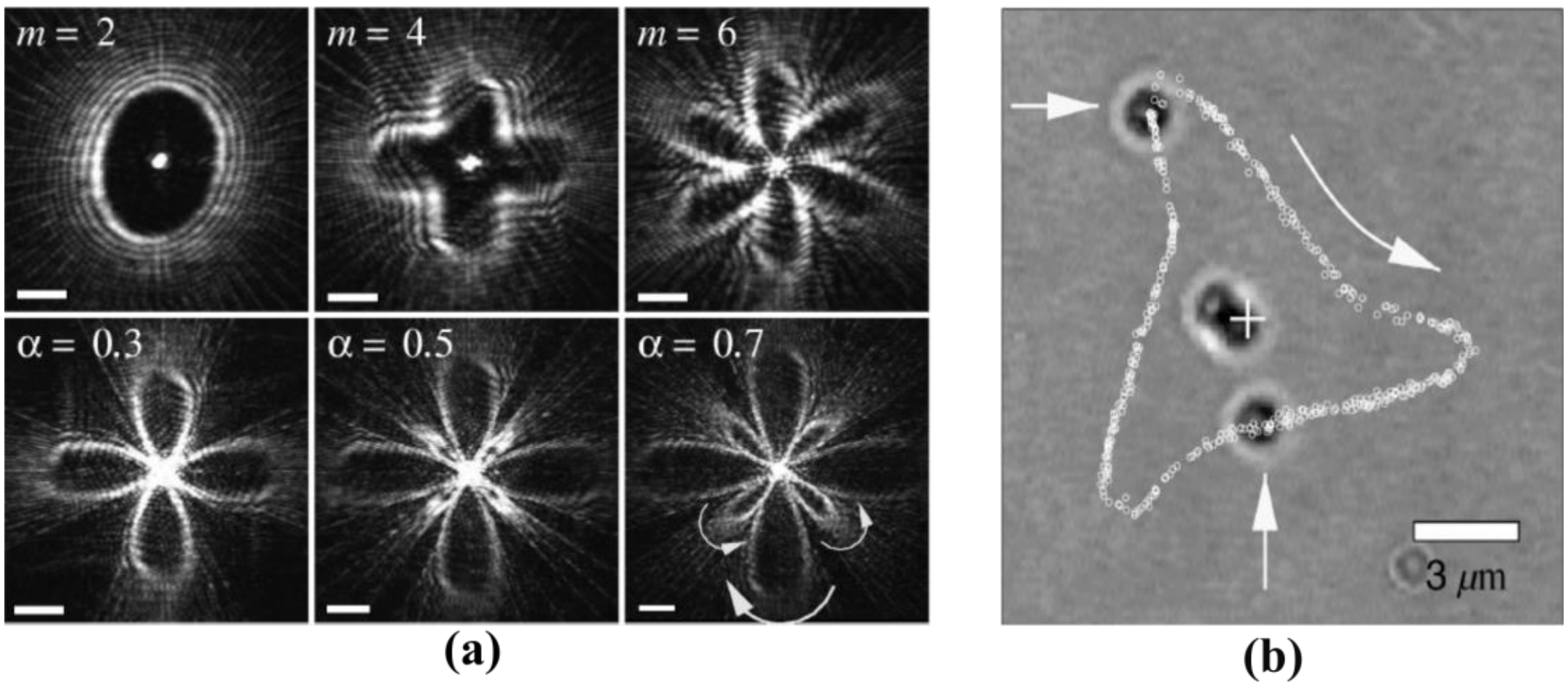

**Fig. 25** Optical trapping with complex optical patterns. (a) Experimentally generated beam patterns with different modes. (b) Two particles being guided along the trajectory shown by the dotted line. Adapted from Ref. [142]

Recently, a new class of optical tweezers along with 2D and 3D trajectories was demonstrated, driving particles based on phase gradients of the light field. This novel method employs a technique to create light fields with an arbitrary (and programmable) transverse phase profile $\phi(\mathbf{r})$. A particular case[76] of such a transverse phase gradient is the azimuthally varying phase $\phi(\mathbf{r}) = \ell\varphi$. Similar phase profiles have been used in the field of laser remote sensing to directly measure the velocity component along the transverse plane[143-145]. When trapped in a beam with a phase gradient, particles can move along the beam. In the case of linear phase gradient $\phi(x) = |q|x$ where |q|=12 radians/$\mu$m, colloidal silica microspheres (1.53 $\mu$m in diameter) dispersed in water were transported along with a transverse linear trap of 5 $\mu$m in length at a speed of 2 $\mu$m/s. The second case of a parabolic gradient $\phi(\mathrm{x}) = \pm(qx)^2$ is more interesting [76]. **Figures 26 (a1)** and **(b1)** illustrate that two colloidal spheres were trapped with negative $\phi(\mathrm{x}) = -(qx)^2$ and positive $\phi(\mathrm{x}) = +(qx)^2$ parabolic phase profiles, respectively. In the negative case, the particles were pushed towards the ends of the linear trap, whereas in the positive case they are pulled towards the center of the trap. **Figures 26 (a2)** and **(b2)** show the linear intensity profiles at the trapping plane. A cross-section of the trapping beams in the x-z plane shows the phase gradient, which originates from either light rays diverging or converging towards the optical trapping plane, as shown in Figs. **26 (a3)** and **(b3)**.

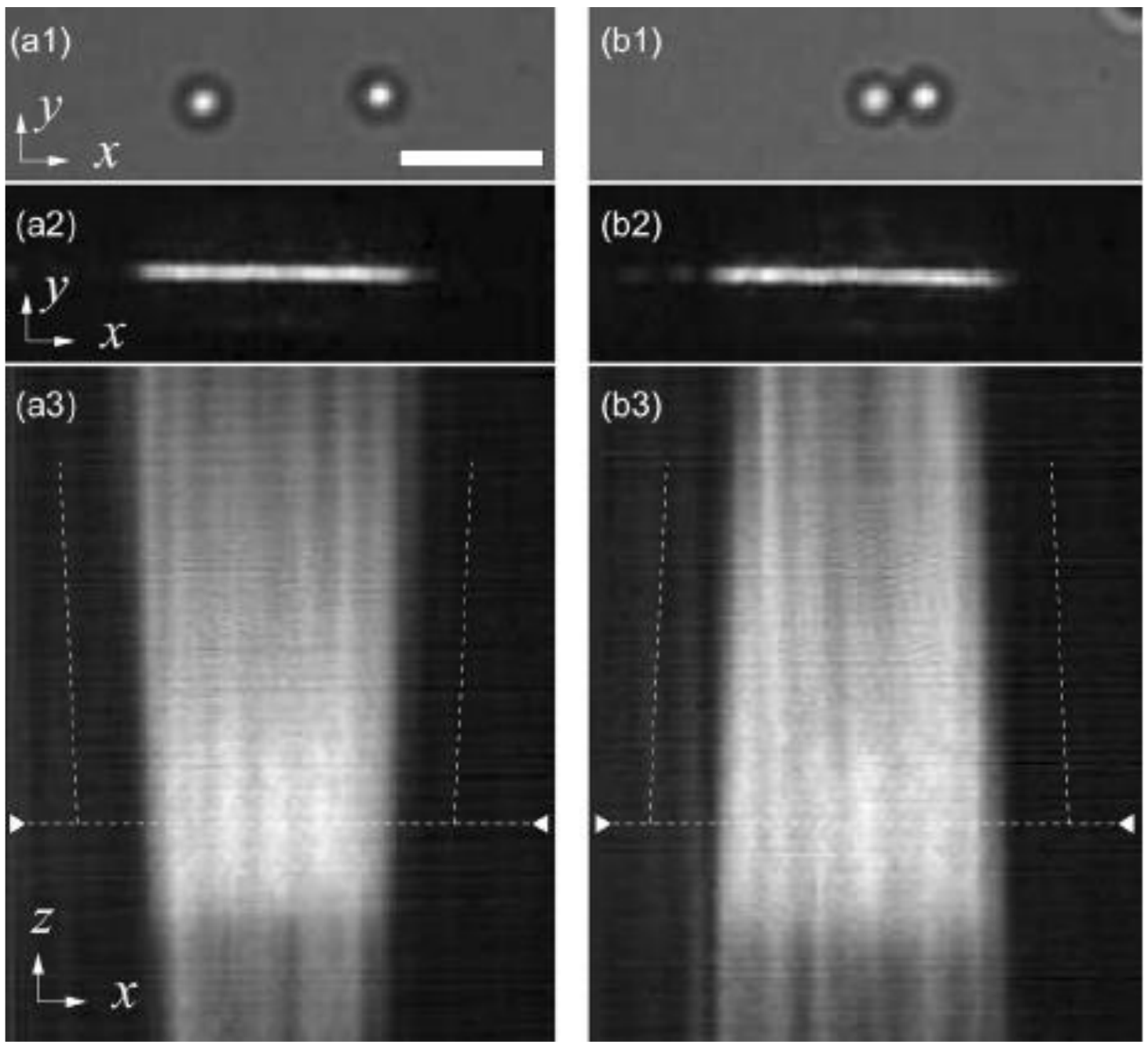

**Fig. 26** Optical trapping with a parabolic phase gradient. Two silica beads (1.5 $\mu$m in diameter) trapped in (a1) positive and (b1) negative parabolic phase gradients and the corresponding beam intensity profiles (a2 and b2) at the trapping plane. A cross-section of the beams in the x-z plane, showing (a3) the divergence and (b3) convergence of the phase gradient. Adapted from Ref. [76]

### *4.2.3 Optical trapping with optical solenoids*

In 2010 a new class of beams was demonstrated, so-called the optical solenoid, which is non-diffracting solutions of the Helmholtz equation in cylindrical coordinates[146]. Perhaps the most notable property of these beams is their 3D spiralling intensity profiles, whose wavefronts carry an independent helical pitch. Moreover, their radial intensity distribution remains invariant in the spiralling frame of reference. These beams are capable of trapping microparticles along with the helical beam intensity profile and transporting them along the spiral trajectory via a phase gradient. Mathematically, these beams can be synthesized as a superposition of m-th Bessel beams as,

$$u_{\gamma,\ell}(\mathbf{r},z) = \sum_{m=[\ell-\gamma k]}^{|\ell|} \frac{\ell-m}{\gamma^2} J_m(q_m R) \exp\left[\frac{i(\ell-m)}{\gamma} z\right] \exp[im\theta] J_m(q_m r), \quad (33)$$

where, $q_m^2 = k^2 - (\ell - m)^2/\gamma^2$ and $[\ell - \gamma k]$ represents the integer part of $\ell - \gamma k$. Figure 27 (a) shows the 3D intensity profile of $I_{\gamma,\ell}(\mathbf{r},z) = |u_{\gamma,\ell}(\mathbf{r},z)|^2$ for $kR = 10$, $\theta = 30^{\circ}$ and $\ell = 10$. To generate such an optical solenoid beam, the required hologram was created on a phase-only liquid crystal SLM and projected into a sample cell containing colloidal silica beads (1.5 μm in diameter) immersed in water, through a 100x microscope objective. The transmitted beam was reflected by a mirror mounted on a translation stage and imaged onto a CCD camera. Figure 27 (b) shows the experimental 3D intensity profile of the beam, where the radius of the helical trajectory transverse to the beam axis was $R = 5$ μm. Microspheres can be trapped in the solenoid beam and transported along its helical path (Fig. 27 (c)). Importantly, the direction of their motion upwards or downwards along the beam axis can be controlled by alternating the sign of the topological charge. The grey-scale image in Fig. 27(c) was created by the superposition of six video frames capturing the same microsphere at different time instances. This experiment demonstrates that a suitable combination of phase- and intensity-gradients can exert retrograde forces on microparticles, which can be transported against the beam propagation direction.

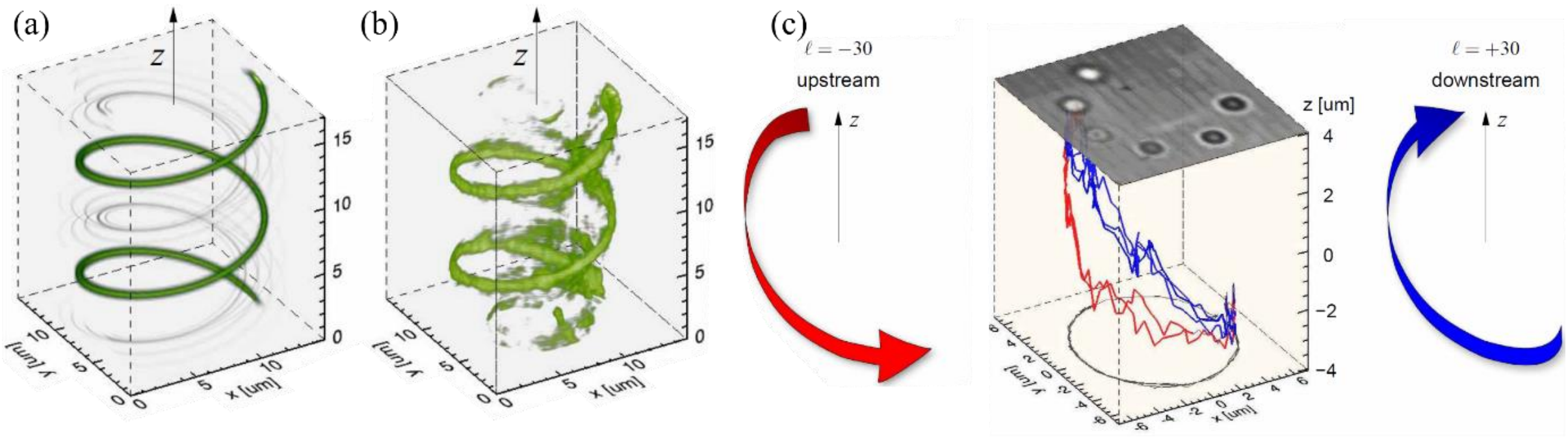

**Fig. 27** Optical manipulation with 3D solenoid beams propagating parallel to the z axis. Theoretical (a) and experimental (b) profiles of the beam in 3D. (c) Experimental trajectories of trapped particles transported downwards ($\ell = +30$) or upwards ($\ell = -30$) along with the helical intensity profile in the optical solenoid beam. Adapted from Ref. [146].

### *4.2.4 Optical trapping and transporting of particles along 3D trajectories*

The use of $LG_0^\ell$ modes to orbit particles confines the motion of the trapped particles to the plane perpendicular to the propagation direction moving around the annular intensity profile, where the azimuthal phase variation drives the particles via scattering forces. Therefore, the trajectory of trapped particles is determined by the intensity and phase profiles of the beam, both of which are always interlinked. Newer approaches proposed advanced techniques in which the intensity and phase profiles of the beam are independent from each other. These allow the creation of 3D parametrized curves (trajectories) of trapped particles in the form of $\boldsymbol{R_0}(s) = (x_0(s), y_0(s), z_0(s))$ as a function of the arc length $s$[147]. To achieve the movement of particles along 3D curves, it is required to generate a hologram that produces the 3D parametrized curve at the focal plane[147],

$$\boldsymbol{R_0}(s) = R\left(\cos\left(\frac{s}{R}\right)\cos\beta, \sin\left(\frac{s}{R}\right), \cos\left(\frac{s}{R}\right)\sin\beta\right), \quad (34)$$

where $R$ is the radius of the curve projected on the x-z plane, rotated by an angle $\beta$ about the $y$ axis with $s \in [0, 2\pi R]$. Figure 28 (a) shows an experimental example of 9 microspheres (5.17 $\mu$m in diameter) trapped along the curve of radius $R = 9$ $\mu$m tilted at an angle $\beta = \pi/4$ rad. Fig. 27 (b) shows the corresponding schematic representation in 3D. More complex structures of light can be generated by the same principle. For example, two knotted rings of light, tilted at opposite angles $\beta = \pm\pi/8$, are shown in Fig. 28 (c). Here, the radius of the ring is $R$ = 12.9 $\mu$m and their centers are separated by $R$/2. Figure 28 (d) is an experimental demonstration showing that the rings act as two 3D optical traps and are capable of organizing colloidal silica spheres. Interestingly, the trapped spheres could freely move between these knotted rings, which are schematically shown in Fig. 28 (e).

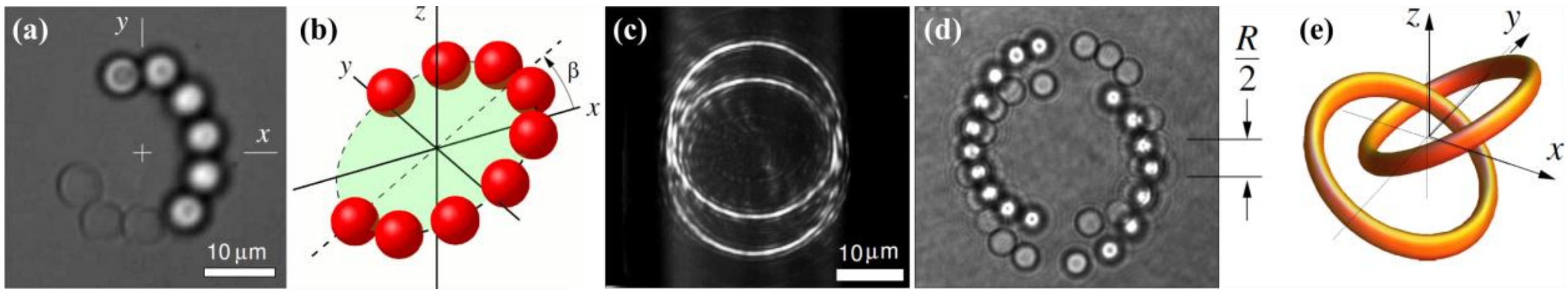

**Fig. 28** Optical trapping and transporting of microparticles along 3D parametrized trajectories**.** (a) Particles trapped along a single ring in 3D. (b) Schematic representation of (a). (c) Experimental intensity distribution of two tilted ring traps with opposite inclination. (d) Colloidal silica spheres trapped in the two rings of (c). (e) Schematic 3D representation of the knotted rings of (c,d). Adapted from Ref. [147].

More recently, alternative techniques have been proposed for the generation of arbitrary 3D parametrized curves of the beam with high intensity and independent control of the phase, also known as freestyle optical traps[148-152]. The main idea behind the generation of a light beam that focuses into a parametric curve of the form $\mathbf{c}(t) = (\mathbf{R}(t), u_z(t))$, with $\mathbf{R}(t) = (R(t)\cos t, R(t)\sin t)$, is to display on an SLM a hologram that generates the polymorphic beam[151],

$$E(\mathbf{r}_0) = \int_0^T g(t)\exp\left[-\frac{ik}{2f^2}u_z(t)\mathbf{r_0^2}\right]\exp\left[\frac{ik}{f}\mathbf{r_0}\mathbf{R}(t)\right]\mathrm{d}t, \quad (35)$$

where, $\mathbf{r}_0 = (x_0, y_0)$ represent the transverse spatial coordinates and the parameter $T$ represents the maximum value the azimuthal angle can take. Further, the function $g(t)$ is a complex value function which, as we will show later, plays a major role in the design of the curved laser trap. The desired optical field is generated in the far field, which is achieved by means of a lens of focal length $f$, acquiring the specific complex amplitude distribution,

$$\tilde{E}(\mathbf{r}, \mathbf{z} = \mathbf{u}_z(t)) = -\frac{i\lambda \exp[iku_z(t)]}{f} \int_0^T g(t)\, \delta\left(\frac{1}{f}(\mathbf{R}(t) - r)\right) \mathrm{d}t. \quad (36)$$

Here, the function $g(t) = |g(t)| \exp[i\Phi(t)]$ accounts for the phase of the trapping beam, which provides the mechanism behind the propulsion of the trapped particles along the 3D parametric curve. More precisely, the variation of the phase of $\tilde{E}(t)$ is controlled by the function $\Psi(t) = \Phi(t) + ku_z(t)$, which for convenience can be expressed as,

$$\Psi(t) = \frac{2\pi\ell}{S(T)} S(t), \quad (37)$$

where S(T) is a real function. Further, the parameter $\ell$ is associated to the accumulation of the phase along the entire parametric curve and for closed curves coincides with the topological charge of the beam. Crucially, since the force exerted on the particles is proportional to the phase gradient, their speed can be controlled on-desired by means of the function $S(t)$. For example, to achieve a uniform propelling force and therefore a uniform speed this function can be set to,

$$S(t) = \int_0^t |\mathbf{c}'(\tau)| \mathrm{d}\tau. \quad (38)$$

Here $|\mathbf{c}'(\tau)| = \sqrt{R(t)^2 + R'(t)^2 + u_z'(t)^2}$ and $\mathbf{c}'(t) = \mathrm{d}\mathbf{c}/\mathrm{d}t$. The direction of motion of the particles can also be controlled by simply reversing the sign of the topological charge $\ell$. See Ref. [149-151] for a detailed description of the mechanism behind the optical trapping and the propulsion forces involved in this technique. By way of example let's consider the case

$$E(\mathbf{r}) = \frac{1}{T} \int_0^T \mathbf{\Phi}(\mathbf{r}, t) \varphi(\mathbf{r}, t) |\mathbf{c}'_2(t)| dt, \quad (39)$$

to be encoded on an SLM[150], where $t \in [0, T]$ and $\mathbf{c}_2(t) = \left(x_0(t), y_0(t), z_0(t)\right)$ is a 3D parametrized curve and $\mathbf{r} = (x, y)$. Further, the derivative $\mathbf{c}'_2(t) = \mathrm{d}\mathbf{c}_2(t)/\mathrm{d}t$ is associated with the length $L = \int_0^t |\mathbf{c}'_2(t)| dt$, where $|\mathbf{c}'_2(t)| = (x'_0(t)^2 + y'_0(t)^2)^{\frac{1}{2}}$ of the 3D curve. The function $\mathbf{\Phi}(\mathbf{r}, t)$ can be written in the form of[150],

$$\mathbf{\Phi}(\mathbf{r}, t) = \exp\left(\frac{i}{\rho^2}[y x_0(t) - x y_0(t)]\right) \exp\left(\frac{i2\pi m}{S(T)} S(t)\right), \quad (40)$$

where

$$S(t) = \int_0^t [x_0(\tau) y_0'(\tau) - y_0(\tau) x_0'(\tau)] \mathrm{d}\tau, \quad (41)$$

with $m$ and $\rho$ being the free parameters to control the phase gradient and the ring radius $R = \lambda f / 2\pi\rho$ at the focal plane, respectively. The term $\varphi(\mathbf{r}, t)$ is a quadratic phase term of the form,

$$\varphi(\mathbf{r}, t) = \exp\left(\mathrm{i}\pi \frac{[x - x_0(t)]^2 + [y - y_0(t)]^2}{\lambda f^2} z_0(t)\right), \quad (42)$$

where, $z_0(t)$ is a defocusing parameter defined along $\boldsymbol{c}_2(t)$. Figure 25 shows an example of the phase and intensity distributions that are obtained. More precisely, **Fig. 29(a)** illustrates the phase profile of a ring trap with a radius of $R$=5 $\mu$m with topological charge $m$=30. The corresponding intensity profile is shown on the right column. **Figure 29(b)** shows the phase profile of a hypocycloid-shaped trap, parametrized as $x_0(t) + iy_0(t) = \rho[2\exp(it) + \exp(-i2t)/2]$, where the phase gradient varies along with the trap, as shown in **Fig. 29 (b1).**

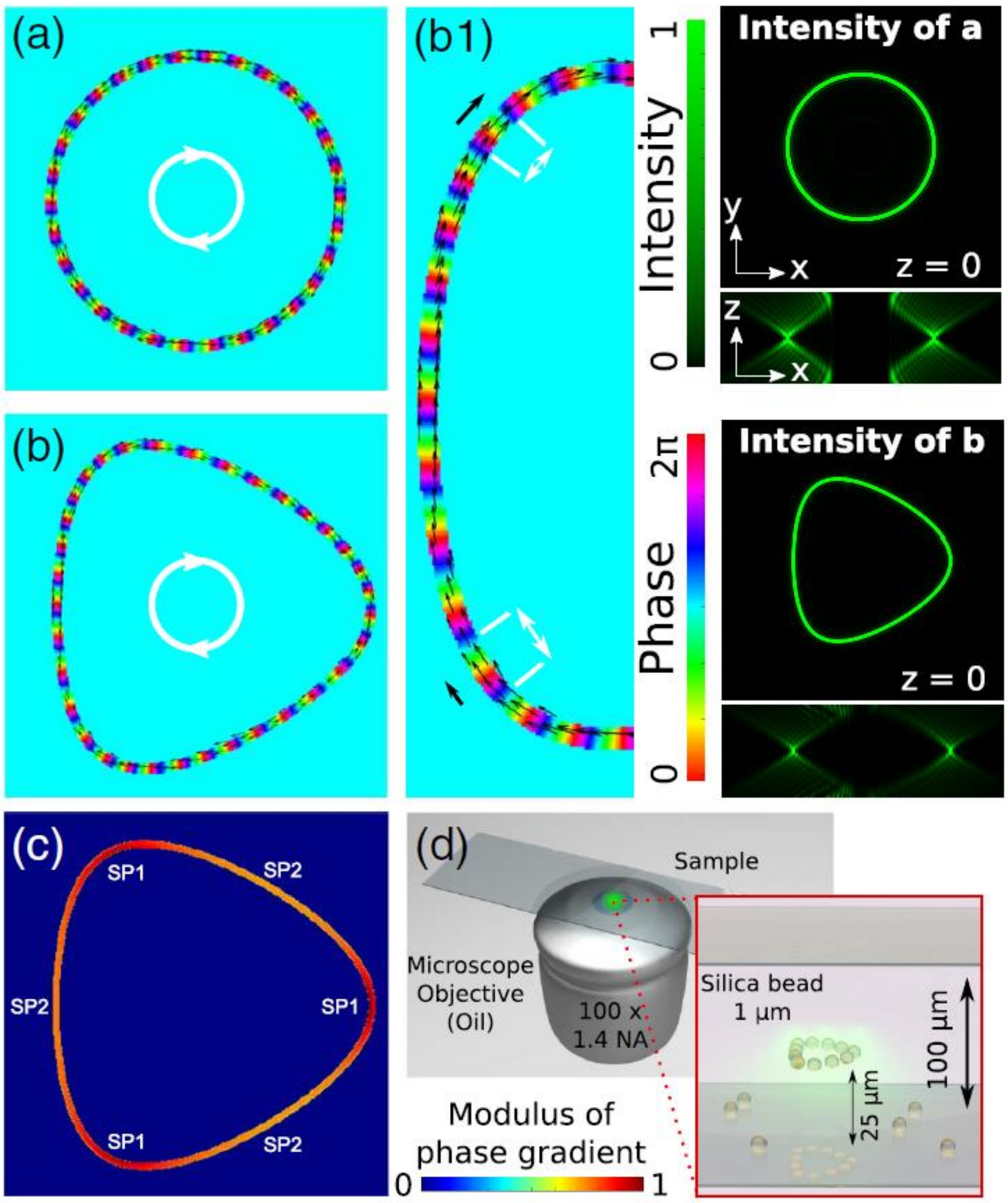

**Fig. 29** Optical trapping with arbitrary 3D parametrized curves of the beam. (a) Phase profile of a ring trap with topological charge m=30 and (b) a triangle trap with topological charge m=34. (b1) An expanded view of a section in (b), where the black arrows indicate the vector field gradient. (c) Phase gradient modulus corresponding to the beam in (b), where SP1 and SP2 indicate stationary points where the modulus is maximum and minimum, respectively. The top right images show a 2D intensity profiles of the focused beam in the x-y and x-z planes. (d) Schematic representation of the optical tweezers, where the beam is focused into a sample cell containing silica beads of 1 μm through a microscope objective with NA=1.4. Adapted from Ref. [149].

**Figure 30** illustrates the complex motion of particles propelled along a helically modulated toroidal surface[149]. Figure 30 (a) shows the time-lapse images of trapped particles over a period of seven seconds along the decagon trajectory in Fig. 30 (b), which is a projection of the 3D toroidal beam (see Fig. 30 (c)) onto the *x*-*y* plane. Fig. 30 (d) shows experimental intensity profiles of the 3D toroidal beam at two different axial planes of $z = \pm 1\mu$m.

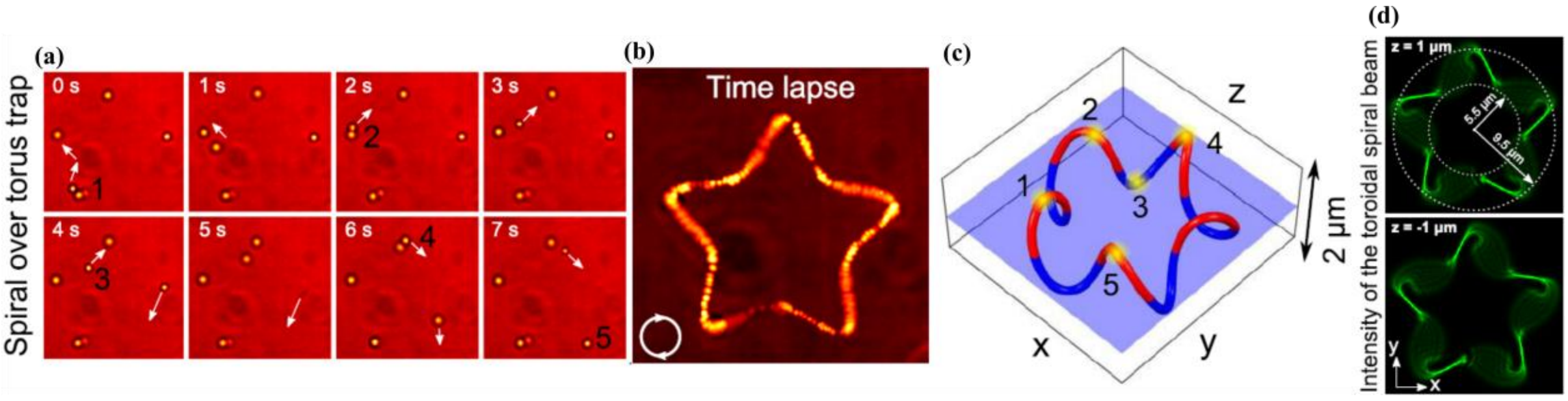

**Fig. 30** Optical trapping with 3D toroidal-spiral beams. (a) Time-lapse images of trapped microparticles moving along the beam over 7 seconds, which results in (b) a decagon trajectory. (c) 3D schematic representation of the toroidal-spiral curve, where the color scheme indicates the axial *z* position of the curve. (d) Intensity profiles of the toroidal beam at two different axial planes. Adapted from Ref. [149].

The general approach to determine the 3D parametric curve consist of first defining a set of $m$ points $\mathbf{P}^{(n)}$, $n = 1,2,3,..,m$, through which it must pass. The problem then reduces to finding a piecewise parametric curve that passes through these points. This problem was solved by J. A. Rodrigo *et. al.*, by using a set of parametric curves, known in computer graphics theory as Beziér Splines $\mathbf{b}_n(\tau)$, that join the points $\mathbf{P}^{(n)}$. As explained in Ref. [151], Each Bezieér spline is defined by four points, two knot points and two associated points, the former denoted by $\mathbf{P}_s^{(n)}$ and $\mathbf{P}_e^{(n)}$ and the later by $\mathbf{T}_s^{(n)}$ and $\mathbf{T}_e^{(n)}$, where the subscripts $s$ and $e$ denote the starting and ending points. Each Beziér Spline is defined in terms of these points as,

$$\mathbf{b}_n(\tau) = (1-\tau)^3\mathbf{P}_s^{(n)} + 3\tau(1-\tau)^2\mathbf{T}_s^{(n)} + 3\tau^2(1-\tau)^2\mathbf{T}_e^{(n)} + \tau^3\mathbf{P}_e^{(n)}, \quad (43)$$

where $\tau \in [0,1]$. The parametric curve is then given as,

$$\mathbf{c}(t) = \{\mathbf{b_1}(\tau), \mathbf{b_2}(\tau), \dots, \mathbf{b_m}(\tau)\}, \quad (44)$$

which has to be continuous and differentiable, see Ref. [151] for further details. An example of a parametric curve constructed in this way is shown in the left panel of **Fig. 31 (a)**, where six Beziér splines, represented by different colors, were used. The intensity and phase profile of a laser beam focused along this curve is shown in the middle and left panel, respectively of the same figure. Crucially, this construction method allows the real-time reconfiguration of $\mathbf{c}(t)$ by simply shifting the knots points of the Beziér splines. By way of example, the left panel of **Fig. 31 (b)** illustrates the progressive change of $\mathbf{c_1}(t)$ into $\mathbf{c_N}(t)$ by shifting one knot point.

Here, four micro-particles, labelled as **A**, **B**, **C** and **D**, trapped along the curve, are illustrated, where the arrows show their motion direction. The actual performance of such real-time reconfigurable optical trap was demonstrated experimentally. To illustrate this, the middle panel of **Fig. 31 (b)** shows an example of the transition from $\mathbf{c_1}(t)$ into $\mathbf{c_{22}}(t)$ in an actual optical trap experiment, where multiple micro-particles were set to move along the different curves. Here, only the cases defined by the curves $\mathbf{c_1}(t)$, $\mathbf{c_{12}}(t)$, $\mathbf{c_{16}}(t)$ and $\mathbf{c_{22}}(t)$ are shown. A time lapse of the whole transition is shown in the right panel of the same figure, where the motion direction of the particles is indicated by the direction of the arrows.

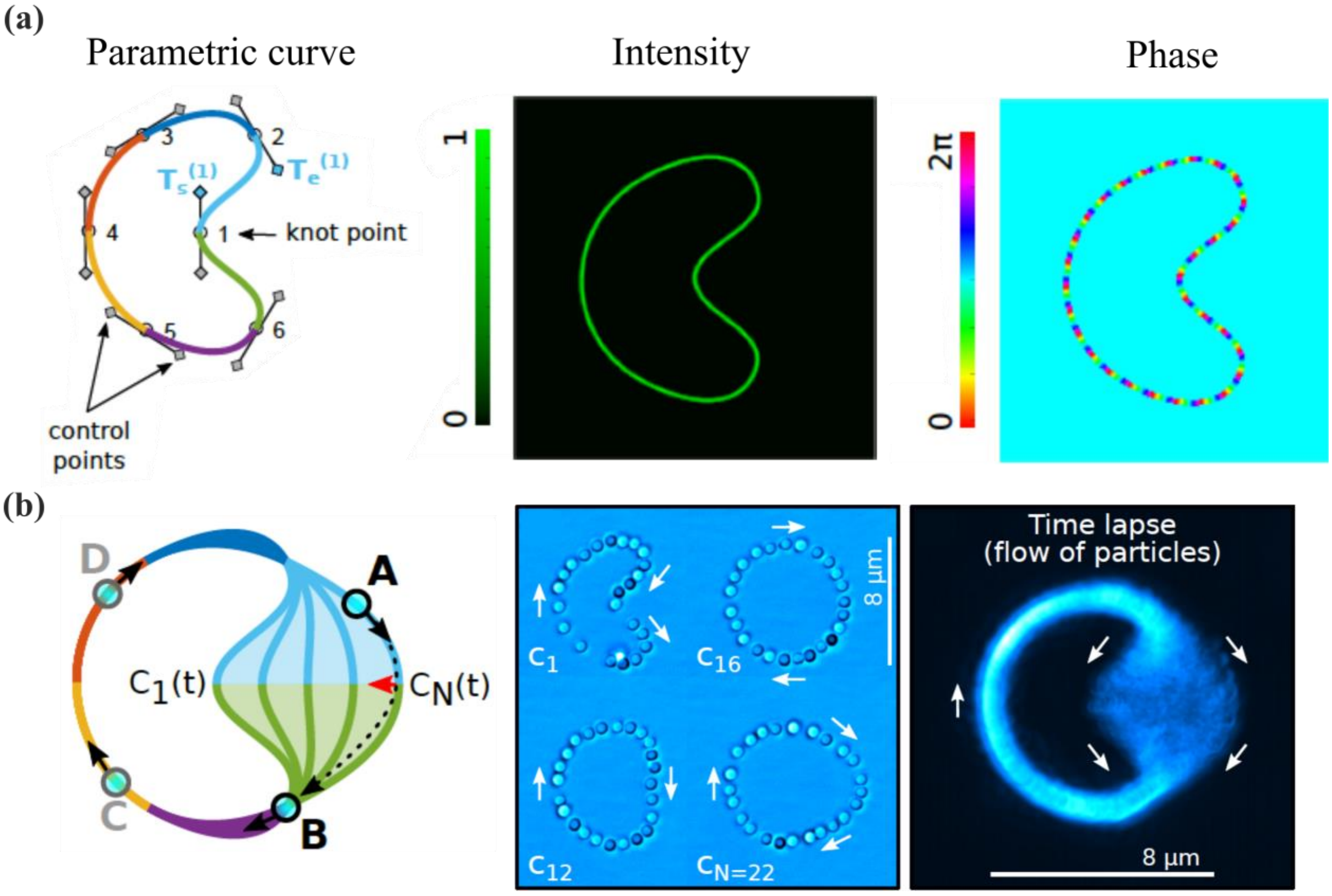

**Fig. 31** Example of a Beziér parametric curve and its application to reconfigure in real-time the trajectory of microparticles. (a) Construction of a parametric curve using Beziér splines (left), intensity (middle) and phase (left) of a laser beam following this parametric curve. (b) Example of real-time reconfiguration of the curve shown in (a) and its application in a real-time reconfigurable optical trap (middle). Adapted from Ref. [151].

### *4.3 Optical trapping with vector beams*

#### *4.3.1 Engineering the optical forces via vector's beam polarization*

Controlling the polarization state of light is of paramount importance in many fields of pure and applied sciences, however, it is only in recent years that increasing attention has been given to complex vector fields, often referred to as classically entangled beams[153,154]. In optical tweezers, the degree of polarization plays a crucial role in vector beams. For example, when tightly focused, different states of polarization of the input beams can lead to distinctive optical forces at the trapping plane. Vector beams with radial and azimuthal polarization are two special cases. Radial polarization possesses a strong field component along the propagation direction, while this is absent from azimuthal polarization[155-158]. Figure 32(a) schematically illustrates this behavior. In a radially polarized vector beam, the electric field oscillates radially in the transverse plane, with no longitudinal component. However, when tightly focused through a high numerical aperture lens, the beam is strongly refracted towards the focus, giving rise to longitudinal electric field components, as illustrated in Fig. 32(b). In the case of an azimuthally polarized beam (see Fig. 32(c)), the electric field oscillates circularly along a transverse plane to the beam propagation that cannot generate a longitudinal electric field component under tight focusing conditions. Indeed, radially polarized vector beams are known for their ability to produce the smallest focal spot size among the family of vector beams[159,160].

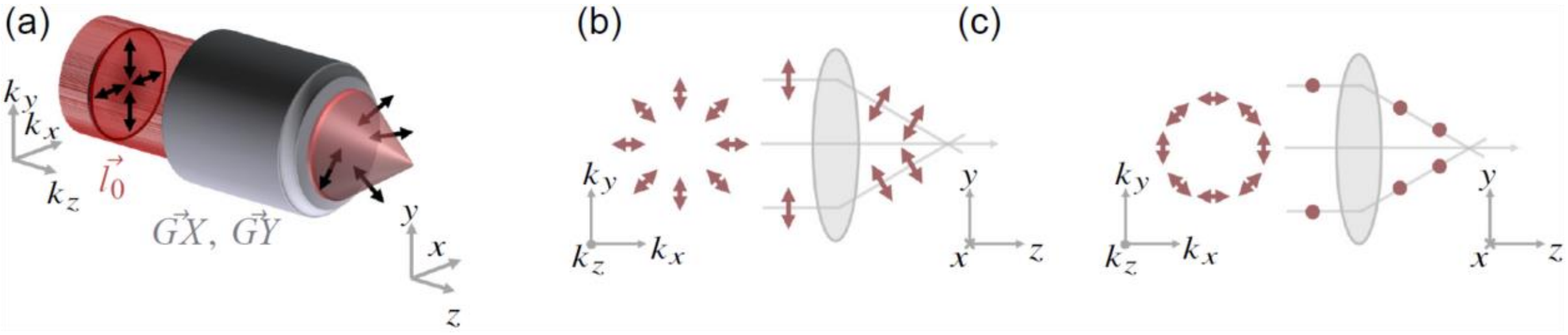

**Fig. 32** (a) Schematic representation of cylindrical vector beams under tight focusing conditions. Tightly focused vector beams with (b) radial polarization and (c) azimuthal polarization. Adapted from Ref [159].

Crucially, the use of complex states of polarization in tightly focusing systems allows engineering of light fields, such as polarization and intensity. For example, the shape of the focused beam can be designed to flat-top by creating appropriate superpositions of modes with specific polarization states[161]. The addition of diffractive optical elements such as SLMs at the input plane, allow for the generation of exotic beam shapes, such as chains of light fields containing 3D dark volumes (low intensity spots) along the light field distributions[162,163]. Precise control over the beam shape to create a perfect spherical spot can be achieved using counterpropagating vector beams, which are constructively interfering at the common focal plane[164-166]. In this configuration, radially polarized vector beams can produce the spherical focal spot containing solely longitudinal electric fields, in contrast to the use of azimuthally polarized vector beams, which only contain the transverse components in the electric field[167]. It is worth noting also that the arbitrary control of the polarization state of a focused field can be achieved by creating coaxial superpositions of two vector beams with radial and azimuthal polarization, each of which has a different weighting factor[168].

*4.3.2 Enhancement of optical forces*

In recent years, it has been widely reported that the state of polarization plays a crucial role in optical tweezers, which was ignored in the early days of the study. Recent reports indicate that radially polarized beams can enhance the axial trapping efficiency up to twice that of linearly polarized light beams at the expense of reducing the transverse trapping efficiency by up to a half[169]. Furthermore, azimuthally polarized vector beams have demonstrated higher transverse trapping efficiencies than those of radially polarized ones[170]. These features of cylindrical vector beams are particularly useful for trapping metallic particles exhibiting high levels of scattering and absorption of light. For example, the enhanced axial trapping efficiency in radially polarized beams due to the strong axial field component with a null axial Poynting vector allows the trapping of metallic particles[171-173]. In addition, the concurrent use of $\pi$-phase radially and azimuthally polarized vector beams can enhance optical forces to a trapped particle by tuning the relative phase between the eigenmodes comprising the beams[174].

*4.3.3 Optical trapping with an array of multiple vector beams*

Recently, the generation of multiple vector beams from a multiplexed hologram displayed on a single SLM was demonstrated[175]. Each vector beam was generated by a pair of independent holograms superimposed on an SLM. As they propagate along independent optical paths using different diffraction gratings, their polarization states can be modulated independently prior to a coaxial recombination through a Polarizing Beam Splitter (PBS). In this way, an array of multiple vector beams can be generated from multiple pairs of multiplexed holograms with independent control over the transverse spatial positions of the beams and their polarization states. Figure 33 (a) shows the intensity profile of nine vector beams with their polarization distribution in Fig. 33 (b). This approach allows the manipulation of multiple particles at different vector beams possessing different polarization states. To experimentally deliver a desired array of vector beams, a 4$f$ lens system relays the SLM plane to the entrance pupil of the microscope objective, as shown in Fig. 33 (c). With appropriate diffraction angles, the generated beam pairs can be projected to the same position in the trapping plane[175]. Figure 33(c) inset 2 shows an experimental image of 4 vector beams at the trapping plane, after passing through the linear polarizer.

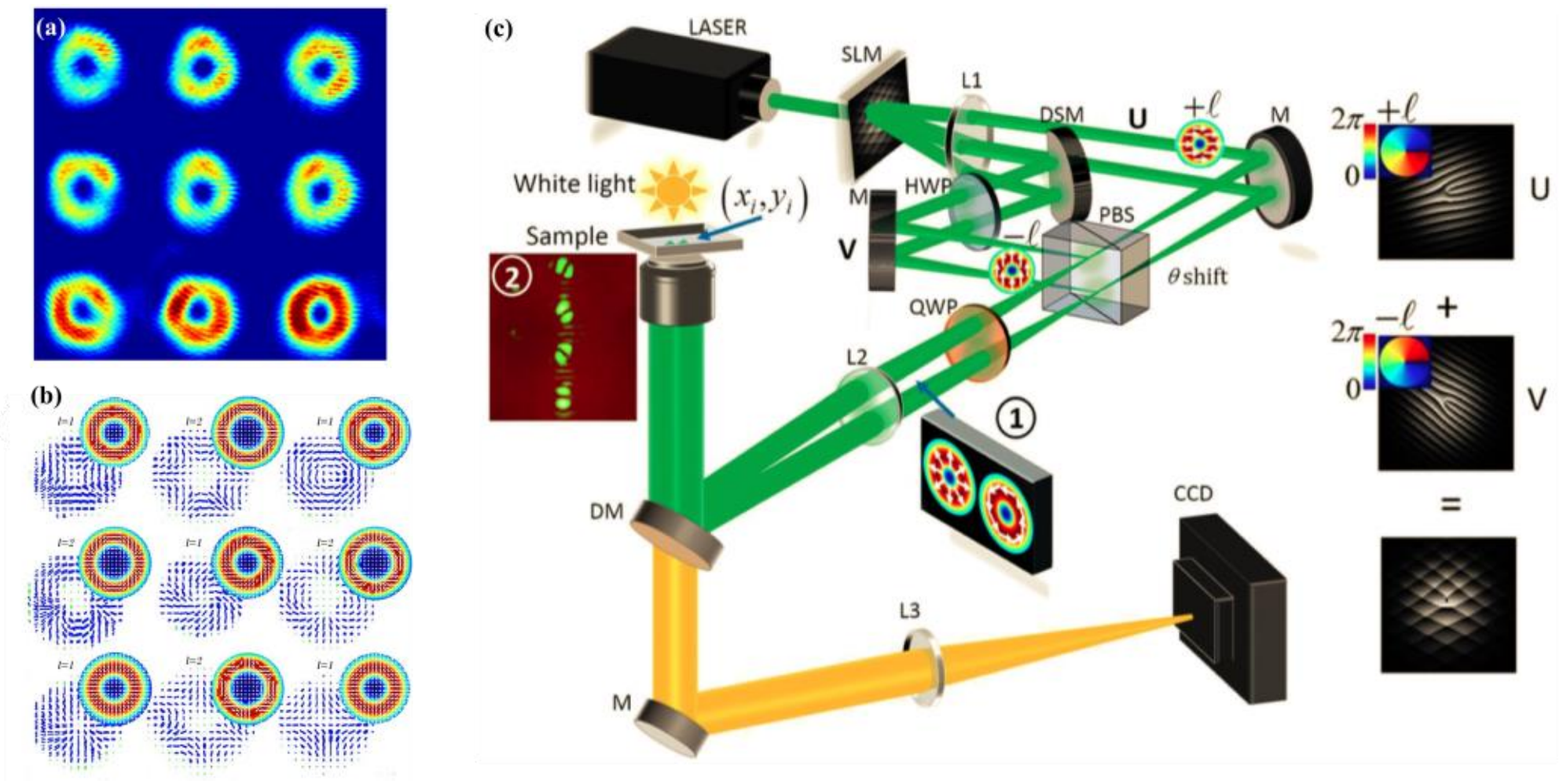

**Fig. 33** Generation of vector beam arrays. Experimental intensity profiles (a) and polarization distribution (b) of 9 vector beams generated from a single hologram. (c) Schematic representation of the experimental setup to generate multiple vector beams. Insets on the right illustrate the multiplexed hologram pair for the generation of two scalar beams travelling along two separate optical paths. Inset 2 shows the generated four vector beams in the trapping plane. Adapted from University of the Witwatersrand [175].

### *4.3.4 Optical trapping with tractor beams*

In contrast to the above discussed optical tweezers, where the optical force exerted on the micro and nanoparticles produces an acceleration along the same direction of the photon flow, in an optical tractor beam, the optical forces drag small microparticles against this flow, *i.e.*, towards the photons source[28, 176-180]. In contrast to a conveyor belt that requires gradient forces to pull microparticles towards the source of the photons [118], the pulling forces of a tractor beam originate from a momentum conservation law. In essence, when the forward scattered light is more collimated than the incident light, and if the backward scattering is weak enough, the momentum conservation law predicts the existence of a pulling force[177, 181]. This can be generated for example by the interference of two plane waves (see **Fig. 34a**). As illustrated here, each beam propagates along the wave vector $\mathbf{k_1}$ and $\mathbf{k_2}$, whereas the resulting beam propagates in the forward direction along $\mathbf{k_1} + \mathbf{k_2}$. Under these conditions, the majority of the scattering is produced in the forward direction, which causes a pushing force $\boldsymbol{F_z}$ in the opposite direction, as also indicated here. Crucially, this effect polarization-dependent allowing to switch between a pushing and a pulling force, by simply changing the polarization of the incident waves, as illustrated in **Fig. 34b**, and **Fig. 34c**, respectively. Crucially, the pulling effect has been also explained in terms of the optical singularity of the Pointing vector around the scatterer[182].

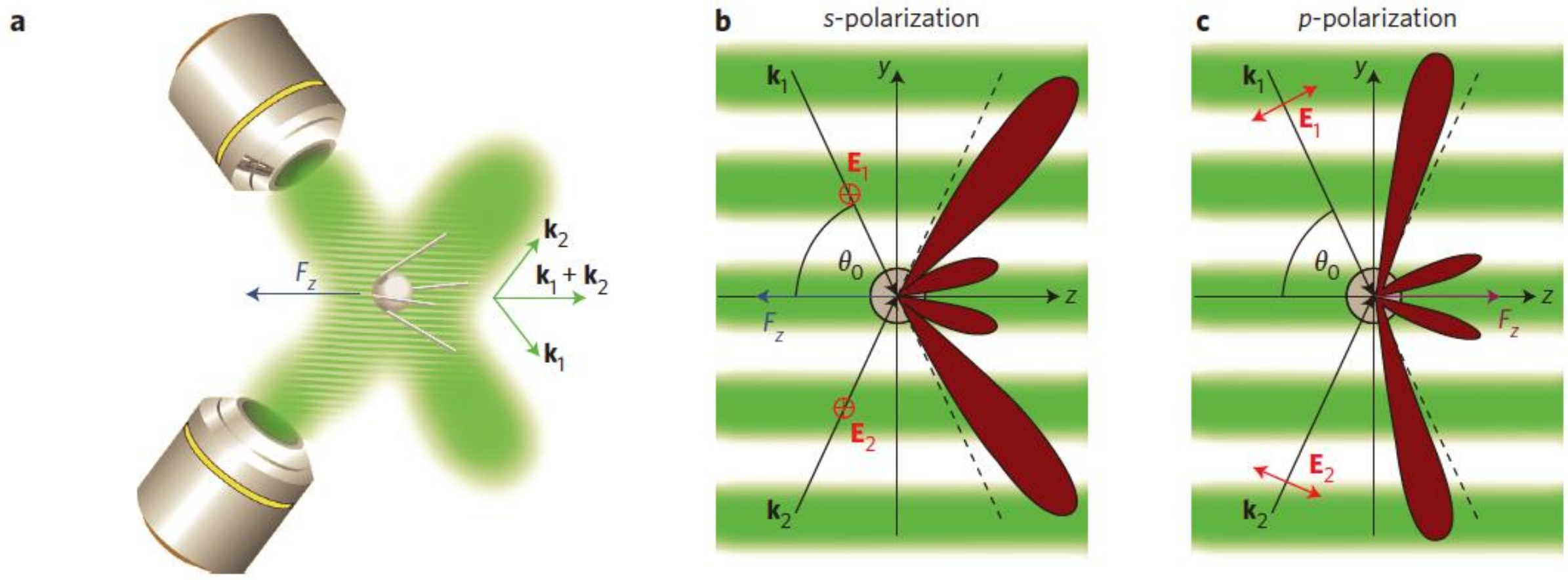

**Figure 34** In **a** it is shown a conceptual representation of a tractor beam generated from the superposition of two waves propagating along the wave vectors $\mathbf{k_1}$and $\mathbf{k_2}$. The beam generated from the superposition propagates in the forward direction where the scatting is stronger, generating a pulling force in the opposite direction. Crucially, this effect is polarization-dependent allowing to switch from a pulling to a pushing force, by simply changing the polarization of the incident beams, from *s*- to *p*-polarized, as indicated in **b** and **c**. Adapted from reference [181]

The optical pulling force can be achieved by utilizing structured beams, objects with specific opical parameters, structured background media, and photophoresis effect as well[180]. The tractor beam rely on photophoretic force is shown in Fig. 35. A long-distance, stable and switchable optical transport was achieved by the use of cylindrical vector beams[183]. Here, thin-walled spherical glass shells with a radius from 25 to 35 µm and a thickness from 200 to 400 nm were used, in which light can be absorbed either in the vicinity of the front side (outside) or in the rear side (inside) of the particle. In the former case, the photophoretic force pushes the particle along the direction of light propagation, whereas in the latter case the particle may be pulled against the beam propagation direction. Importantly, these two situations could be exchanged by switching from a radially polarized beam to an azimuthally polarized one or *vice versa*.

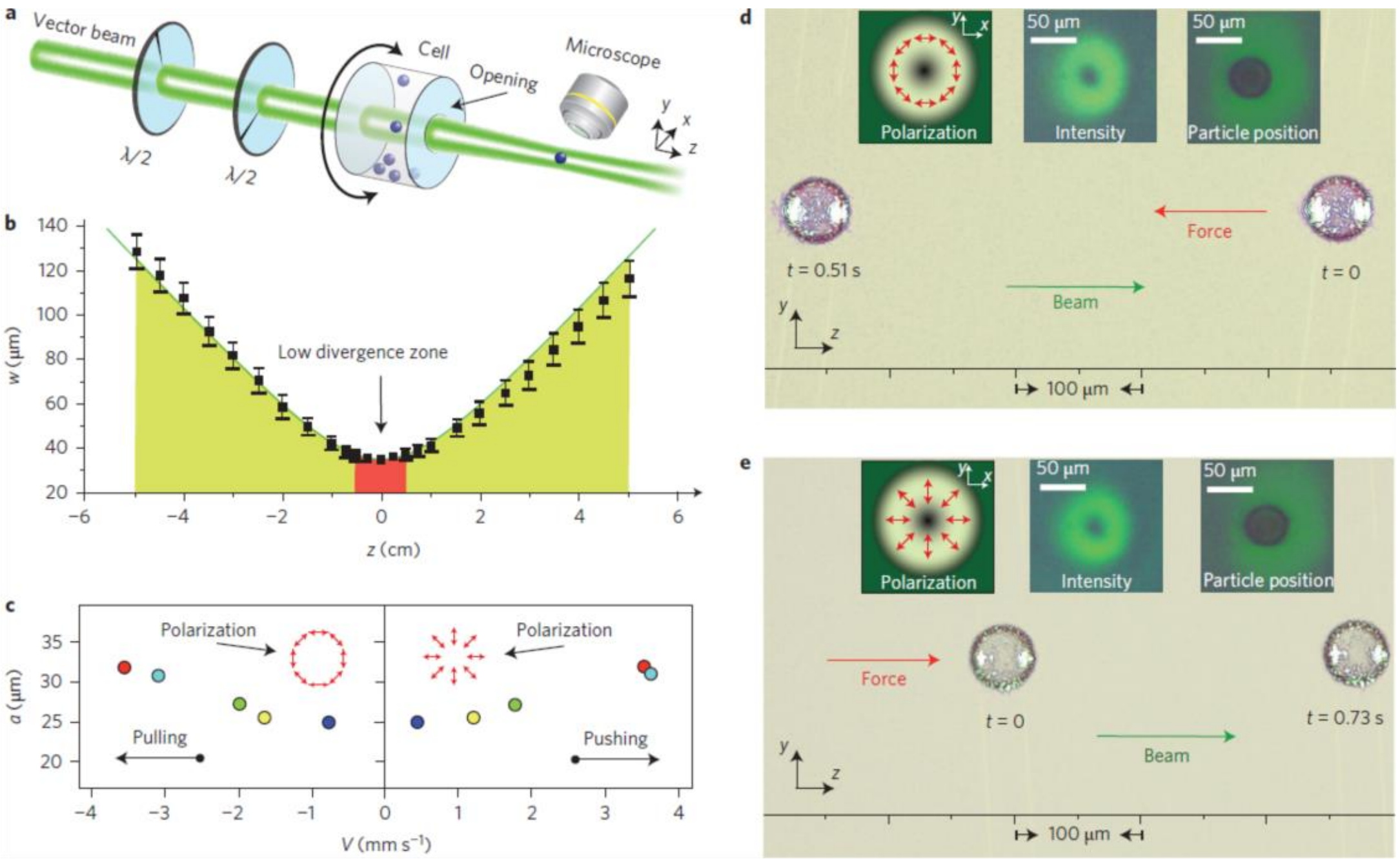

**Fig. 35** Demonstration of optical tractor beams. (a) Experimental setup showing the beam convertor and the particle dispenser. Half-wave plates (λ/2) are used to change the state of polarization of the vector beam. (b) Profile of the vector beam along the propagation direction, where the beam waist is represented in red and the region of stable trapping in yellow. (c) Velocity of glass shells as a function of their external diameter for both azimuthal (left) and radial (right) polarization, where colors indicate data obtained for the same shell size. (d) Snapshots of a shell (25μm in radius), illuminated by an azimuthally polarized vector beam, moves against the beam propagation direction (pulling) at a speed of v = 0.8 mm $s^{-1}$. (e) The same particle illuminated by a radially polarized beam moves towards the beam propagation direction (pushing) at a speed of V = 0.4 mm s−1. Adapted from Ref. [183].

Figure 35(a) schematically illustrates a part of the experimental setup for this effect. To enhance the photophoretic forces the glass shells were coated with a thin layer of Au (7-15 nm) and were placed inside a rotating cylindrical cuvette, in which the rotating axis was aligned to be parallel to the beam axis. Figure 35(a) shows the glass shells falling by gravity are trapped and pushed outside the rotating cuvette by the vector beam. Here, two half-wave plates allowed to switch the vector beam from radial to azimuthal polarization in order to generate a pushing or a pulling force. Fig. 35(c) shows the particle velocity depending on the radius of the shell, for both azimuthally (pulling) and radially (pushing) polarized beams. Figure 35(d) schematically illustrates the experimental observation of a particle pulled by an azimuthally polarized beam, whereas Fig. 35(e) illustrates the case of a particle pushed by a radially polarized beam.

### *4.4 Optical trapping and transporting of metal nanoparticles*

We know that the refraction and the dominant gradient force are the bases for stable optical trapping of dielectric particles, while the laser beam can induce relatively large scattering and absorption forces on metal particles close to their localized surface plasmon resonances. Therefore, in the early days, it was generally accepted that the stable optical trapping of metal nanoparticles cannot be obtained readily[184]. However, in 1994 Svoboda and Block demonstrated that metal Rayleigh particles can be trapped stably in three dimention[185]. In their experiment, the near-infrared laser beam was used to trap the gold nanoparticle, viz., the optical trapping was achieved off-resonance. Interestingly, in 2008, Dienerowitz et al[186] experimentally demonstrated the metal nanoparticles can be trapped by laser beam close to their plasmonic resonance. They showed that a vortex beam (Laguerre-Gaussian beam with annular profile) can confine the metal nanoparticles in the dark region of the beam centre. Since the vortex beam carries OAM, they observed the rotation of particles as well (Fig. 36 (a)). Later, Lehmuskero et al.[187] optically trapped gold nanoparticles and set them into orbital rotation with orbiting frequency of 86Hz using an LG vortex beam carrying OAM. In their experiment, to achieve a stable trap, a *q*-plate (Fig. 36(b)) was adopted to produce vortex beam with uniform intensity distribution and a circularly polarized laser beam was used to induce circular symmetric optical gradient forces. Recently, the dynamics of electrodynamically coupled metal nanoparticles in an optical ring vortex trap was studied[77]. Figure 36 (c) shows the schematic of the focused Bessel-Gauss optical vortex beam and ring trap over a gold nanoplate. They used a retroreflection geometry with a gold nanoplate mirror to generate a constant-radius optical vortex. They demonstrated that, compared with a glass coverslip, the retroreflection geometry can significantly increase the spatial confinement and optical drive force, and a superior trap can be created accordingly.

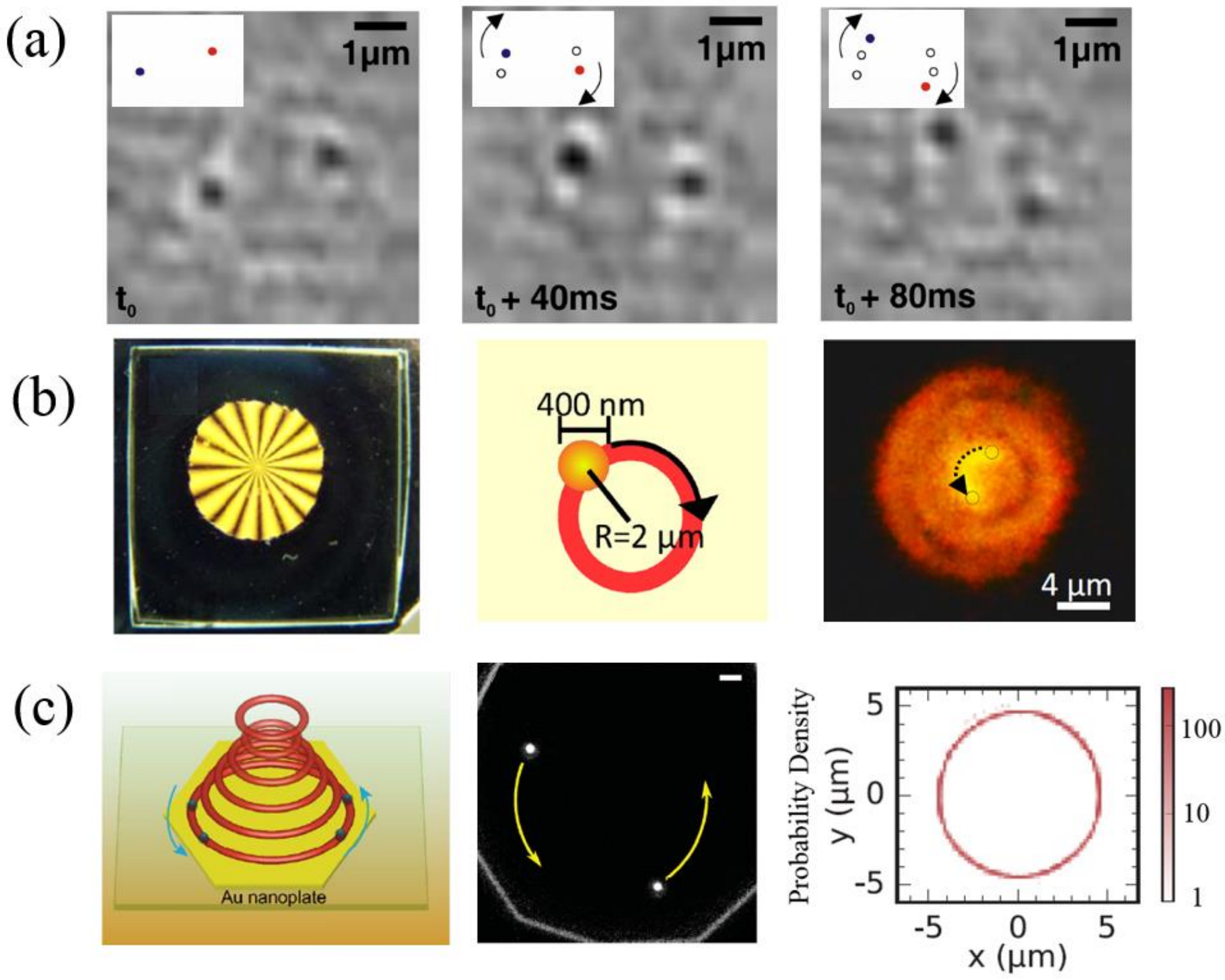

**Fig.36**. Optical trapping of metal particles using structured beams. (a) The confinement and manipulation of gold nanoparticles by LG vortex beams. The inserts indicate the locations of these two particles for different moments. Adapted from Ref. [186]. (b) Fast orbital rotation of metal nanoparticles using circularly polarized vortex beam. Left: The image of the *q*-plate; Middle: The illustration of optically trapped metal particle rotates along a circular orbit; Left: Image of the trapped particle rotates along a circular orbit. Adapted from Ref. [187]. (c) Optical manipulation of metal particles using a retroreflection geometry with a gold nanoplate mirror. Left: Schematic of the ring vortex trap over a gold nanoplate; Middle: Image of two silver nanoparticles trapped over the gold nanoplate mirror; Right: The corresponding probability densities of silver nanoparticles in the ring traps. Adapted from Ref. [77].

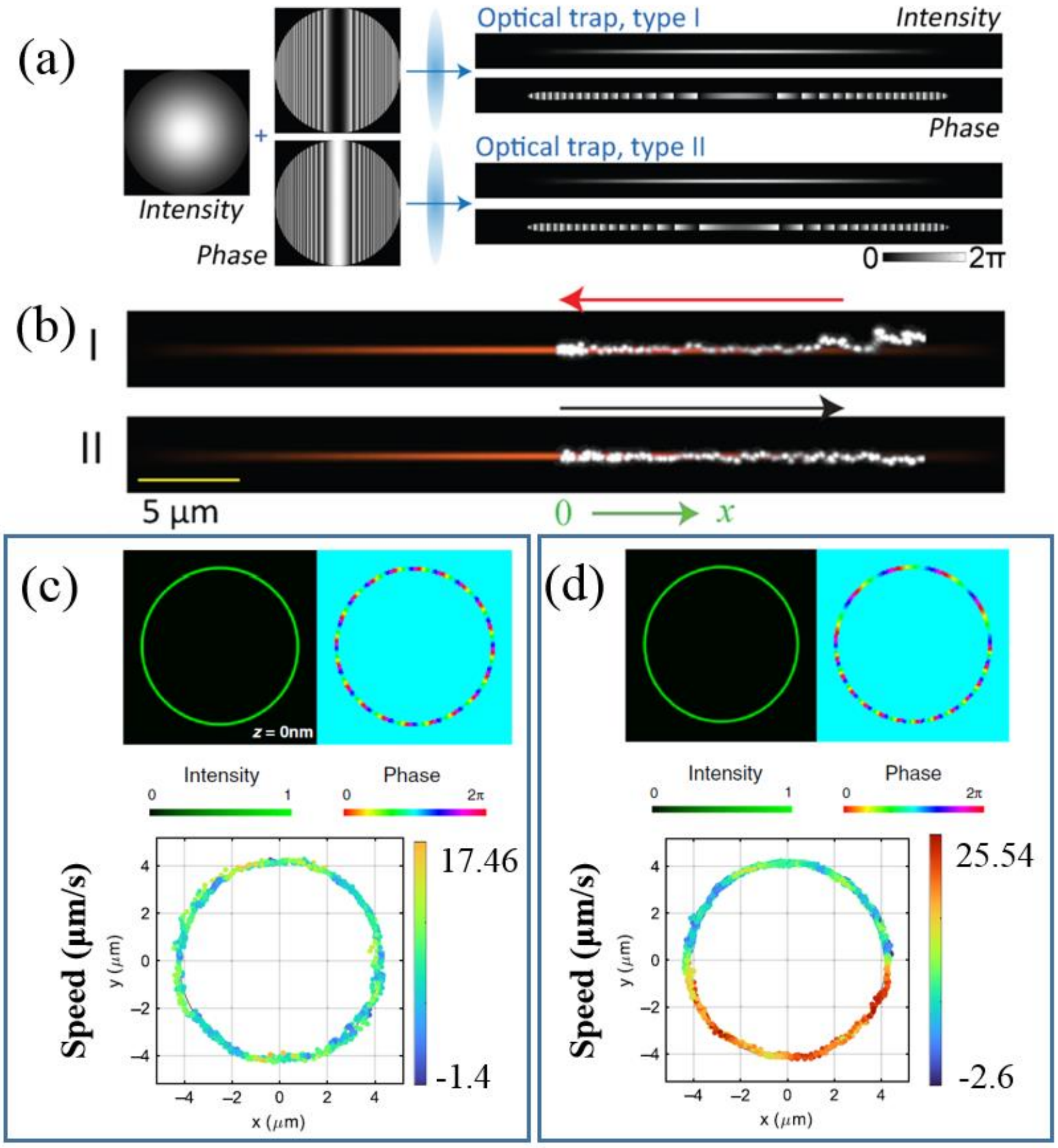

**Fig. 37**. 2D optical trap of metal particle using a structured beam with phase gradient. (a) Schematic diagram of the generating of a structured beam with phase gradient for the optical line trap. Left: The intensity profile and the designed phase masks for the optical traps of type I (top) and type II (bottom), respectively; Right: The intensity profiles and the corresponding phase profiles of the structured beams with phase gradient for the two different types of line traps. (b) Trajectory images of a single silver nanoparticle in the optical traps of type I (top) and type II (bottom), respectively. The white dots denote the silver nanoparticles. (c) Intensity and phase profiles of a vortex beam with uniform phase gradient (top) and the corresponding trajectories of an optically transported gold nanoparticle around the optical ring traps (bottom). (d) Same with those in (c) but for tailored nonuniform phase gradient. (a) and (b) adapted from Ref. [188]. (c) and (d) adapted from Ref. [190].

Moreover, similar to the case of trapped dielectric particles, the trapped metal particles can move along the beam when trapped in a structured laser beam with a phase gradient as well. Figure 37 (a) shows the intensity profile of a Gaussian beam the two different phase masks for producing the phase gradient with opposite sign, namely, the type I and type II denote the phases are modulated by a convex and concave cylindrical lens, respectively[188]. Figure 37(b) shows that that for a line optical trap the type I phase structure, the single silver nanoparticle moves from the end to the center of the line, and for a type II trap, the single silver nanoparticle moves from the center to the end. Such line optical traps were implemented for the assembly of multiple silver nanoparticles. Besides, it was demonstrated that these nanoparticles arrays could be assembled and disassembled on desired by simply changing the sign of the phase gradient. Recently, the pairwise interactions between metal nanoparticles in a vortex ring trap with transverse phase gradients was studied[189]. This study focused on the deep understanding of the multiparticle dynamics during the self-assembly of optical matter. They revealed that for small phase gradients, the total force is modulated by a separation-dependent interference effect. On the contrary, for strong phase gradients, the symmetry of the interaction between two nanoparticles brakes, resulting in a change of the distance for which a stable binding can be achieved. More recently, Rodrigo et al studied how to control the speed of metal nanoparticles by tailored phase-gradient propulsion force[190]. Figures 37 (c) and (d) show two ring vortex traps with uniform phase gradient and tailored nonuniform phase gradient, respectively. From Figs. 37 (c) and (d), we can see that the metal nanoparticle in the second trap rotates much faster than that in the first trap.

It is worth noting that the aforementioned works are limited to 2D particle rotation in contact with a surface of the coverslip, and the 3D trapping and motion control of metal nanoparticles are not achieved until 2021. Rodrigo et al. experimentally demonstrated the transport of metal nanoparticle in an optical vortex ring trap along 3D trajectory on demand[152]. To obtain 3D optical manipulation of metal nanoparticles along an arbitrary curve, they adopted a freestyle laser trap[151], which can induce a propulsion optical force due to the phase gradient of the beam. The 3D ring trap of metal nanoparticles with uniform phase gradient is shown in Fig.38 (a) shows, which indicates that the propulsion force is constant along the ring. While Fig. 38(b) shows that the propulsion force increases with the increasing angle. Besides, the tilted ring optical trapping and transport of metal nanoparticles was demonstrated as well[152].

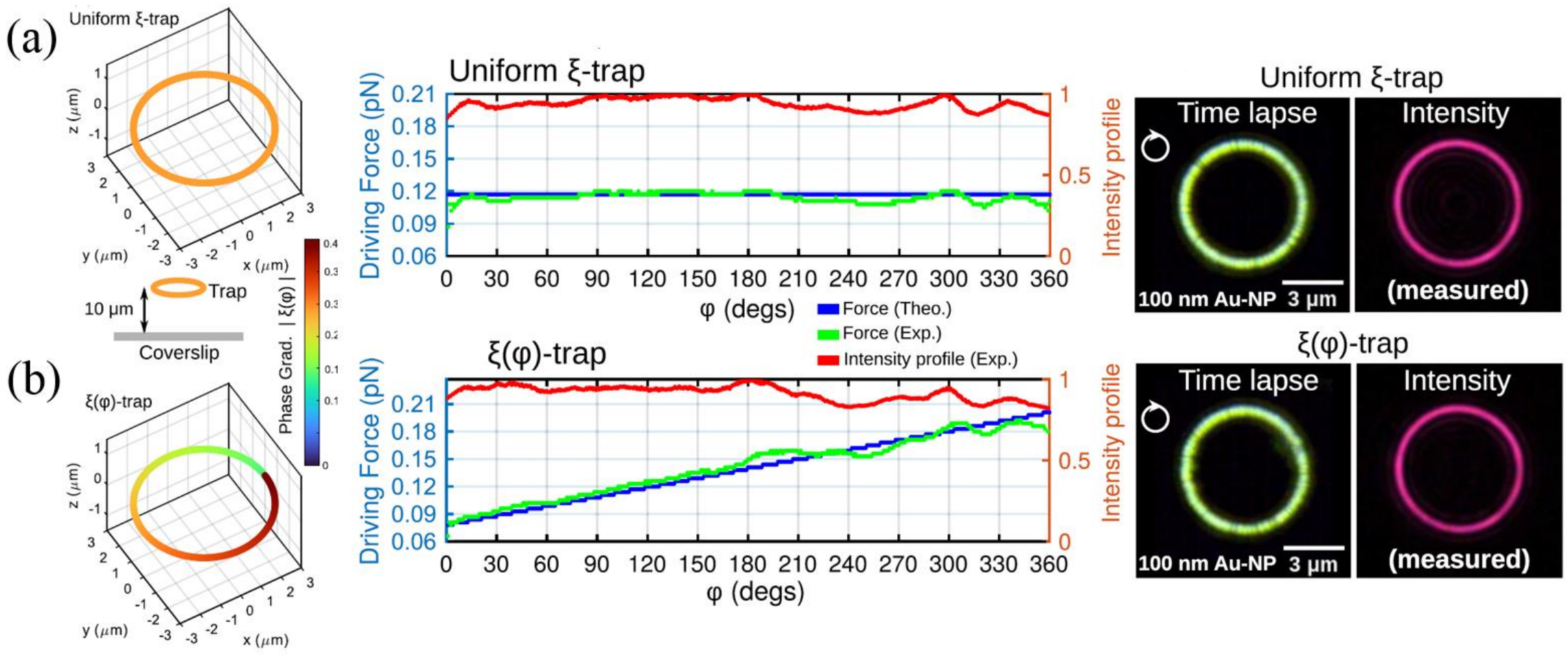

**Fig. 38**. 3D trapping and transporting of metal nanoparticles. (a) Left: Ring trap with uniform phase gradient. The inset shows that the location of the trap is 10μm from the chamber wall (coverslip); Middle: Theoretical and experimental estimated optical propulsion force along the ring of the trap. Right: The time-lapse image of the nanoparticles in the ring trap. (b) Same with those in (a) but for tailored phase gradient. Adapted from Ref. [152]

### *4.5 Levitated optomechanics with structured light beams*

Structured light has demonstrated its extensive applications in optical tweezers for micromanipulation in a liquid environment. Only recently there has been a surge of interest in levitated optomechanics using optical tweezers in air/vacuum[30, 191-217]. The dynamics of a levitated particle in a dilute gas or vacuum environment (underdamped) is fundamentally different from those trapped in a viscous medium like water (overdamped) due to the particle's finite inertia, which is typically ignored in overdamped systems (see Section 2.2). Including the inertial term, the equation for the center-of-mass motion of an optically trapped Brownian particle in one dimension (x-direction) is described by the Langevin equation

$$m\ddot{x} = -\Gamma_0\dot{x} + \kappa_0 x + F_{th}, \quad (45)$$

where $x$ is the particle position, $m$ is its mass, $\Gamma_0$ is the friction coefficient, $\kappa_0 = m\Omega_0^2$ is the trap stiffness at the trap frequency $\Omega_0/(2\pi)$. The fluctuating force $F_{th}$ is due to the random collisions from the surrounding fluid molecules. The corresponding form of the power spectral density (PSD) of the oscillation from Eq. (45) yields the Lorentzian response

$$S_x(\omega) = \frac{k_B T}{\pi m}\frac{\Gamma_0}{(\omega^2-\Omega_0^2)^2+\omega^2\Gamma_0^2}, \quad (46)$$

where $k_B$ is Boltzmann's constant and $T$ is the absolute temperature. In the underdamped regime ($\Gamma_0 \ll \Omega_0$), the particle undergoes high-quality mechanical oscillations with a quality factor $Q = \Omega_0/\Gamma_0$ at the oscillation frequency $\Omega_0$. In the experiment, the damping $\Gamma_0$ can be controlled by changing the pressure inside the vacuum chamber while $\Omega_0$ is determined by $\sqrt{\kappa_0/m}$, i.e. dependent on the optical power for trapping and mass of the particle. At atmospheric pressures, $\Gamma_0(\gg 2\Omega_0)$ is typically larger than the critical damping ($\Gamma_0 = 2\Omega_0$) and the PSD is similar to those observed in liquids, exhibiting a roll-off frequency at $\Omega_0$. At a gas pressure <10 mbar, the levitated particle is typically underdamped ($\Gamma_0 \ll \Omega_0$), where a strong resonant peak evolves at $\Omega_0$, which is the signature of high-quality mechanical oscillations. Levitated nanomechanical oscillators with high quality factors have been identified as promising candidates for ultrasensitive force and acceleration detectors[218,219], achieving a force sensitivity in the order of zepto-Newton in high vacuum [220].

A great driving force in the development of levitated optomechanics has been their potential for the realization of ground state cooling of nanomechanical oscillators. Integrating both sides over ω in Eq. (46) yields the mean square displacement (MSD)

$$\langle x^2 \rangle = \frac{k_B T_{CM}}{m\Omega_0^2} = \frac{k_B T_{CM}}{\kappa_0}. \qquad (47)$$

Thus, the centre-of-mass motion temperature $T_{CM}$ of the trapped particle is proportional to its MSD. In levitated systems, due to the presence of inertia, stochastic forces can be well controlled by the external optical forces as a means of artificial damping. Tremendous progress has been made in cooling $T_{CM}$ i.e. minimizing $\langle x^2 \rangle$ towards the ground state for the demonstration of the quantum behaviour of a mesoscopic object, using feedback cooling[191-195] and cavity cooling schemes[196-198], including the first demonstration of the ground state cooling [199]. The exquisite control achieved in these experiments also opens up a broad range of exciting new experiments for testing fundamental theories of physics both in the quantum and classical regimes as well as novel platforms towards developing next-generation sensing technologies (see, for example, Ref. [30, 200] for comprehensive reviews). Optical tweezers are capable of confining objects in vacuum ranging in size from tens of nanometres up to several micrometres, covering materials of silica[221,222], diamonds[206,223] and birefringent crystals, such as vaterite[224,225] using either single[191,226] or counter-propagating beam geometries[227]. The use of structured light in levitated optomechanics has found many applications so far[226-229], yet it has great potential to advance this emerging field. In this section, we will discuss recent developments in levitated optomechanics based on the use of structured light.

We first discuss the exchange of orbital angular momentum between light and matter in the underdamped regime. Mazilu *et al.* demonstrated an optically levitated silica microparticle (5 μm in diameter) placed within a Laguerre-Gaussian beam (optical power of 81.6 mW), where the beam's annular diameter is larger than the particle diameter in a vacuum environment (gas pressure of ~150 mbar)[202]. Through light scattering, the OAM of light is transferred to the levitated microparticle, which orbits around the annular beam profile with increasing angular velocity as the gas pressure is reduced. Figure 39 shows numerical simulations and experimental observations of particle trajectories for different topological charge ℓ. Both orbital radius and velocity increase with ℓ, where the outward inertial force (centrifugal in this case) can be counteracted by the radial trap only up to a maximal orbital velocity with ℓ up to 14 [Fig. 39(d,e)]. On the other hand, no orbital motion is observed for ℓ smaller than 5 as the diameter of the trapping beam is smaller than the diameter of the particle [Fig. 39(d,e)]. This leads to rotation of the particle at a stable particle position in the center of the beam.

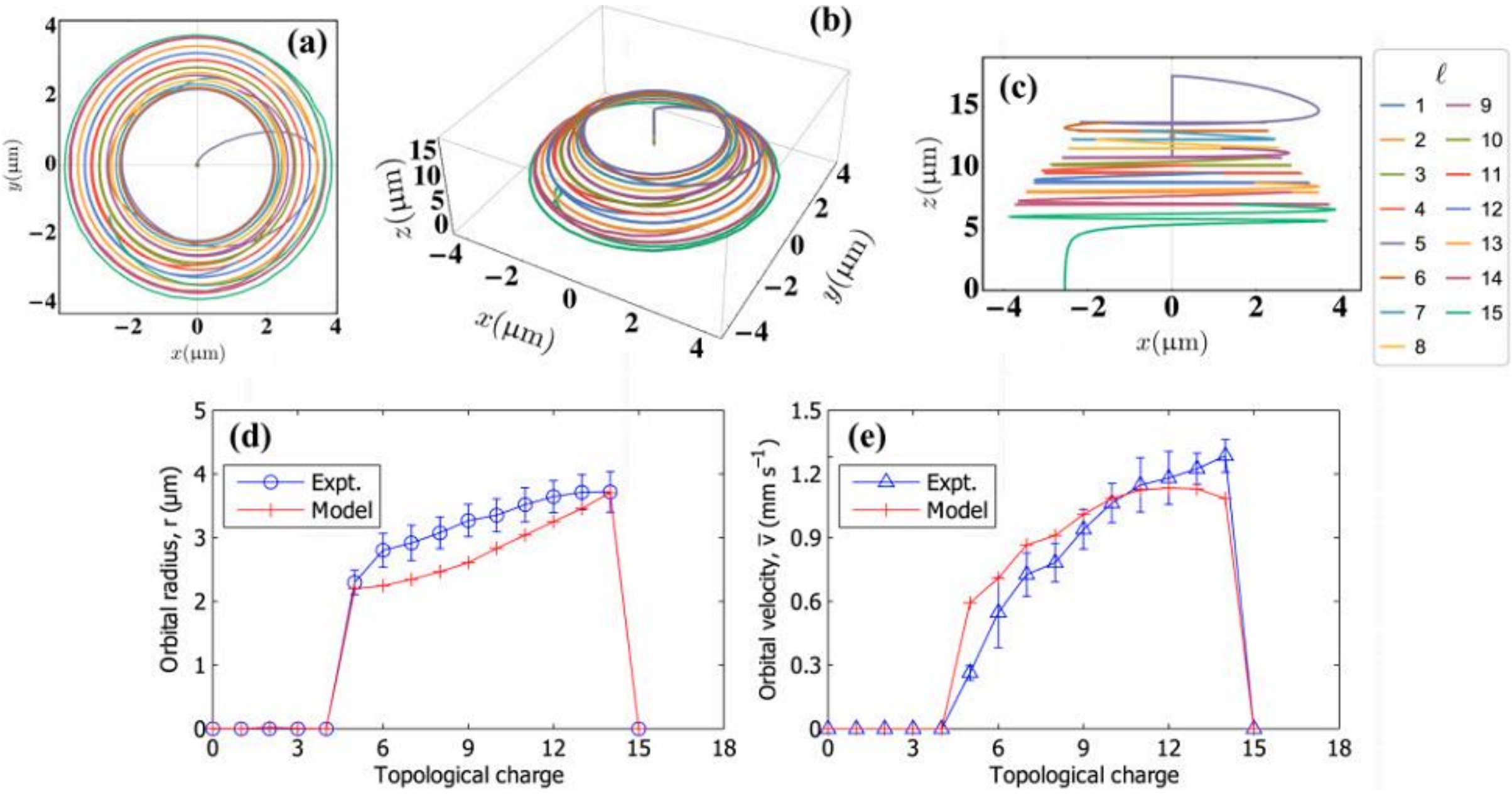

**Fig. 39** Particle trajectories of a silica microparticle levitated in LG beams with different topological charge. (a-c) Numerical simulations for ℓ from 4 up to 14. (d–e) Measured versus calculated orbital radius and velocity as a function of ℓ. Adapted from [202].

Levitated systems based on LG beams thus rely on a delicate balance between the optical gradient and scattering forces, with contributions from inertial forces and gravity. Importantly, the study demonstrated that there is a fundamental limit to the magnitude of OAM that can be transferred to a levitated particle in the underdamped regime.

It is interesting to consider the case of a particle placed within a perfect vortex beam (see section 4.1.3) [123,124], where the beam radius is independent of ℓ. Arita *et al.* explored vacuum trapping of a silica microparticle (5 μm in diameter) placed within an optical potential comprised of a perfect vortex beam, which is the Fourier transform of a Bessel beam[203]. In such an optical landscape, the trapped particles exhibit a complex 3D orbital motion that includes a periodic radial motion around the perfect vortex beam. Figure 40(a) shows the particle trajectories with different topological charge ($\ell$ = 3, 10, 30). To understand the complex motion, a 3D topography of the perfect vortex beam was investigated when $\ell$ = 15 [see Fig. 40(b)].

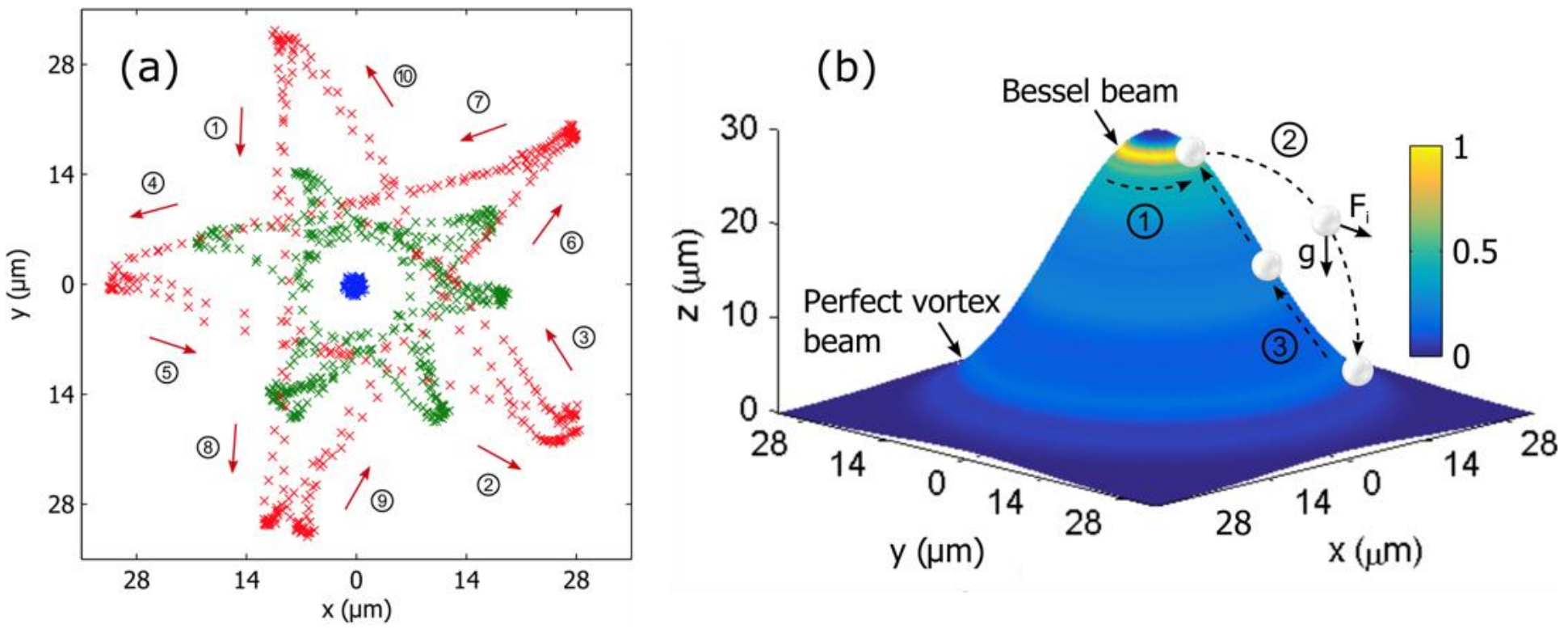

**Fig. 40** Perfect vortex traps in vacuum. (a) Particle trajectories with different topological charge $\ell$ = 3, 10 and 30 for blue, green and red crosses, respectively. Circled numbers indicate the order of the walked path when $\ell$ = 30 (red crosses). (b) Topography of the measured perfect vortex and Bessel beams ($\ell$ = 15) around the beam axis ($x = y = 0$ μm) with a schematic of particle motion. Adapted from Ref. [203].

Here we note that a perfect vortex beam is the Fourier transform of a Bessel beam. This means that over the three-dimensional space, the perfect vortex ($z$ = 0 μm) evolves on propagation to the Bessel beam within tens of micrometers ($z$ = 30 μm), which has implications for the particle dynamics. Figure 2(b) includes a schematic of the particle motion that depicts the particle trajectory: (1) trapped and set into rotation at the Bessel beam; (2) horizontally launched into free space and lands on the perfect vortex beam; (3) driven by both the scattering and gradient forces towards the Bessel beam, where the particle restarts its orbital cycle, but its radial excursion can be branched into different directions with different radial ranges depending on $\ell$ [see Fig. 40(a)]. In underdamped systems, perfect vortex and accompanied Bessel beams can produce a rich variety of orbital motions, where the optical gradient and scattering forces interplay with the inertial and gravitational forces acting on the trapped particle. OAM transfer in levitated systems offers novel perspectives for testing nonlinear dynamics of nanomechanical rotors and resonators beyond translational or harmonically bound motion [204] and for mesoscopic quantum studies analogous to quantum gases interacting with light fields possessing OAM[205].

Zhou *et al.* also proposed the use of LG beams for the levitation of nanodiamonds with nitrogen-vacancy (NV) centers in high vacuum [112]. Nanodiamonds are promising candidates for the realization of hybrid quantum systems featuring high-quality mechanical oscillations with long spin coherence time (~100 μs) of the electron spins of NV centers [206]. However, the light absorption of nanodiamonds from the trapping laser beam typically causes the thermal damage and prevents experiments in high vacuum. Trapping of nanodiamonds in a core-shell structure (a nanodiamonds-core in a less absorptive silica shell) with LG beams minimizes the overlap of the beam with the trapped nanodiamonds, thus can avoid significant heating of the system. Figure 41 shows the schematic of the dual beam trap with linearly polarized $LG_{03}$ beams and the geometry of the trap relative to the trapped core-shell particle in the focal region. The low-absorptive silica shell acts as a sample-holder, which interacts with the trapping beam and provides the scattering force for levitation. Thus, the nanodiamonds-core has negligible interaction with the annular shaped trapping beam.

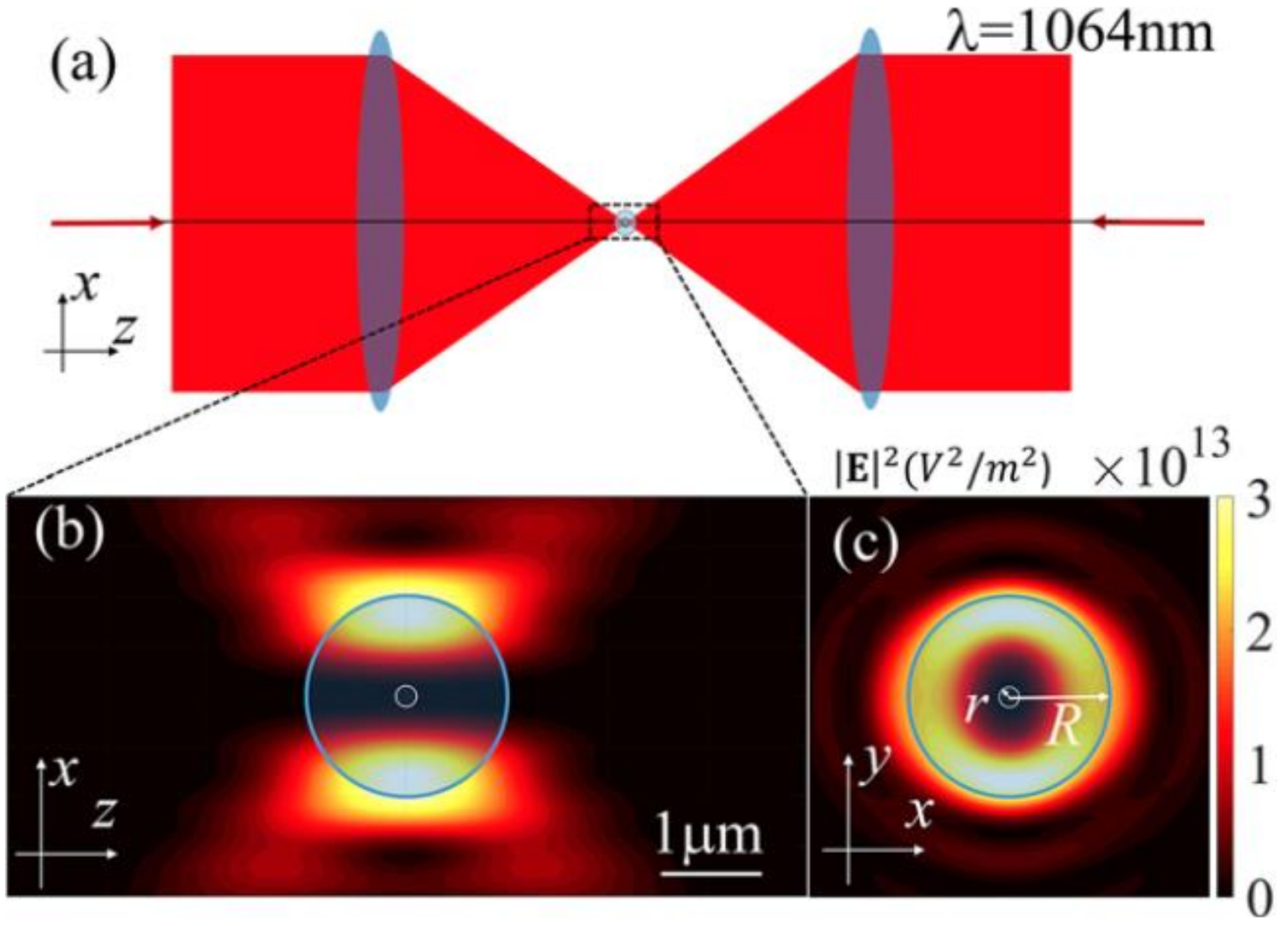

**Fig. 41** Trapping of nanodiamonds with linearly polarized $LG_{03}$ beams. (a) Schematic of the dual beam trap. (a) Geometry of the trap relative to a nanodiamonds-core (core radius $r$ = 100 nm) coated with a silica shell (shell radius $R$ = 1 μm) (b) from the front and (c) side views, respectively. Adapted from Ref. [112].

Their numerical study revealed that the stability of the trap and the rate of light absorption are dependent on the azimuthal index and the polarization of the LG beams, in which the azimuthally polarized Gaussian ($LG_{00}$) and the linearly polarized ($LG_{03}$) beams are the optimal choices to trap a core-shell particle while avoiding significant heating (<500 K) of nanodiamonds.

Other key areas of levitated optomechanics relevant to structured light include the use of non-standard Gaussian optical potentials for studying thermodynamics and non-equilibrium physics of small systems. Levitated systems are well suited to studying Brownian motion of a well isolated single particle with high temporal and spatial resolution. By modulating the spatial profile of the optical potential, the motion of the particle can be controlled, providing the potential for constructing new kinds of heat engines based on levitated nanoparticles [207,208] as well as new insights into stochastic processes in the underdamped regime[209].

With a double-well optical potential, Rondin *et al.* addressed the Kramers turnover problem[210]. It describes the transition between two local potential minima as the surrounding gas friction is varied. The double well potential was created by two tightly focused laser beams [Fig. 42(a)], where the intensity and exact relative position of the two foci determine the height of the energy barriers [Fig. 42(b,c)]. The transition rates between the two wells are determined by the potential profiles at the extrema and by the surrounding gas viscosity or pressure. Figure 42(d) shows the Kramers turnover rate at different gas pressures using an optically levitated nanoparticle, showing an excellent agreement with theory (solid line). Here the structured light field offers a novel experimental platform, where the levitated nanoparticle serves as a statistical simulator, allowing us to test fundamental theories of thermodynamics in small systems.

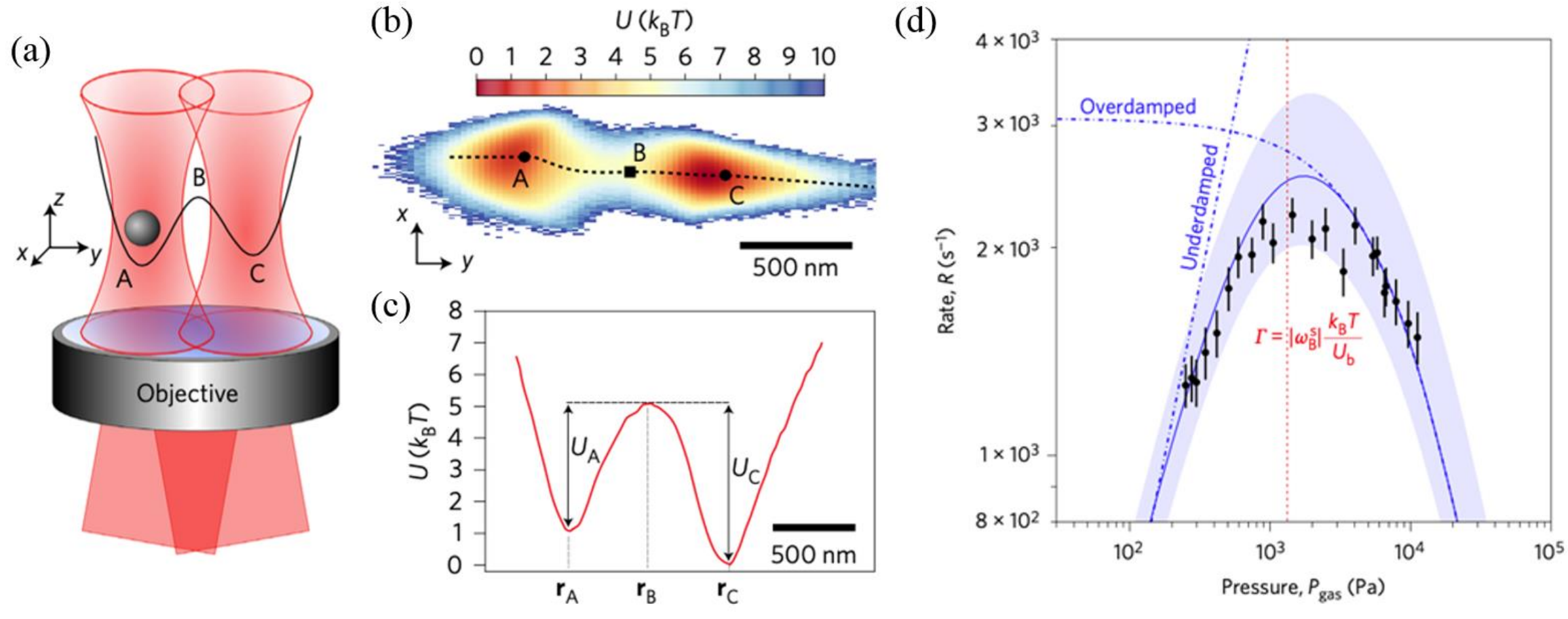

**Fig. 42** Testing Kramers turnover with a double-well potential. (a) Two focused infrared beams forming two potential wells (A and C) linked by a saddle point B. (b) Potential profile in the transverse (x–y) plane, where the dotted line represents the minimum energy path. (c) Potential energy profile with energy barriers $U_A$ and Uc at (A and C). (d) Kramers turnover rate depending on gas pressure. Adapted from Ref. [210].

The double-well potential can be used to study collective particle dynamics, e.g. optical binding[21] and forces between levitated particles [211]. Arita *et al.* demonstrated trapping and rotation of two microparticles in vacuum using a spatial light modulator-based approach, allowing individual control over the rotation direction and rate to each trap and the interparticle separation [212]. By trapping and rotating two vaterite birefringent microparticles with circularly polarized light, they observed macroscopic Raman-like modulation of the incident light field at the sum and difference frequencies with respect to the individual rotation rates. This first demonstration of optical interference between two microparticles in vacuum provided a strong foundation to explore optical binding of two micro-gyroscopes in the underdamped regime [213]. Figure 43(a) shows a stroboscopic image of two vaterite microparticles optically levitated and rotated in vacuum. Depending on the particle separation *R*, these particles are optically bound through light scattering, forming a shallow optical potential related to the collective center-of-mass motion of the two-particle system [dashed line in Fig. 43(b)]. The two-particle array was trapped in vacuum by the two foci of the trapping laser beams of 1070 nm with varied inter-particle separation *R* [Fig. 43(c)]. Figure 43(d) shows the experimental data from Arita *et al.*, which for the first time measured the optical binding strength depending on the particle separation *R*, using optically trapped and rotated microparticles [213].

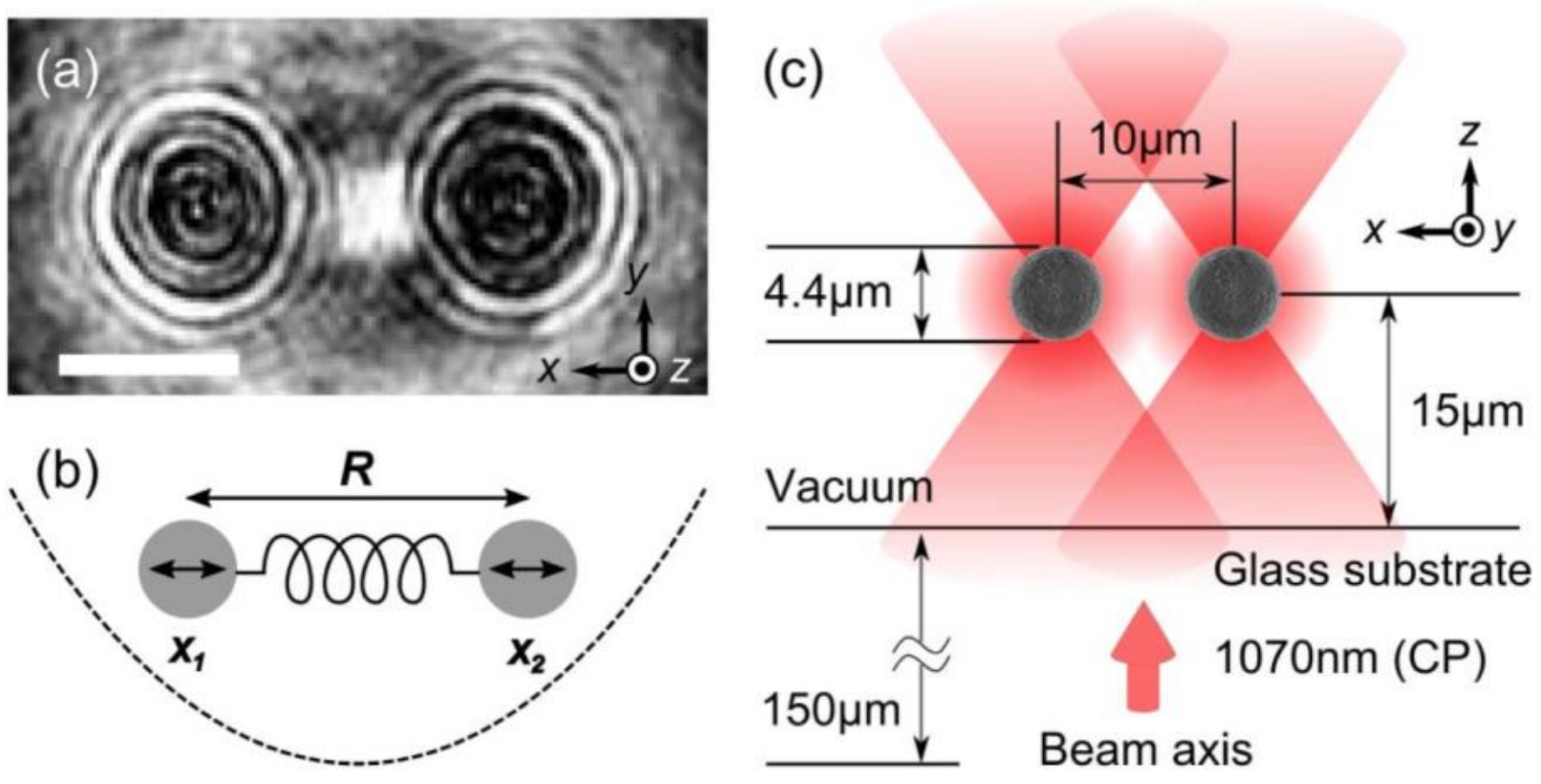

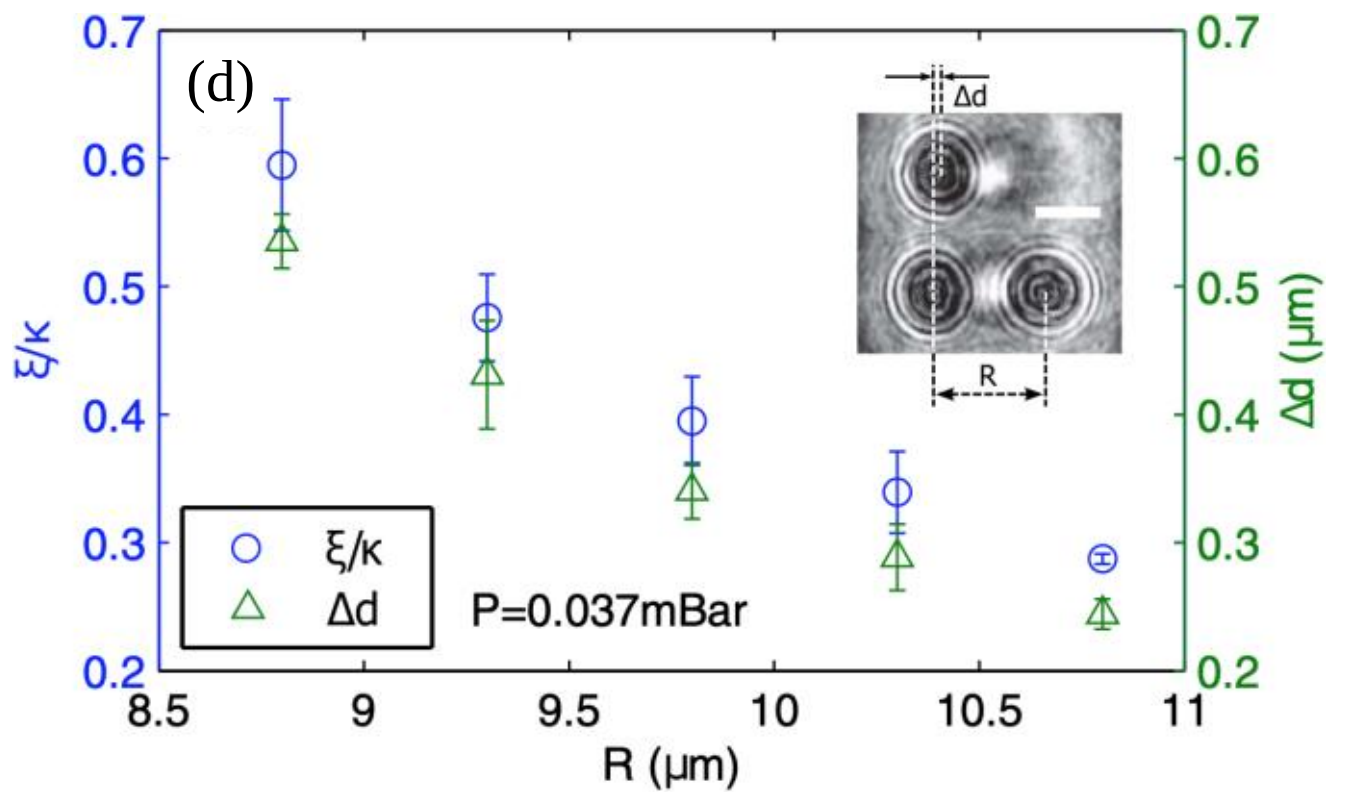

**Fig. 43** Optical binding between two rotating microparticles in vacuum. (a) Two vaterite birefringent microparticles optically levitated and rotated in vacuum with the scale bar of 5 μm. (b) Two normal modes of the bound array with the potential (dashed line) related to the center-of-mass motion of the two-particle system. (c) Double trap formed by two foci of the trapping laser beams (1070 nm). (d) Optical binding strength $\xi$ relative to the trap stiffness $\kappa$ of individual particles and particle displacement $\Delta d$ as a function of the particle separation $R$. Insets show the particle displacement $\Delta d$ induced by the presence of the other particle with $R$ = 9.8 μm. Adapted from Ref. [213].

Levitated multiparticle systems demonstrated in these experiments open up a promising scope for addressing mesoscopic quantum entanglement [214], and by including the rotational degrees of freedom, quantum/vacuum friction[215-217], provided interparticle cooling or sufficiently high rotation rates can be achieved. We note that, at the time of finalizing this manuscript, Svak *et al.* also demonstrated optical binding of multiple particles levitated in vacuum using the counter-propagating beam geometry[230].

### *4.6 Biomedical application of optical trapping with structured light*

In 1987, Ashkin demonstrated the trapping and manipulation of tobacco mosaic viruses and *Escherichia coli* bacteria using the single-beam optical tweezers[231]. A single cell had also been trapped in the single optical tweezers using infrared laser beam[18], as shown in Fig. 44 (a) and (b). Figure 44 (a,b) shows the division of yeast cell in the optical trap with an initial clump of (a) two cells divided into (b) six cells after an elapsed time of about 3 h. They also demonstrated the manipulation of particles within the cytoplasm of a spirogyra, and subcellular organelle[18,232]. Later on, the optical tweezers has been applied to study a series of single molecule protein, nucleic acid, and enzymes, and provide an important and fine tool in the single molecule arsenal[17, 233, 234]. There are a series of review articles on high-resolution optical tweezers and single molecule biophysics. Although the single beam optical tweezers have great contributions in biology studies, especially single molecule biophysics[234], the complex manipulation of the biological cells are in high demand. Here, we only focus on the application of structured light in optical tweezers, which also covers a broad area in biomedical applications of the optical tweezers. Holographic optical tweezers with structured light allows the manipulation and patterning of multiple cells both of the same or different species to form building blocks similar to primary tissue[235]. Fig.44 (c–d) shows the mouse embryonic and mesenchymal (arrow) stem cells trapped in the holographic optical tweezers with a structured beam. Fig. 44 (e–f) demonstrates the mouse primary calvarae cells (arrow) surrounded by embryonic stem cells[236]. These examples suggest that the optical tweezers with structured light has been playing important role in the single-/multiple-cell manipulation for tissue engineering and cell cluster studies.

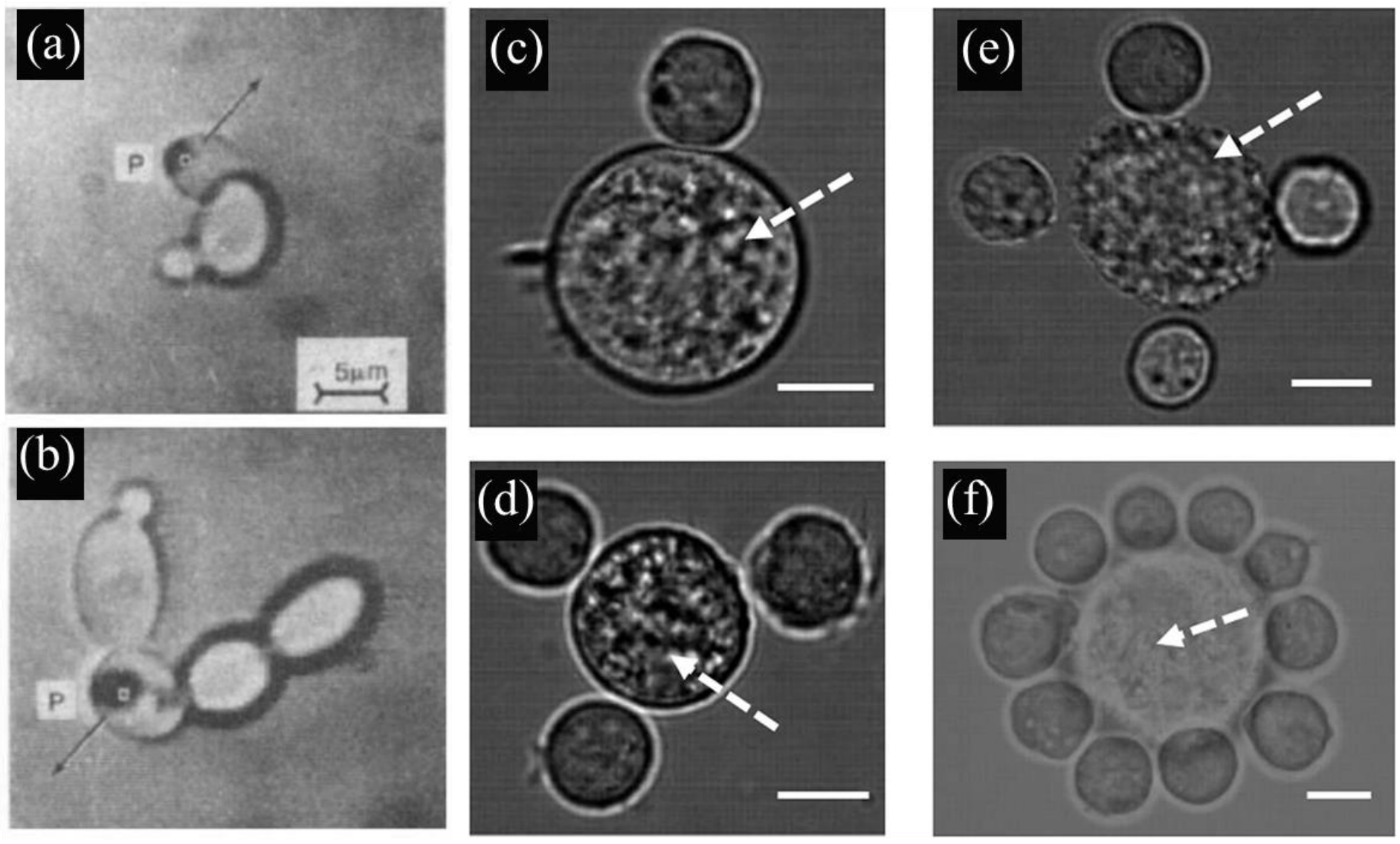

Fig. 44. Trapping of biological cells with structured beams. (a) Trapping of yeast cell in IR trap. (b) division of yeast cell in single trap. (c-f) Patterning of multiple cell types using holographic optical tweezers. (c–d). Mouse embryonic and mesenchymal (arrow) stem cells. (e–f). Mouse primary calvarae cells (arrow) and embryonic stem cells. Scale bar = 12 μm. (a) and (b) adapted from Ref. [18]. (c)- (f) adapted from Ref. [235].

It is worth to mention that the rod-shaped bacteria caught in the optical trap often align its orientation with the light propagation[236]. On occasion of bacteria cluster, it will be clearer to algin the rod-shaped bacteria parallel to the observation plane. The holographically shaped tug-of-war tweezers with elongated focus provides the possibility to orient the bacteria and better trapping stability over conventional dual-beam tweezers on the bacteria1[237]. Ultimately, the optical tweezers would be able to manipulate the cells in vivo[238]. Fig. 45 shows the optical trapping of red blood cells in the blood vessels in (a-d) mouse ear[239] and (e,f) the zebrafish [240]. The optical tweezers can either induce a clot inside the blood vessel or clear the jammed blood vessel with optical manipulation. With injected nanoparticle, the particle (green arrowhead) adhered to the endothelium of the caudal vein (blue dotted lines in Fig. 45 (e)) is pulled away from the endothelium into the fast blood flow (purple arrow) using optical tweezers (black crosshairs). At time 5.5 s an erythrocyte is drawn into the trap. This replaces the particle in the trap which is subsequently pulled back towards the original adhesion point of the endothelium. Four separate particles (numbered) are fished out of the blood flow and moved towards a sheltered region at the tip of the tail. The structured light may in future provide even complex manipulation of multiple particles in the patterned trajectories in vivo.

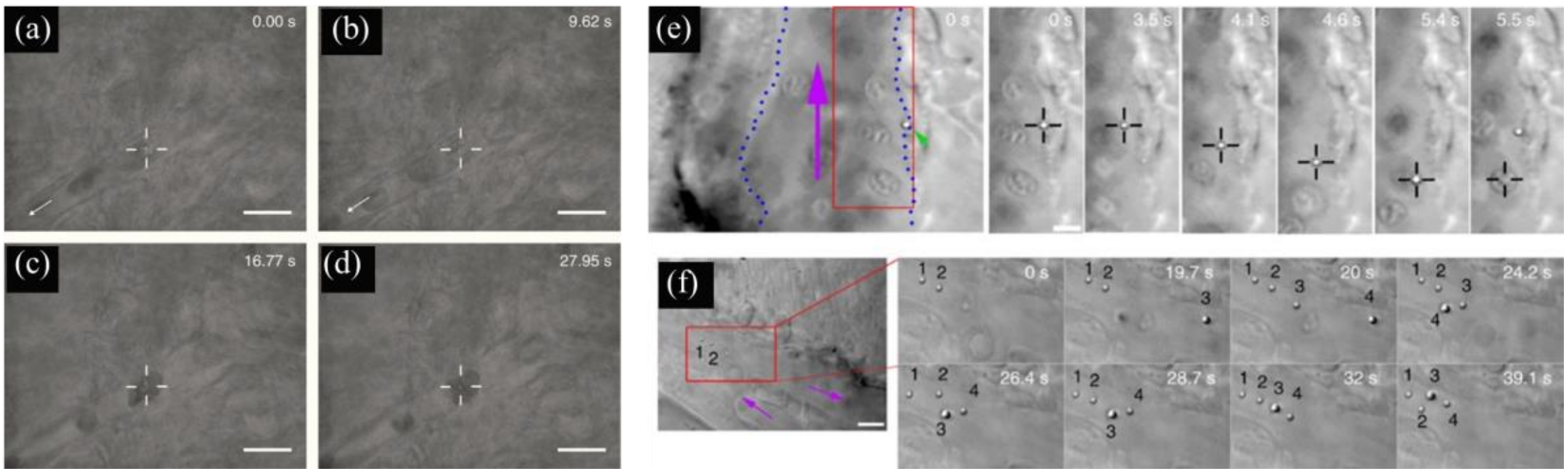

Fig.45 optical trapping of the red blood cells in vivo. (a-d) trap and manipulate the red blood cells in vivo in the ear blood vessel of the mouse. (a-d) adapted from Ref. [239]. (e,f) trap and manipulate the nanoparticle in vivo. Purple arrows indicate flow direction. Experiment was repeated at least 10 times. Scale bar, 5 μm. (e,f) adapted from Ref. [240].

Although the biological cells have a refractive index above that of the surrounding environment (positive polarizability), the refractive index contrast is very weak. Hence, the trapping of biological cells is less stable as compared with the standard polymer sphere. However, the cells may assume different geometry, i.e., rod-shape or irregular shape, holographic optical tweezers could provide multiple traps to hold the biological cells with weak refractive index contrast. The dual-trap holographic tweezers could hold two ends of a bacteria and align the bacteria horizontally (see schematics in Fig. 46(a), in contrast, the bacterium is aligned vertically in a single beam optical trap (schematics in Fig. 46(b)). Figure 46(c) shows the T-cell trapped in the single beam optical trap, which clearly suggests that the cell is rotating while the stage is moving. Figure 46(d) shows the position distribution of the cell (black) and a polymer sphere (red). The T-cell is more scattered around the tweezers in contrast to the polymer sphere. The holographic dual trap was able to hold the T-cell more stably than a single optical trap[241]. One benefit of using the multiple optical tweezers is to align the orientation of the cell for specific applications, e.g., super-resolution nanoscopy towards a certain direction[242]. This is another advantage to combine the holographic optical tweezers with structured light and the interdisciplinary methods, e.g., super-resolution microscopy and Raman spectroscopy.

Figure 46(e) shows the super-resolution image of the rod-shaped bacteria hold in a dual trap optical tweezers[242]. Noticeably, Figure 46(f) shows that the holographic tweezers also allow the study of bacteria cluster disassembly using the tug-of-war tweezers[237].

In summary, the optical trapping with structured light has been changing the style of optical manipulation in a more complicated fashion and extending the application in cell manipulation with finer possibilities.

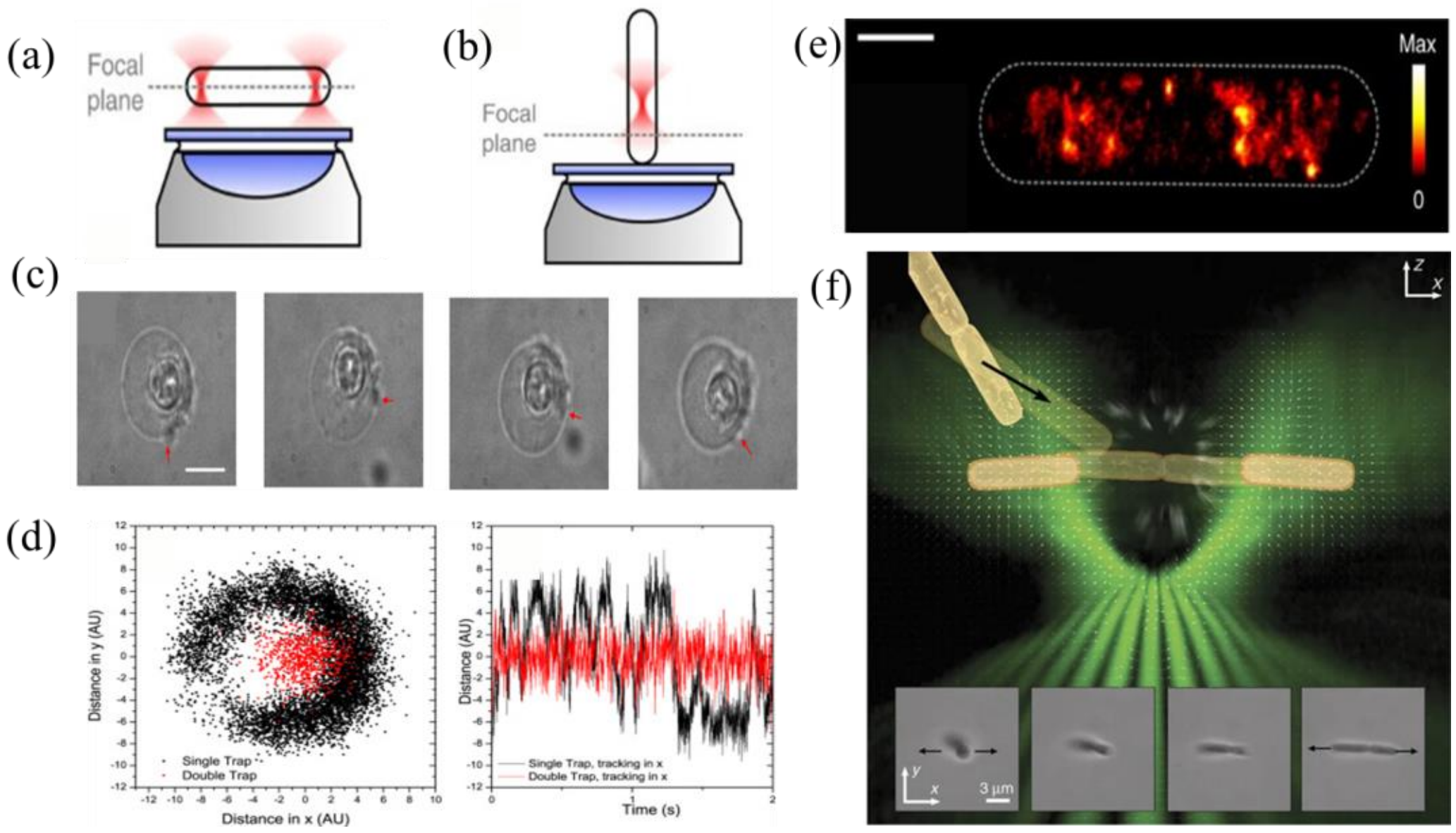

Fig.46 Strategies for stable optical trapping of rod-shaped bacteria. Schematics for (a) holographic dual-trap optical tweezers and (b) conventional single-trap optical tweezers. (c) The T-cell under a single beam optical tweezers experience rotation in the presence of stage motion. (d) the locations of a single cell (black) and a standard polymer sphere (red) in a single optical tweezers, and the positional traces (right). (e) The combination of *d*STORM and optical trapping allows isotropic super-resolution of 2D localization microscopy for each orientation of the rod-shaped bacterium. (f) The schematics of tug-of-war tweezers for the study of bacteria disassembly. Adapted from Refs. [242, 241, 237].

## 5 Conclusions and outlook

Structured light, custom light fields with tailored phase, intensity or polarization, can help us understand the nature of the light. It is hard to overstate the importance of the structured optical beams for the astonishingly wide range of applications. In this review, we have covered many seminal contributions and the recent advances in optical tweezers using structured light beams, from single to arrays of traps, and for both scalar and vector light beams. Since the invention of optical tweezers, the technique has evolved into a variety of forms, which are the most powerful and indispensable tools for the study of light-matter interactions at the micro and nanoscales. We have reviewed the remarkable progress that has been made in this field, including the optical trapping in suspension and air/vacuum, the optical transporting of metal nanoparticles and the biomedical application of optical trapping. It is clear that optical tweezers will evolve as novel beam shaping technologies evolve, for example, tractor beams[177-183], anomalous vortex beams[243, 244], partially coherent vortex beams[245,246] and grafted vortex beams[247,248], and structured light with photonic and plasmonic structures[249-253]. There is no doubt that new developments in the structured light beams will continuously boost the fields of optical manipulation[254].

Moreover, among the diverse optical trapping schemes discussed in this review, in our view, several topics have great potential to find exciting future applications, such as the optical trapping of metal particles[184-190] and chiral particles[52,53], vacuum levitation[191-230], structured light in waveguides[255-259], optical binding and other collective motions in structured light fields[111,186,260-262], quantum optomechanics[30], and optical trapping for multidisciplinary applications[32]. In the future, besides SLMs and DMDs, more flexible, efficient and much less expensive devices will be developed to produce structured beams, which can help to build the next generation of optical trapping technology. Furthermore, acoustic trapping[30] and plasmonic trapping[252] have developed rapidly in recent years, and trapping of nanoparticles with electron beams[263,264] has emerged as well, which can find promising applications in biosciences, biosensors and engineering as well.

*Acknowledgements*

We thank Prof. Kishan Dholakia for his instructive advice and help on the preparing of this manuscript. Y.Y thank Dr. Leiming Zhou for the helpful discussion. This work was supported by the National Natural Science Foundation of China (Grant No. 11874102 and 61975047), the Sichuan Province Science and Technology Support Program (2020JDRC0006), and the Fundamental Research Funds for the Central Universities (ZYGX2019J102). M.C. and Y.A. thank the UK Engineering and Physical Sciences Research Council for funding.